\pdfoutput=1

\documentclass[12pt,a4paper]{article}

\usepackage{ifthen} 
\newboolean{pdflatex}
\setboolean{pdflatex}{false} 

\newboolean{articletitles}
\setboolean{articletitles}{true} 

\newboolean{uprightparticles}
\setboolean{uprightparticles}{false} 

\def\paperauthors{Vladimir Gligorov, Stephanie Hansmann-Menzemer, Conor Fitzpatrick} 
\def\paperasciititle{HLT1 Technology Comparison Document} 
\def\papertitle{A Comparison of CPU and GPU implementations for the LHCb Experiment Run 3 Trigger} 
\def\paperkeywords{{High Energy Physics}, {LHCb}} 
\def\papercopyright{\the\year\ CERN for the benefit of the LHCb collaboration} 
\def\paperlicence{CC-BY-4.0 licence}
\def\paperlicenceurl{https://creativecommons.org/licenses/by/4.0/}

\usepackage[top=1in, bottom=1.25in, left=1in, right=1in]{geometry}

\usepackage{microtype}
\usepackage{lineno}  
\usepackage{xspace} 
\usepackage{caption} 

\usepackage{graphicx}  
\usepackage{color}
\usepackage{colortbl}
\graphicspath{{./figs/}} 

\usepackage{amsmath} 
\usepackage{amssymb}
\usepackage{amsfonts}
\usepackage{upgreek} 

\usepackage{multirow}

\newcommand*\patchAmsMathEnvironmentForLineno[1]{%
\expandafter\let\csname old#1\expandafter\endcsname\csname #1\endcsname
\expandafter\let\csname oldend#1\expandafter\endcsname\csname
end#1\endcsname
 \renewenvironment{#1}%
   {\linenomath\csname old#1\endcsname}%
   {\csname oldend#1\endcsname\endlinenomath}%
}
\newcommand*\patchBothAmsMathEnvironmentsForLineno[1]{%
  \patchAmsMathEnvironmentForLineno{#1}%
  \patchAmsMathEnvironmentForLineno{#1*}%
}
\AtBeginDocument{%
\patchBothAmsMathEnvironmentsForLineno{equation}%
\patchBothAmsMathEnvironmentsForLineno{align}%
\patchBothAmsMathEnvironmentsForLineno{flalign}%
\patchBothAmsMathEnvironmentsForLineno{alignat}%
\patchBothAmsMathEnvironmentsForLineno{gather}%
\patchBothAmsMathEnvironmentsForLineno{multline}%
\patchBothAmsMathEnvironmentsForLineno{eqnarray}%
}

\usepackage{hyperxmp}

\usepackage[pdftex,
            pdfauthor={\paperauthors},
            pdftitle={\paperasciititle},
            pdfkeywords={\paperkeywords},
            pdfcopyright={Copyright (C) \papercopyright},
            pdflicenseurl={\paperlicenceurl}]{hyperref}

\usepackage[colorinlistoftodos,textsize=scriptsize]{todonotes}

\usepackage[all]{hypcap} 

\usepackage{xspace} 
\usepackage{upgreek}

\def\MagUp {\mbox{\em Mag\kern -0.05em Up}\xspace}

\ifthenelse{\boolean{uprightparticles}}%
{

 \def\PDelta      {\ensuremath{\Delta}\xspace}                 
 \def\PXi         {\ensuremath{\Xi}\xspace}                 
 \def\PLambda     {\ensuremath{\Lambda}\xspace}                 
 \def\PSigma      {\ensuremath{\Sigma}\xspace}                 
 \def\POmega      {\ensuremath{\Omega}\xspace}                 
 \def\PUpsilon    {\ensuremath{\Upsilon}\xspace}

 \def\PB      {\ensuremath{\mathrm{B}}\xspace}                 
                  
 \def\PD      {\ensuremath{\mathrm{D}}\xspace}

 \def\PK      {\ensuremath{\mathrm{K}}\xspace}

 \def\Pi      {\ensuremath{\mathrm{i}}\xspace}

 \def\Ps      {\ensuremath{\mathrm{s}}\xspace}

 \def\thebaroffset{0.0em}
}
{

 \mathchardef\PDelta="7101
 \mathchardef\PXi="7104
 \mathchardef\PLambda="7103
 \mathchardef\PSigma="7106
 \mathchardef\POmega="710A
 \mathchardef\PUpsilon="7107
                  
 \def\PB      {\ensuremath{B}\xspace}                 
                  
 \def\PD      {\ensuremath{D}\xspace}

 \def\PK      {\ensuremath{K}\xspace}

 \def\Pi      {\ensuremath{i}\xspace}

 \def\Ps      {\ensuremath{s}\xspace}

 \def\thebaroffset{0.18em}
}
\newcommand{\offsetoverline}[2][\thebaroffset]{\kern #1\overline{\kern -#1 #2}}%

\makeatletter
\ifcase \@ptsize \relax
  \newcommand{\miniscule}{\@setfontsize\miniscule{4}{5}}
\or
  \newcommand{\miniscule}{\@setfontsize\miniscule{5}{6}}
\or
  \newcommand{\miniscule}{\@setfontsize\miniscule{5}{6}}
\fi
\makeatother

\DeclareRobustCommand{\optbar}[1]{\shortstack{{\miniscule (\rule[.5ex]{1.25em}{.18mm})}
  \\ [-.7ex] $#1$}}

\def\squark    {{\ensuremath{\Ps}}\xspace}

\def\KorKbar {\kern \thebaroffset\optbar{\kern -\thebaroffset \PK}{}\xspace}

\def\D       {{\ensuremath{\PD}}\xspace}

\def\DorDbar {\kern \thebaroffset\optbar{\kern -\thebaroffset \PD}\xspace}

\def\Dp      {{\ensuremath{\D^+}}\xspace}
\def\Dm      {{\ensuremath{\D^-}}\xspace}

\def\DpDm    {\ensuremath{\Dp {\kern -0.16em \Dm}}\xspace}

\def\B       {{\ensuremath{\PB}}\xspace}

\def\BorBbar {\kern \thebaroffset\optbar{\kern -\thebaroffset \PB}\xspace}

\def\Bd      {{\ensuremath{\B^0}}\xspace}

\def\BdorBdbar {\kern \thebaroffset\optbar{\kern -\thebaroffset \Bd}\xspace}

\def\Bs      {{\ensuremath{\B^0_\squark}}\xspace}

\def\BsorBsbar {\kern \thebaroffset\optbar{\kern -\thebaroffset \Bs}\xspace}

\def\Y#1S{\ensuremath{\PUpsilon{(#1S)}}\xspace}

\def\LorLbar     {\kern \thebaroffset\optbar{\kern -\thebaroffset \PLambda}\xspace}

\def\AT#1     {\ensuremath{A_{\mathrm{T}}^{#1}}\xspace}           

\def\C#1      {\ensuremath{\mathcal{C}_{#1}}\xspace}                       
\def\Cp#1     {\ensuremath{\mathcal{C}_{#1}^{'}}\xspace}                    
\def\Ceff#1   {\ensuremath{\mathcal{C}_{#1}^{\mathrm{(eff)}}}\xspace}        
\def\Cpeff#1  {\ensuremath{\mathcal{C}_{#1}^{'\mathrm{(eff)}}}\xspace}       
\def\Ope#1    {\ensuremath{\mathcal{O}_{#1}}\xspace}                       
\def\Opep#1   {\ensuremath{\mathcal{O}_{#1}^{'}}\xspace}                    

\newcommand{\aunit}[1]{\ensuremath{\text{\,#1}}}       

\newcommand{\tev}{\aunit{Te\kern -0.1em V}\xspace}
\newcommand{\gev}{\aunit{Ge\kern -0.1em V}\xspace}
\newcommand{\mev}{\aunit{Me\kern -0.1em V}\xspace}
\newcommand{\kev}{\aunit{ke\kern -0.1em V}\xspace}
\newcommand{\ev}{\aunit{e\kern -0.1em V}\xspace}
\newcommand{\mevc}{\ensuremath{\aunit{Me\kern -0.1em V\!/}c}\xspace}
\newcommand{\gevc}{\ensuremath{\aunit{Ge\kern -0.1em V\!/}c}\xspace}
\newcommand{\mevcc}{\ensuremath{\aunit{Me\kern -0.1em V\!/}c^2}\xspace}
\newcommand{\gevcc}{\ensuremath{\aunit{Ge\kern -0.1em V\!/}c^2}\xspace}

\def\gsim{{~\raise.15em\hbox{$>$}\kern-.85em
          \lower.35em\hbox{$\sim$}~}\xspace}
\def\lsim{{~\raise.15em\hbox{$<$}\kern-.85em
          \lower.35em\hbox{$\sim$}~}\xspace}

\def\pt         {\ensuremath{p_{\mathrm{T}}}\xspace}

\def\tell1  {TELL1\xspace}
\def\ukl1   {UKL1\xspace}

\usepackage{cite} 
\usepackage{mciteplus}

\usepackage{longtable} 

\begin{document}

\renewcommand{\thefootnote}{\fnsymbol{footnote}}
\setcounter{footnote}{1}


\begin{titlepage}

\vspace*{-1.5cm}
\centerline{\large EUROPEAN ORGANIZATION FOR NUCLEAR RESEARCH (CERN)}
\vspace*{1.5cm}
\noindent
\begin{tabular*}{\linewidth}{lc@{\extracolsep{\fill}}r@{\extracolsep{0pt}}}
\ifthenelse{\boolean{pdflatex}}
{\vspace*{-1.2cm}\mbox{\!\!\!\includegraphics[width=.14\textwidth]{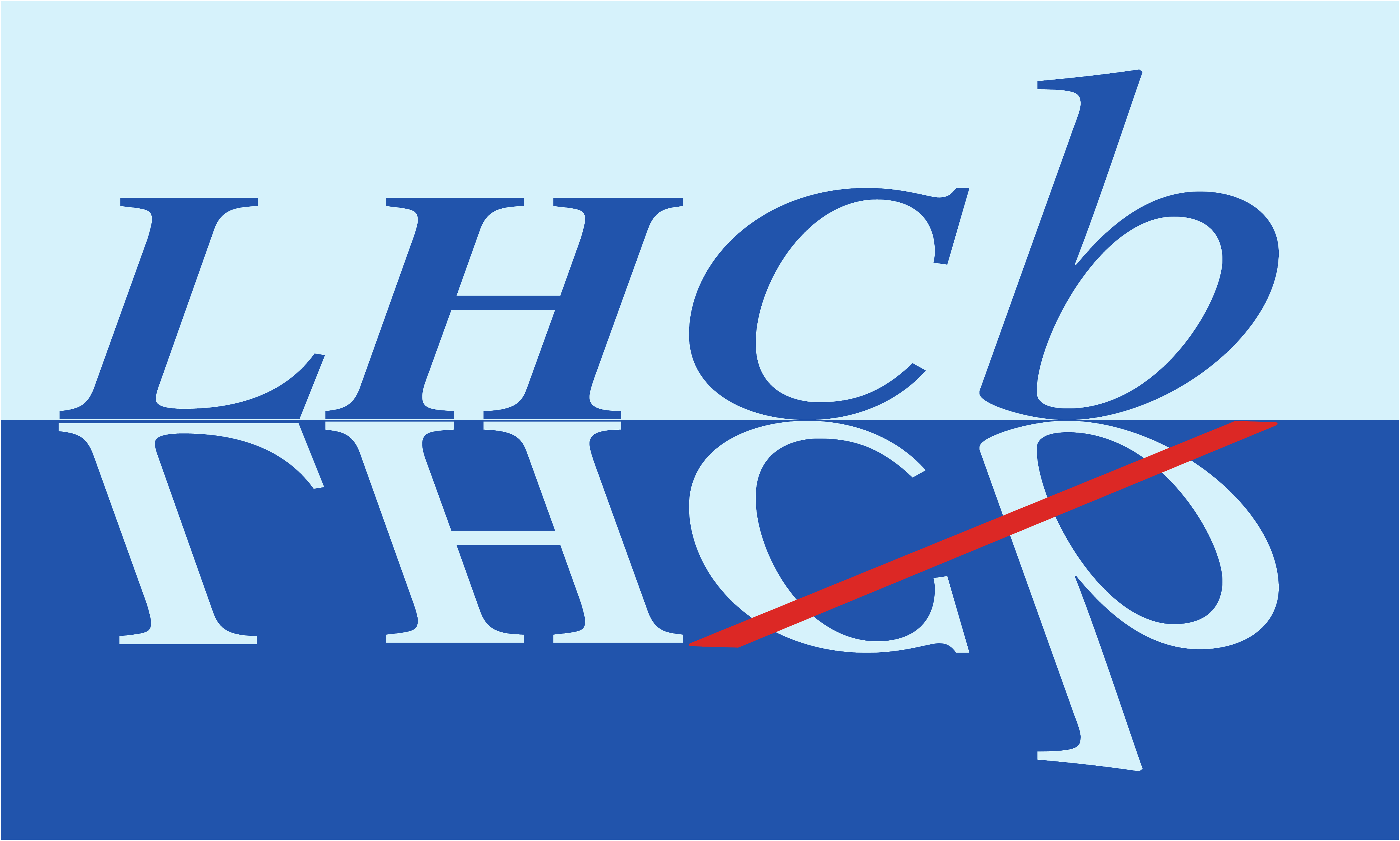}} & &}%
{\vspace*{-1.2cm}\mbox{\!\!\!\includegraphics[width=.12\textwidth]{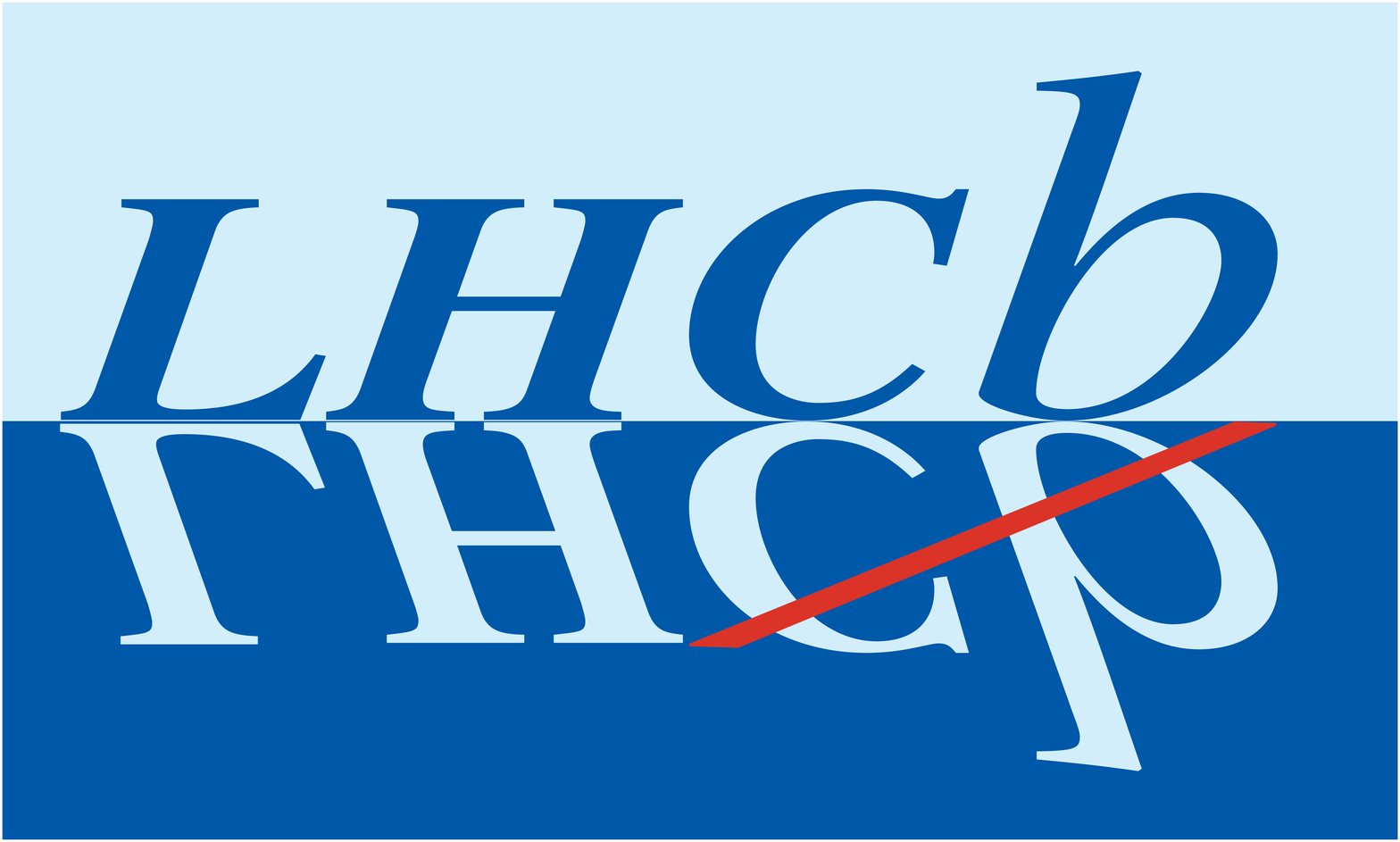}} & &}
 \\
  & & LHCb-DP-2021-003 \\
 & & \today \\ 
\hline
\end{tabular*}

\vspace*{4.0cm}

{\normalfont\bfseries\boldmath\huge
\begin{center}
  \papertitle
\end{center}
}

\vspace*{2.0cm}

\begin{center}
    Authors are listed on the following pages
\end{center}

\vspace*{2.0cm}

\begin{abstract}
  \noindent
The LHCb experiment at CERN is undergoing an upgrade in preparation for the Run 3 data taking period of the LHC. As part of this upgrade the trigger is moving to a fully software implementation operating at the LHC bunch crossing rate. We present an evaluation of a CPU-based and a GPU-based implementation of the first stage of the High Level Trigger. After a detailed comparison both options are found to be viable.  This document summarizes the performance and implementation details of these options, the outcome of which has led to the choice of the GPU-based implementation as the baseline. 
  
\end{abstract}

\vspace*{1.0cm}
\begin{center}
    Published in Computing Software for Big Science 6, Article number: 1 (2022)
\end{center}

\vspace*{2.0cm}

{\footnotesize 
\centerline{\copyright~\papercopyright. \href{\paperlicenceurl}{\paperlicence}.}}

\begin{center}

\begin
{flushleft}
\small
R.~Aaij$^{21}$,
M.~Adinolfi$^{34}$,
S.~Aiola$^{17}$,
S.~Akar$^{43}$,
J.~Albrecht$^{10}$,
M.~Alexander$^{39}$,
S.~Amato$^{1}$,
Y.~Amhis$^{8}$,
F.~Archilli$^{12}$,
M. ~Bala$^{25}$
% ...

~\\
{\footnotesize \it
$^{1}$Universidade Federal do Rio de Janeiro (UFRJ), Rio de Janeiro, Brazil\\
% ...
$^{8}$Universit{\'e} Paris-Saclay, CNRS/IN2P3, IJCLab, Orsay, France\\
% ...
$^{10}$Fakult{\"a}t Physik, Technische Universit{\"a}t Dortmund, Dortmund, Germany\\
% ...
$^{12}$Physikalisches Institut, Ruprecht-Karls-Universit{\"a}t Heidelberg, Heidelberg, Germany\\
% ...
$^{17}$INFN Sezione di Milano, Milano, Italy\\
% ...
$^{21}$Nikhef National Institute for Subatomic Physics, Amsterdam, Netherlands\\
% ...
$^{25}$National Center for Nuclear Research (NCBJ), Warsaw, Poland\\
% ...
$^{34}$H.H. Wills Physics Laboratory, University of Bristol, Bristol, United Kingdom\\
% ...
$^{39}$School of Physics and Astronomy, University of Glasgow, Glasgow, United Kingdom\\
% ...
$^{43}$University of Cincinnati, Cincinnati, OH, United States\\
% ...
\medskip
}
\end{flushleft}

\bigskip
\end{center}

\vspace{\fill}

\end{titlepage}

\pagestyle{empty}  


\newpage
\setcounter{page}{2}
\mbox{~}


\renewcommand{\thefootnote}{\arabic{footnote}}
\setcounter{footnote}{0}

\tableofcontents
\cleardoublepage


\pagestyle{plain} 
\setcounter{page}{1}
\pagenumbering{arabic}


%

\section{Introduction}
\label{sec:Introduction}
The LHCb experiment is a general purpose spectrometer instrumented in the forward direction based at the LHC.~\cite{LHCb-DP-2008-001} Although optimized for the study of hadrons containing beauty and charm quarks, LHCb's physics programme gradually expanded over the course of its first datataking period\footnote{The first datataking period was broken into two ``runs'', with Run~1 taking place from 2009 to 2013 and Run~2 taking place from 2015 to 2018.}, taking in Kaon physics, prompt spectroscopy, Electroweak physics, and searches for putative heavy and long-lived exotic particles beyond the Standard Model.\\
~\\
The LHC provides a non-empty bunch crossing rate of up to 30~MHz, with a variable number of inelastic proton-proton interactions per bunch crossing ($\mu$) which can be adjusted to suit the physics programme of each experiment. During Runs~1~and~2 LHCb took data with a $\mu$ of between $1.1$ and $2.5$, corresponding to an instantaneous luminosity of around $4\cdot 10^{32}$~cm$^{-2}$s$^{-1}$. A fixed-latency hardware trigger (L0), based on calorimeter and muon system information, reduced the 30~MHz LHCb bunch crossing rate to $\sim\!\!\! 1$~MHz at which the detector readout operated. These events were then passed to a two stage asynchronous High Level Trigger (HLT) system entirely implemented in software.
In the first stage, HLT1, a partial event reconstruction selected events based on inclusive signatures, reducing the event rate by an order of magnitude. Accepted events were stored using a 11~PB disk buffer in order to align and calibrate the detector. After this procedure, events were passed to the second stage, HLT2, which had sufficient computing resources to run the full offline-quality detector reconstruction. A multitude of dedicated selections deployed in HLT2 reduced the data to an output rate of 12.5~kHz using a combination of full and reduced event formats \cite{LHCb-DP-2019-001}.\\
~\\
The ambitious goal of the upgraded LHCb experiment in Run~3 (i.e. starting from 2022) is to 
remove the hardware trigger and directly process the full 30 MHz of data at an increased  luminosity of $2\cdot 10^{33}$~cm$^{-2}$s$^{-1}$ in the HLT. At this luminosity, corresponding to a $\mu$ of around $6$, it is no longer possible to efficiently identify bunch crossings of interest based purely on calorimeter and muon system information, as there is too much QCD background generated by the pileup $pp$ collisions.~\cite{Akar:1700272} It is necessary to instead fully read the detector out for every bunch crossing and fully perform the real-time processing in the HLT. This allows the much more sophisticated selections, in particular selections based on charged particle trajectories reconstructed in the whole of LHCb's tracker system, to be deployed already in HLT1. Such a fully software trigger will not only allow LHCb to maintain its Run~1~and~2 efficiencies for muonic signatures, but will lead to a factor two improvement in efficiency for hadronic signatures compared to the calorimeter-based L0 trigger despite the harsher Run~3 environment.  \\
~\\
LHCb's Run~3 datataking conditions imply a data volume of around 32 Terabits (Tb) per second, comparable to what the ATLAS~\cite{CERN-LHCC-2017-020} and CMS~\cite{CERN-LHCC-2017-014} software triggers will be required to process during High-Luminosity LHC running from 2027 onwards. The design and delivery of LHCb's high-level trigger is therefore also one of the biggest computing challenges that the field of High Energy Physics is facing today. The closest current parallel to LHCb's system is that of the ALICE experiment~\cite{Buncic:2011297}, which will also operate a triggerless readout in Run~3 with an objective to reduce an input data rate of roughly 10~Tb/s to a manageable amount by performing a full detector reconstruction in real-time.\\
~\\
The concept of a pure CPU-based solution for this approach was reviewed during the preparation of LHCb's Trigger and Online TDR~\cite{LHCb-TDR-016} in 2014 followed by a systematic rewrite of the LHCb trigger software infrastructure, which enabled data taking in these conditions. In parallel, R\&D efforts have explored a possible usage of GPUs for HLT1~\cite{Aaij:2019zbu}, referred to as the hybrid approach in the following. An intensive effort was launched to demonstrate if such a hybrid system could be delivered in time for Run~3 and concluded in a positive review of its TDR ~\cite{LHCb-TDR-021} in early 2020. After a detailed comparison of both options the collaboration selected the hybrid approach as the new Run~3 baseline. This decision parallels that of ALICE, which pioneered the use of GPUs among LHC experiments during the last decade~\cite{Gorbunov:1605120} and whose Run~3 triggerless readout and full real-time reconstruction mentioned earlier will be mainly implemented on GPUs.\\
~\\
This document compares both options and summarizes the salient points which led to the decision. It reflects the status of both implementations at the time the decision was made, in April 2020. Further significant improvements~\cite{LHCB-FIGURE-2020-014} in both throughput and physics performance, which will not be discussed in this document, have been achieved since then.\\
\\
This document is structured as follows. In Section~1, the DAQ and HLT architecture of both systems is summarized. Section~\ref{sec:BoundaryConditions} describes the boundary conditions within which the implemented HLT1 triggers must operate, including available financial resources and operational constraints.
Section~\ref{sec:ImplementedHLT1} describes the HLT1 trigger sequence and algorithms, which are the basis of the performance comparison. Section~\ref{sec:Throughput} summarizes the performance of the two architectures in terms of throughput, while Section~\ref{sec:PhysicsPerf} presents their physics performance. Section~\ref{sec:CostBenefit} combines the performance assessment into a cost-benefit analysis.

\subsection{DAQ and HLT architectures}
\label{sec:sub:DAQEvolution}
LHCb's DAQ and event building (EB) infrastructure is described in the Trigger and Online TDR~\cite{LHCb-TDR-016}. The full detector is read out for all LHC bunch-crossings, and information from subdetectors is received by around $500$ custom backend FPGA ``TELL40'' boards hosted in a farm of EB servers, with three TELL40 boards per server. These subdetector fragments are then assembled into ``events'', with one event corresponding to one LHC bunch crossing, and sent to the HLT for processing. Both the event building and HLT are fully asynchronous and no  latency requirements exist in the system. 

The following two HLT processing architectures are under consideration.
\begin{enumerate}
    \item \textbf{CPU-only}, which implements both HLT1 and HLT2 using Event Filter Farm (EFF) CPU servers. 
    \item \textbf{Hybrid}, which implements HLT1 using GPU cards installed in the EB servers with the CPU-based HLT2 running as before on the EFF.
\end{enumerate}
The HLT2 software and processing architecture are identical in both cases.

\subsubsection{CPU-only architecture}
\begin{figure}[ht]
\begin{center}
\includegraphics[width=0.9\textwidth]{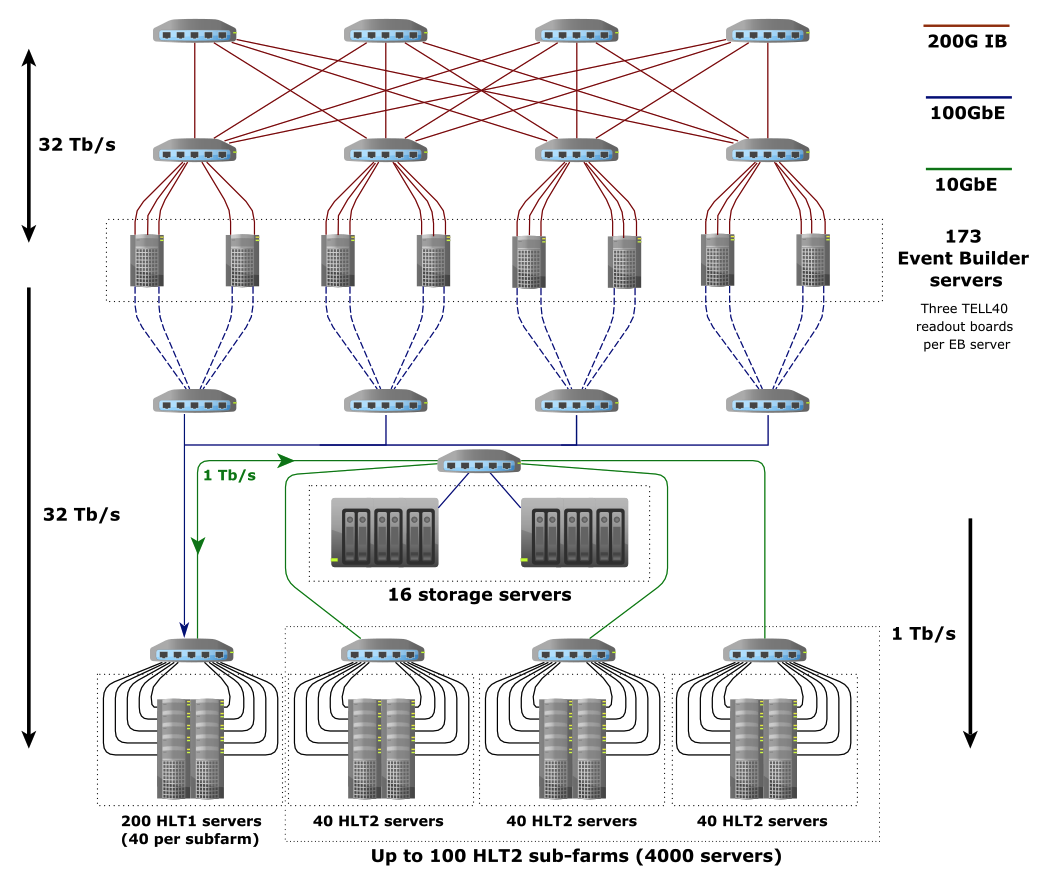}
\caption{CPU-only architecture of the Run~3 DAQ, including the Event Builder, the Event Filter Farm and dedicated storage servers for the disk buffer. The network between the storage servers and HLT1 servers (leftmost green line) is bidirectional, allowing the HLT1 servers to be used for HLT2 processing when there are no LHC collisions happening. The label ``200G IB'' refers to the Infiniband link between the detector and EB servers, while ``100GbE'' and ``10GbE'' refer to Ethernet links of 100~Gb/s and 10~Gb/s, respectively. }
\label{fig:eb_baseline}
\end{center}
\end{figure}

The CPU-only architecture is illustrated in Fig.~\ref{fig:eb_baseline}. 
Briefly, it consists of 
\begin{itemize}
    \item A set of custom FPGA cards, called TELL40, which together receive on average 32~Tb/s of data from LHCb's subdetectors;
    \item A set of EB servers which host the TELL40 cards and implement a network protocol to bring the subdetector data fragments produced in a single LHC bunch crossing (``event'' in LHCb nomenclature) together and then group  O(1000) of these events
    into multi-event packets (MEPs) to minimize I/O overheads further down the line. The EB servers will be equipped with 32-core AMD EPYC (7502) CPUs;
    \item An EFF illustrated on the bottom left of Fig.~\ref{fig:eb_baseline} which receives MEPs from the EB and executes HLT1. The memory available in the EB servers and HLT1 EFF nodes allows data to be buffered for $O(20)$ seconds in case of temporary network or processing issues. The HLT1 EFF servers are assumed to be equipped with the same 32-core AMD EPYC (7502) CPUs as the EB servers. When LHCb is not taking data this HLT1 EFF can also receive events from the disk servers (described below) and run the HLT2 process on them;
    \item A high performance network, including high-speed network interface cards (NICs) located in the EB servers and EFF nodes as well as a large switch, connects the EB and HLT1 EFF and allows to transmit at a rate of 32~Tb/s.
    \item The HLT1 process reduces the event rate by a factor of between 30-60, and an array of disk servers buffers this HLT1 output data while the detector alignment and calibration are performed in quasi-real-time. This disk buffer, whose size is a tunable parameter of the system as discussed later in Section~\ref{sec:CostBenefit}, allows HLT1 output data to be buffered for $O(100)$ hours in case of problems with the alignment and calibration which require specialist intervention;
    \item A second EFF  receives events from the disk servers once the alignment and calibration constants are available and runs the HLT2 process on them. Because of limitations in network bandwidth this second EFF cannot be used to process HLT1.
\end{itemize}
Advances in server technology have permitted a substantial reduction of the number of servers required by the EB, from 500 in the TDR~\cite{LHCb-TDR-016} to around 173. This allows to host three TELL40 cards per EB server rather than the one card foreseen in the TDR. This improvement means that the EB will be much more compact and as a consequence easier to upgrade in future.

\subsubsection{Hybrid architecture}

\begin{figure}[ht]
\begin{center}
\includegraphics[width=0.9\textwidth]{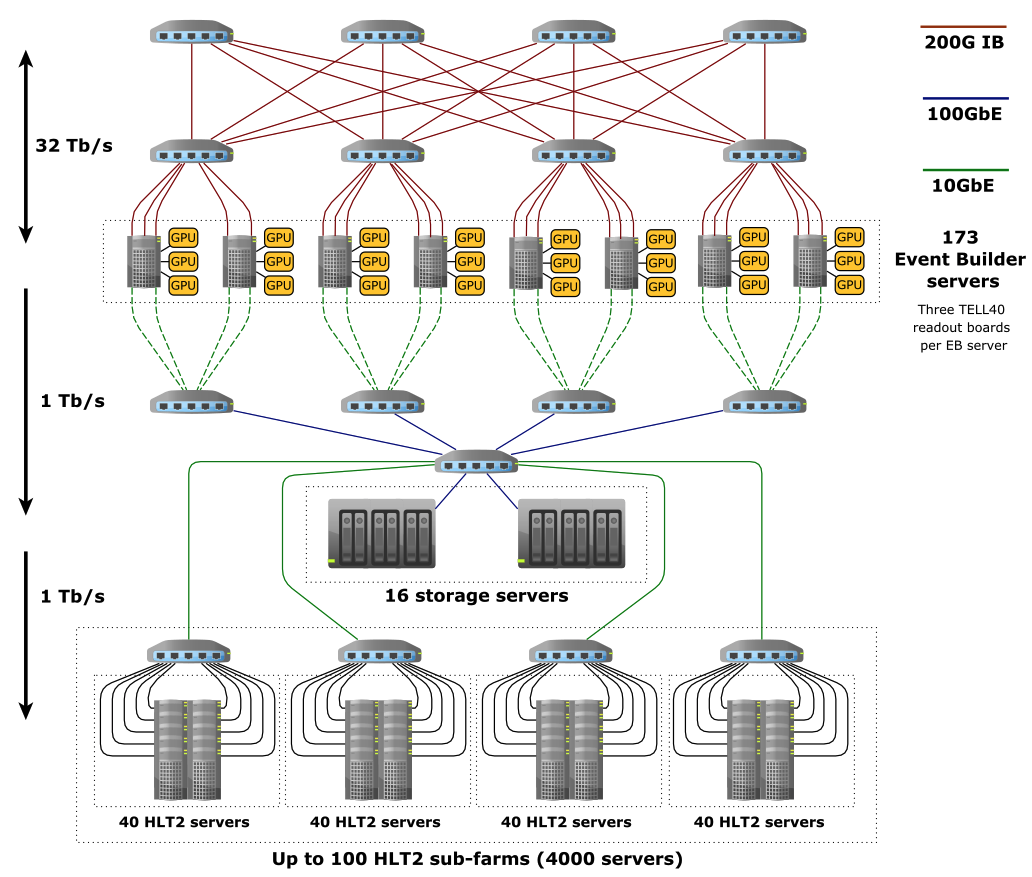}
\caption{Run~3 DAQ architecture in the case of the hybrid solution, with GPUs placed in the EB servers to reduce the data rate. Labels are the same as in Figure~\ref{fig:eb_baseline}.}
\label{fig:eb_hybrid}
\end{center}
\end{figure}

The hybrid architecture is illustrated in Fig.~\ref{fig:eb_hybrid}. It follows the same processing logic as the CPU-only solution: the full detector data is received by TELL40 boards and assembled into MEP packets by the EB servers, those MEP packets are then sent for processing by HLT1 and the events selected by HLT1 are recorded to the disk buffer for later processing by HLT2. Compared to the CPU-only solution it replaces the HLT1 EFF by GPU cards running HLT1 which are installed in the spare PCI express slots available in the EB servers. This allows HLT1 to reduce the data rate at the output of the EB by a factor of 30-60. 
This reduction in turn allows communication between the EB and EFF using a lower bandwidth (and consequently cheaper) network and removes the need to buy and install dedicated NICs in the EB and HLT1 EFF servers as they are already equipped with on-board 10Gb interfaces. For the same reason a much smaller switch is required to handle the data traffic between HLT1 and the disk servers. HLT2 then runs similarly to the CPU-only solution on the EFF.

\subsection{Real Time Analysis}
\label{sec:sub:HLTEvolution}

In Run~2, LHCb successfully adopted a real-time analysis model, which is documented in detail in references~\cite{LHCb-DP-2016-001,LHCb-DP-2019-002}. 
The most important aspects relevant for the comparison presented in this document are summarized here:
\begin{itemize}
    \item To make optimal use of limited offline resources, around three quarters of LHCb's physics programme is written to the TURBO stream~\cite{LHCb-TDR-018}, a reduced format, which on one hand allows flexible  event information to be added in a selective manner and on the other hand can discard a user-specified fraction of both the raw and reconstructed detector data.
    \item Consequently, HLT2 must be able to run the complete offline-quality reconstruction. Therefore HLT2 must use the highest quality alignment and calibration at all times. A substantial disk buffer must therefore be purchased to allow events selected by HLT1 to be temporarily saved while the full detector is aligned and calibrated in real-time. This buffer must be big enough not only to cover the steady-state data taking conditions but also to permit recovery from unforeseen operational issues in a reasonable timescale without loss of data. 
    \end{itemize}

\section{Assumptions and boundary conditions}
\label{sec:BoundaryConditions}
Having described the overall design of LHCb's Run~3 DAQ and HLT, as well as the processing technologies under consideration, we will now describe the boundary conditions which these technologies have to respect, as well as common assumptions relevant to the cost-benefit comparison in Section~\ref{sec:CostBenefit}.
\subsection{Use of storage and computing resources during and outside datataking}
Throughout this document it is assumed that the LHC is in ``data-taking'' for 50\% of the year, and in either the winter shutdown or longer technical stops for the other 50\%. When in data-taking, it is assumed that the LHC is in stable beams 50\% of the time. 
During data-taking, it is assumed that all CPU resources are used to process HLT1 and/or HLT2. Outside data-taking, it is  assumed that all HLT CPU resources are used to produce simulation for LHCb analyses. 
GPU resources can only be used to process HLT1 in-fill, and cannot be used opportunistically during data-taking. They cannot yet be used for producing simulation, and there is no realistic prospect of this changing on a short timescale. However in principle GPU resources could be used out of data-taking if use-cases can be found, as discussed in Ref.~\cite{LHCb-TDR-021}.
When LHCb is not taking data, the EB nodes will be used to produce simulation. However they 
will not be used for any task other than event building while taking data, as all the available memory bandwidth in these nodes will be required for the event building process and transferring data to the HLT1 application.
\subsection{Existing and pledged HLT2 computing resources}
We quantify the computing resources available for HLT2 in terms of a reference QuantaPlex (``Quanta'') server consisting of two Intel E5-2630v4 10-core processors, which was the workhorse of our Run~2 HLT. This reference node corresponds to approximately 380 HEPSPEC.\footnote{The HEPSPEC benchmark is defined at \href{http://w3.hepix.org/benchmarking}{http://w3.hepix.org/benchmarking}. In reality our HLT2 farm will consist of a mixture of servers of different generations and with different numbers of physical cores, but because of the asynchronous nature of HLT2 processing load-balancing between these servers is an implementation detail and it is more convenient to quantify the available resources in units of the reference node as if the system were fully homogeneous.} We currently have roughly 1450 equivalent such servers available for Run~3 processing, with a further 1200 equivalent servers pledged, corresponding to a total capacity of slightly more than one million HEPSPEC. These servers can only be used to process HLT2 as it would not be cost-effective to equip so many old servers with the high-speed NICs required to process HLT1. So far no economical way has been found to reuse the predominantly very small disk-drives in the old Run~2 servers, so there are no free storage resources available.
\subsection{Event building and data-flow}
\label{sec:sub:Dataflow}
In the CPU-only scenario no data reduction happens before the EFF, so all the data collected in the building network has to be distributed to the HLT1 CPU nodes at 32~Tb/s. Using newly available AMD 32-core CPUs a cost-effective implementation is to use a dual-socket server with a total of 64 physical cores and two network interfaces of 100~Gb/s. Each EB node requires two high-speed network connections for sending the event fragments between EB nodes while they are built. 
In addition, the distribution network needs to connect all HLT1 CPU nodes to the EB nodes as well as to at least 30 storage servers. These connections need to be optical.\\
~\\
In the hybrid scenario the GPUs hosted within the EB nodes execute the HLT1 process and reduce the data rate, so that only $0.5-1.0$~Tb/s has to be sent from the EB to the disk buffer servers. The EB servers can therefore use their on-board 10 Gigabit interfaces to send the data and the distribution network needs significantly fewer optical connections.

\subsection{Disk buffer boundary conditions}
\label{sec:sub:DiskBufferBoundaryConditions}
The disk buffer needs to be able to handle at least 1~MHz of events coming from HLT1, with a potential upgrade to be able to handle 2~MHz as Run~3 progresses. A typical minimum bias event in LHCb Run~3 conditions is expected to be around 100~kB, however events selected by HLT1 are bigger than average minimum bias events since they typically contain a hard QCD process leading to the production of a heavy flavour hadron. Therefore assuming an event size of 120~kB to account for this effect, this implies 120~GB/s both for writing events coming from HLT1 and reading events out into HLT2. The nominal rate of a 12~TB disk is 100~MB/s, thus 50~MB/s for writing and 50~MB/s for reading. However the read and write speed of hard disks decreases as they fill up. As the system must be costed so that there is no dead-time even when the buffer is almost full, an effective sustainable write and read rate of a single 12 TB disk is assumed to be 35-40~MB/s.\\
~\\
Since this part of the system is hardware limited and must be able to handle burst data rates, a minimum of 2880 physical disks is required assuming minimal redundancy, and 3120 physical disks with adequate redundancy. A final point to note on the disk buffer is that the usable disk sizes are only around 80\% of the nominal disk size, so the 12~TB disks actually only provide around 9.6~TB of usable storage each.
In practice, the cost of such a minimal disk buffer is so large compared to the overall 
budget discussed earlier that spending money on buying additional disks or bigger disks is not really an interesting option.

\section{HLT1 sequence}
\label{sec:ImplementedHLT1}
The CPU-only and hybrid approaches under study in this document implement an HLT1 configuration which broadly corresponds to the one used in Run~2~\cite{LHCb-DP-2019-001} and whose physics objectives have been described in reference~\cite{LHCb-TDR-021}. The reconstruction consists of the following components, whose performance in terms of efficiency, purity and resolution is discussed in Section~\ref{sec:PhysicsPerf}.
\begin{itemize}
    \item Vertex Locator (Velo) detector decoding, clustering, and reconstruction. Conceptually very similar algorithms are used here in the CPU and the GPU implementation. Only minor differences in physics performance are expected due to a limited number of architecture-specific optimizations. 
    \item The primary vertex (PV) finding with tracks reconstructed in the Velo. Again only minor differences are expected in the physics performance of the CPU- and GPU-based implementations.
    \item Decoding of the UT and SciFi tracking detector raw banks. 
    The raw bank formats are converted to the input of the pattern recognition algorithms. These are specific to the data layout and event model used by each architecture.
    \item Reconstruction of Velo-UT track segments. Different algorithms are used here for the CPU and the GPU implementations, which lead to slight differences in the physics performance. 
    \item Reconstruction of long tracks\footnote{Long tracks are tracks that traverse the entire tracking system from Velo to SciFi. They deliver the best parameter estimate in terms of position and momentum and thus are the most valuable tracks for physics analysis.} starting from reconstructed Velo-UT track segments. Both the CPU and GPU tracking algorithms use a parameterization of particle trajectories in the LHCb magnetic field and the initial Velo-UT momentum estimate\footnote{The momentum resolution of Velo-UT tracks is about 15\% with significant non-Gaussian tails due to the small and inhomogeneous B field between Velo and UT.} to speed up their reconstruction. One major difference in strategy is that the CPU algorithm matches Velo-UT track segments to doublets in the SciFi x-layers, while the GPU algorithm matches them to triplets. In additon, the CPU algorithm applies a 490~MeV transverse momentum threshold when defining its search windows. These choices, and the differences in the Velo-UT algorithms, lead to somewhat different performance. 
    \item Decoding of the muon raw banks and calculation of crossing points in the muon system, as well as implementation of the muon identification algorithm. Both architectures use very similar algorithms here.
    \item A simplified Velo-only Kalman filter which uses the momentum estimate from the forward tracking to calculate a covariance matrix and estimate an uncertainty on the track impact parameter. Again the underlying algorithm is the same for both implementations.
\end{itemize}
The HLT1 sequence described covers most use cases required by bottom and charm physics. There are however a few
dedicated algorithms missing, such as a reconstruction for high momentum muons for electroweak physics, reconstruction of beam gas events or a reconstruction of low momentum tracks down to a transverse momentum of 80~MeV, which is motivated by the strange physics programme. While preliminary versions of these algorithms were ready in time for this comparison, they were not yet fully optimized in the same way as the other described algorithms. We did use these preliminary versions to test whether their inclusion impacted on the relative throughput between the CPU and GPU implementations, and found that the relative slowdown from including these algorithms was similar across the two architectures. It is therefore expected that these missing components will not change the conclusions of this document. \\
~\\
In addition to the reconstruction algorithms listed above, both the CPU and GPU HLT1 implement a representative subset of HLT selections including the finding of displaced vertices, monitoring, as well as writing of decision and selection reports. While the specific components are not exactly the same in both cases, these parts of the system consume comparatively little throughput and hence the differences are not relevant for this comparison. \\
~\\
A global event cut that removes 7\% of the busiest minimum bias events is applied in all selections before the reconstruction is performed for both the CPU and GPU HLT1 implementations. The criterion used is that the total number of hits in the UT and SciFi tracking detectors be below a certain value. This criterion is chosen for historical reasons and because the UT and SciFi reconstructions are the most sensitive to occupancy, especially when trying to reconstruct low \pt signatures. However in practice any subdetector occupancy could be used, and LHCb will likely use the occupancy of whichever subdetector shows the best data-simulation agreement once the new detector is commissioned in Run~3. The impact of this criterion on the physics is given in Section~\ref{sec:geceffs}.

\section{Throughput}
\label{sec:Throughput}
We define throughput as the number of events which can be processed by a given architecture per second in steady-state conditions, that is to say neglecting the time it takes to initialize the HLT1 application at the start of each datataking period. They can be converted into GB/s by multiplying by the average expected Run~3 event size of 100~kB. An event is processed when it is read into HLT1, the HLT1 reconstruction and selection sequences used to decide whether or not to keep this event, and the event is finally written out (or not). The throughput of the CPU and GPU HLT1 implementations are measured using the same minimum bias samples. The GPU throughput is measured directly on the candidate production card. The CPU throughput is measured on both the Quanta reference nodes used for HLT2, and the dual-socket AMD 7502 EPYC nodes used for HLT1.
While the HLT1 throughput measurements include both reconstruction, selection, and saving of trigger candidates, the HLT2 throughput measurement only includes the reconstruction and not the selection and saving of trigger candidates. Additional costs associated with these missing components of HLT2 are expected but we expect that they can be counterbalanced with future performance improvements in the reconstruction.

~\\
The measured throughputs used in the rest of this document are 
\begin{itemize}
    \item CPU HLT1 : 171~kHz;
    \item GPU HLT1 : 92~kHz;
    \item CPU HLT2 on an AMD EPYC node : 471~Hz;
    \item CPU HLT2 on a Quanta node : 134~Hz.
\end{itemize}
Although we will discuss the cost-benefit of the two architectures later in Section~\ref{sec:CostBenefit}, we can already conclude that both the CPU and GPU HLT1 architectures can be implemented using rather compact systems of $O(170)$ CPU servers or $O(330)$ GPU cards. 

\section{Physics performance}
\label{sec:PhysicsPerf}
This section presents key figures which are evaluated in a like-for-like comparison of the GPU and CPU performance.
This includes performance numbers for track  reconstruction and fitting, for PV reconstruction, for Muon-ID and the overall HLT performance for some representative trigger selections.
Identical algorithms are used to fill the histograms and produce the plots based on the output of HLT1. 
The source code is compiled to operate on both the GPU and the CPU. The output is also translated to the same format.
This ensures that the same definitions of physics performance parameters such as efficiencies and resolutions are used when doing the comparison. The CPU compiled version of the GPU code has been checked to give results which agree to within $10^{-4}-10^{-3}$ with results obtained with the GPU version. More details of this comparison can be found in reference~\cite{LHCb-TDR-021}.\\
~\\
An overview of the samples used for these studies is given in Tab.~\ref{tab:MCsamples}. The specific samples used for the individual studies are listed in the corresponding subsections.

\begin{center} 
\begin{table}[h]
\begin{center}
\begin{tabular}{l|c}
            Sample                 & \# Events\\ \hline
 $J/\psi \rightarrow \mu^+ \mu^-$ & 10k\\
 $B^0_s \rightarrow \phi \phi$ & 10k\\
$B^0 \rightarrow K^{*0} e^+e^-$  & 10k\\ 
$B^0 \rightarrow K^{*0}  \mu^+\mu^-$  & 10k\\
$D_s \rightarrow K\pi\pi$ & 10k \\
$Z\rightarrow \mu^+ \mu^-$ & 10k\\
Minimum Bias & 10k \\
\end{tabular}
\end{center}
\caption{\label{tab:MCsamples} Simulated samples used to evaluate the physics performance. All samples are simulated using the same pileup assumed in the Run~3 minimum bias samples.}
\end{table}
\end{center}

\subsection{Tracking efficiencies and ghost rates}
\label{sec:TrackEff}
\begin{figure}\centering
\includegraphics[scale=0.38]{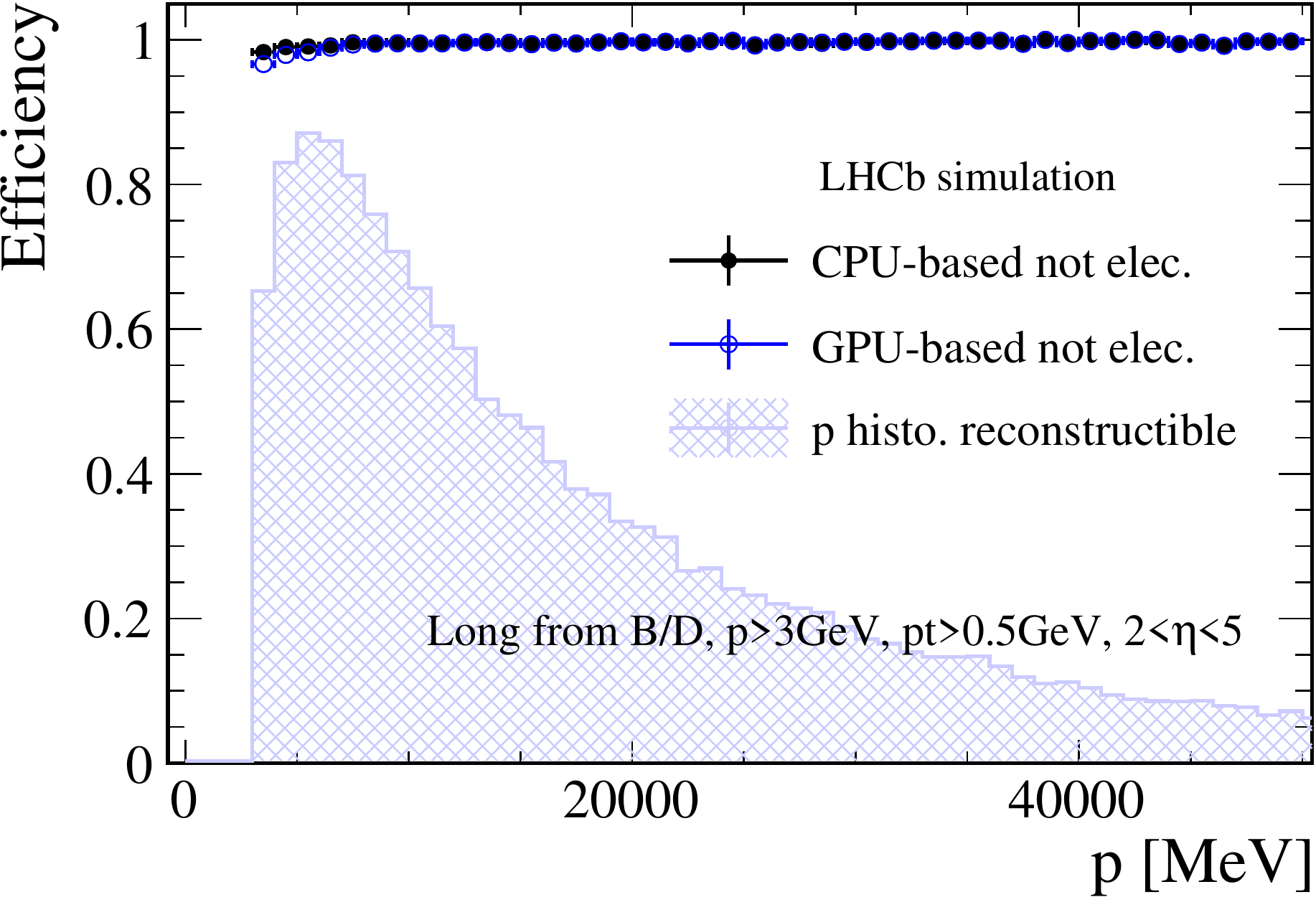}
\includegraphics[scale=0.38]{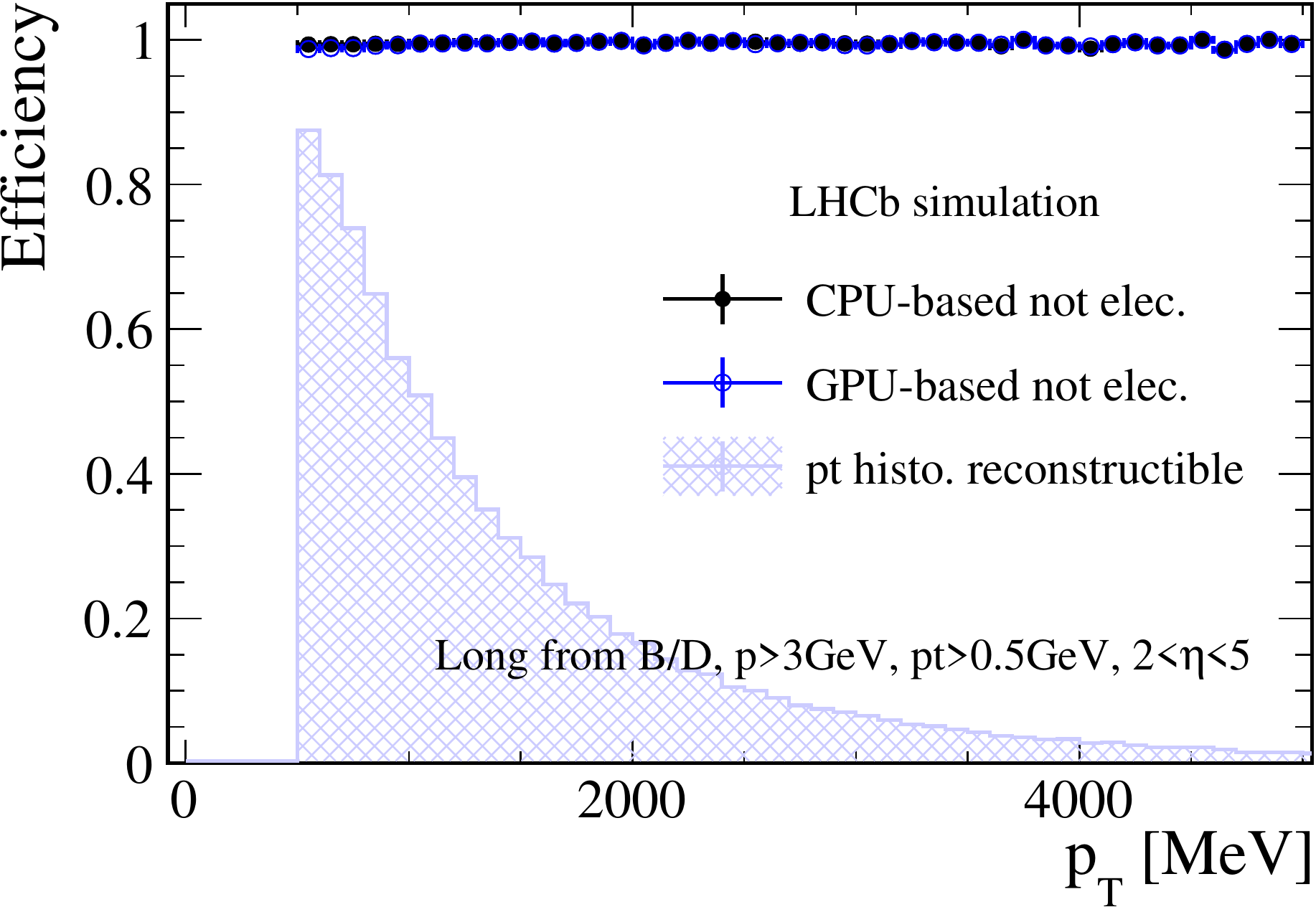}
\includegraphics[scale=0.38]{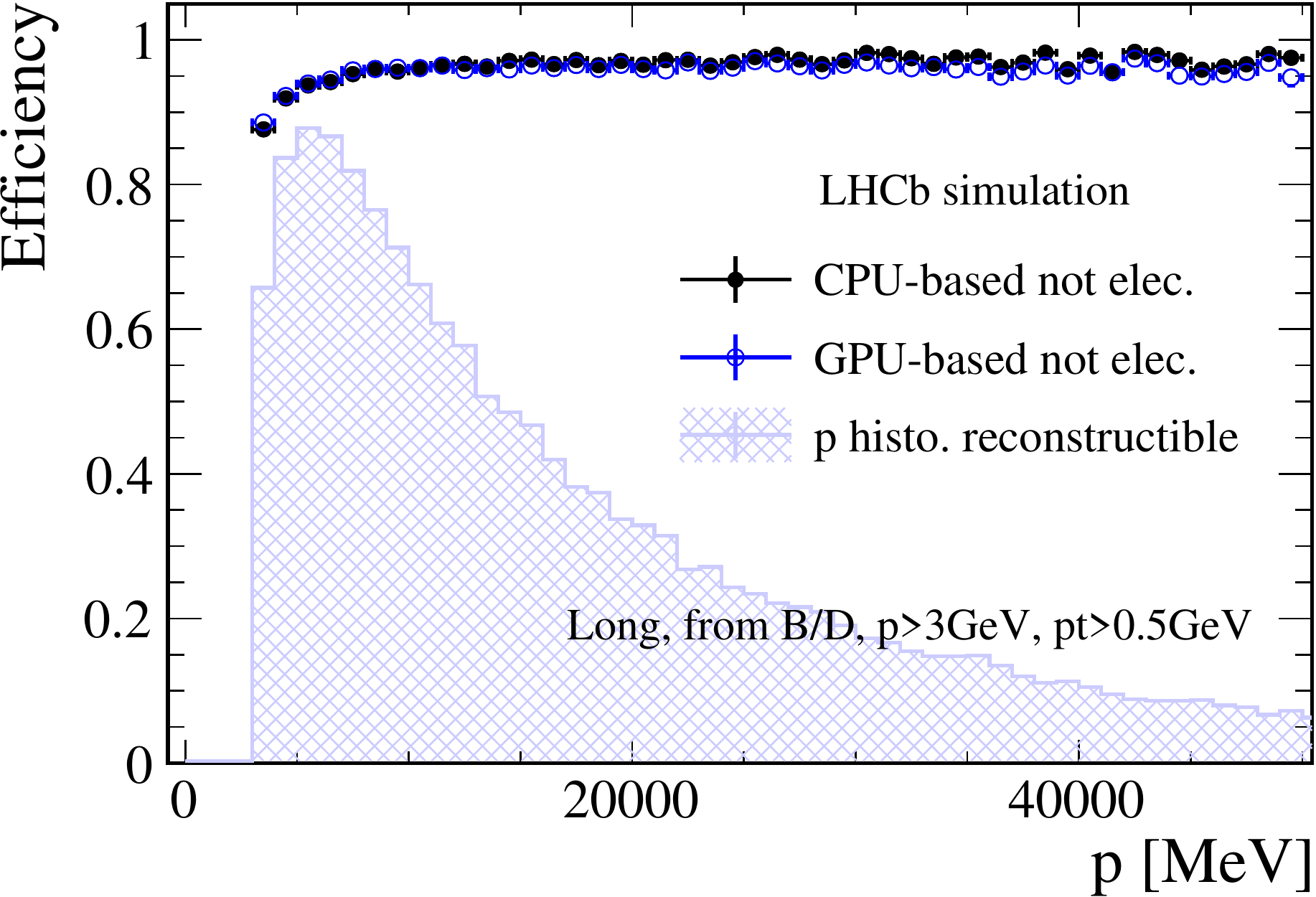}
\includegraphics[scale=0.38]{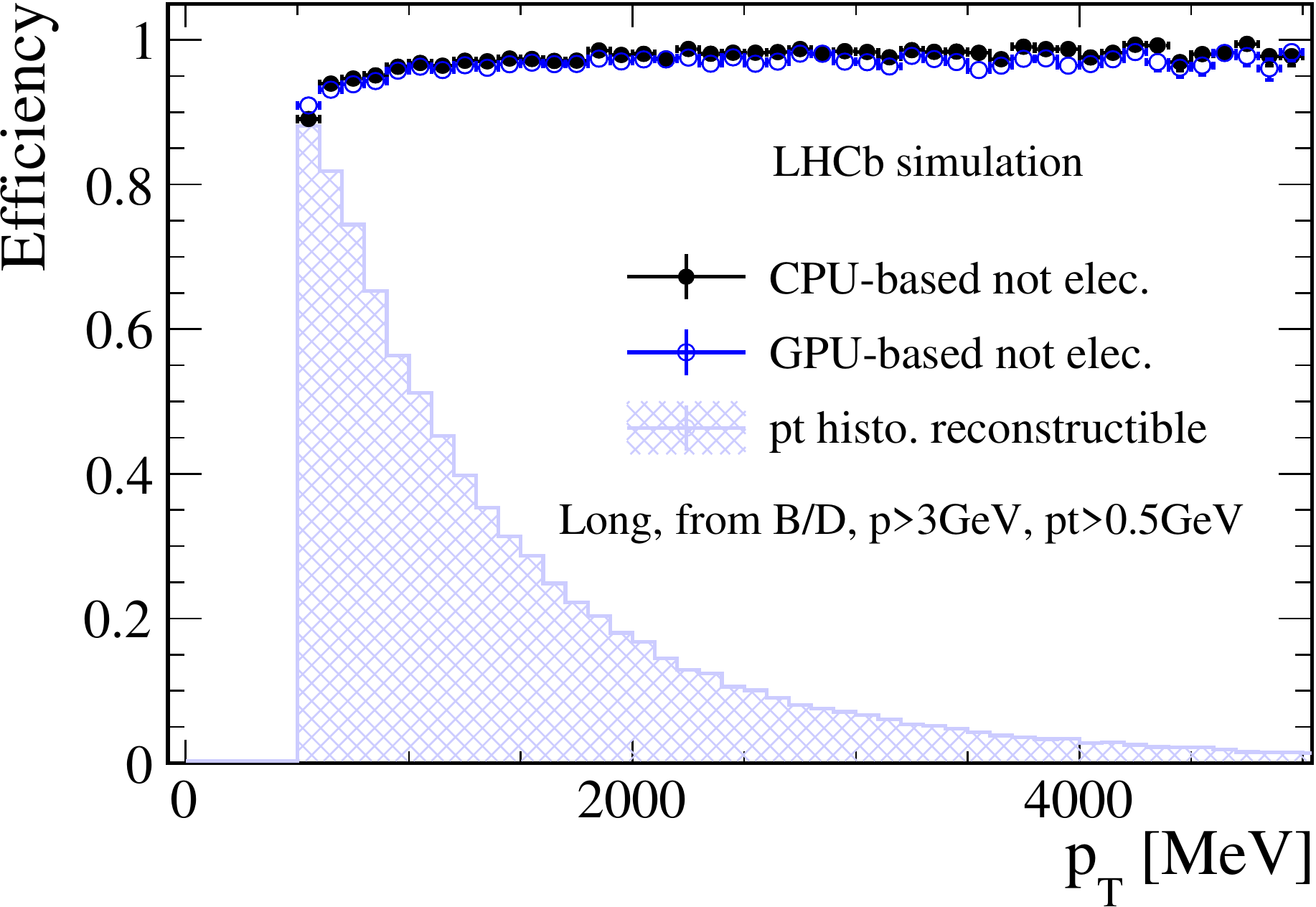}
\includegraphics[scale=0.38]{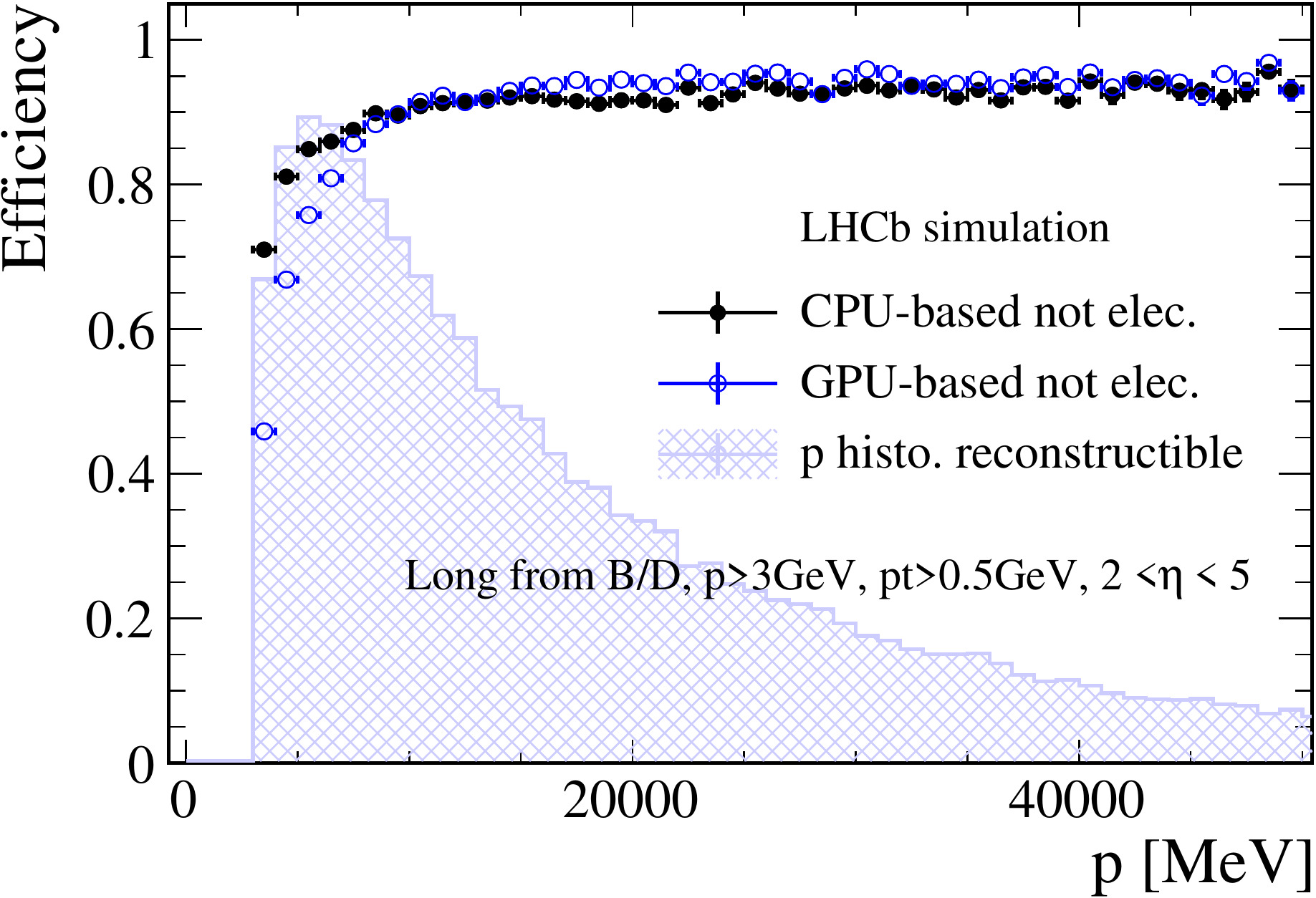}
\includegraphics[scale=0.38]{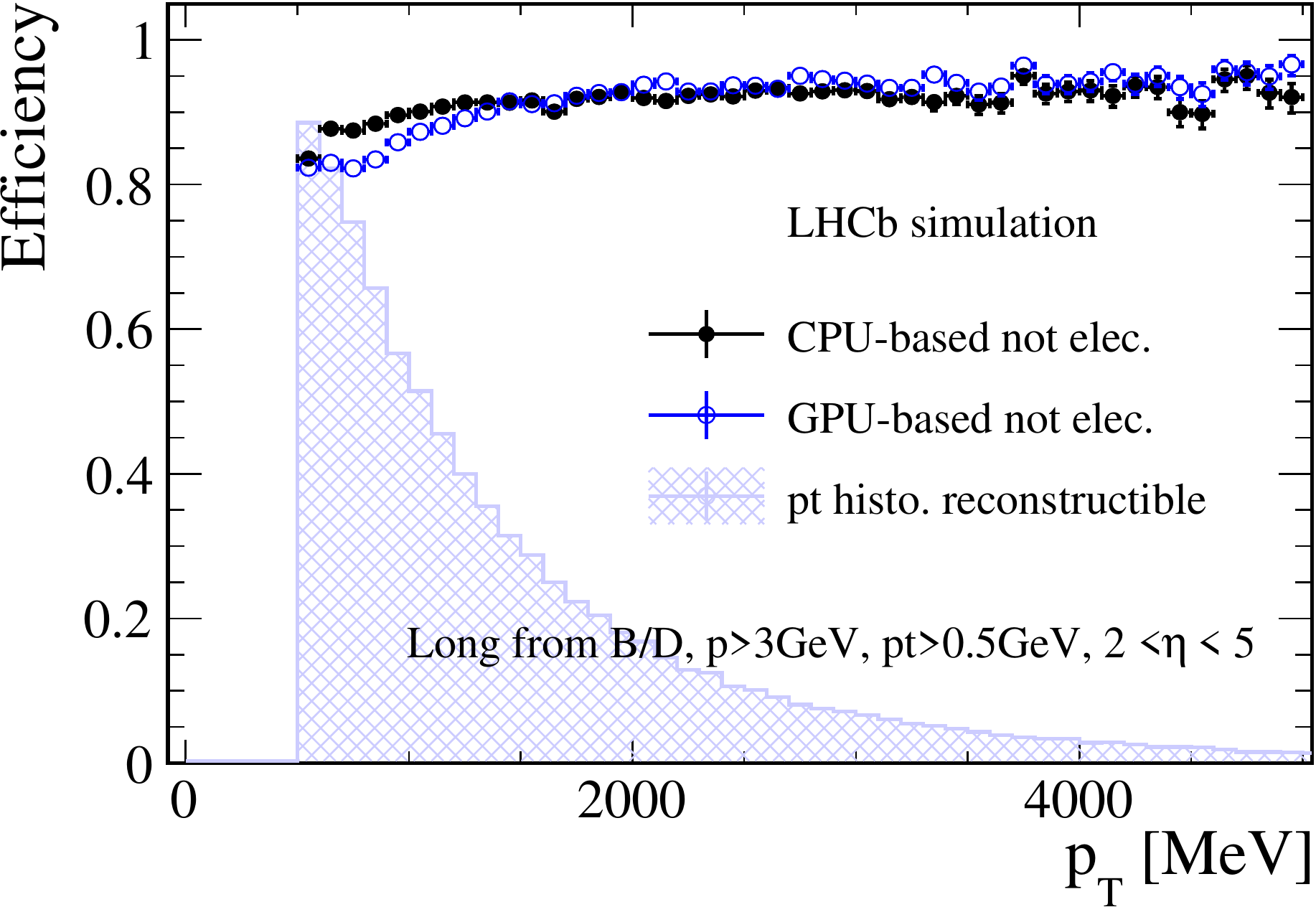}
    \caption{Track reconstruction efficiency versus momentum, $p$ (left) and transverse momentum, $p_T$ (right) for long reconstructible particles from $B$ and $D$ decays within $2<\eta<5$. The reconstruction efficiency in the Velo (top), as Velo-UT track (middle) or as long track (bottom) are shown.}
    \label{fig:tracking_efficiency}
\end{figure}

\begin{figure}[htb]
\centering
\mbox{
\includegraphics[scale=0.27]{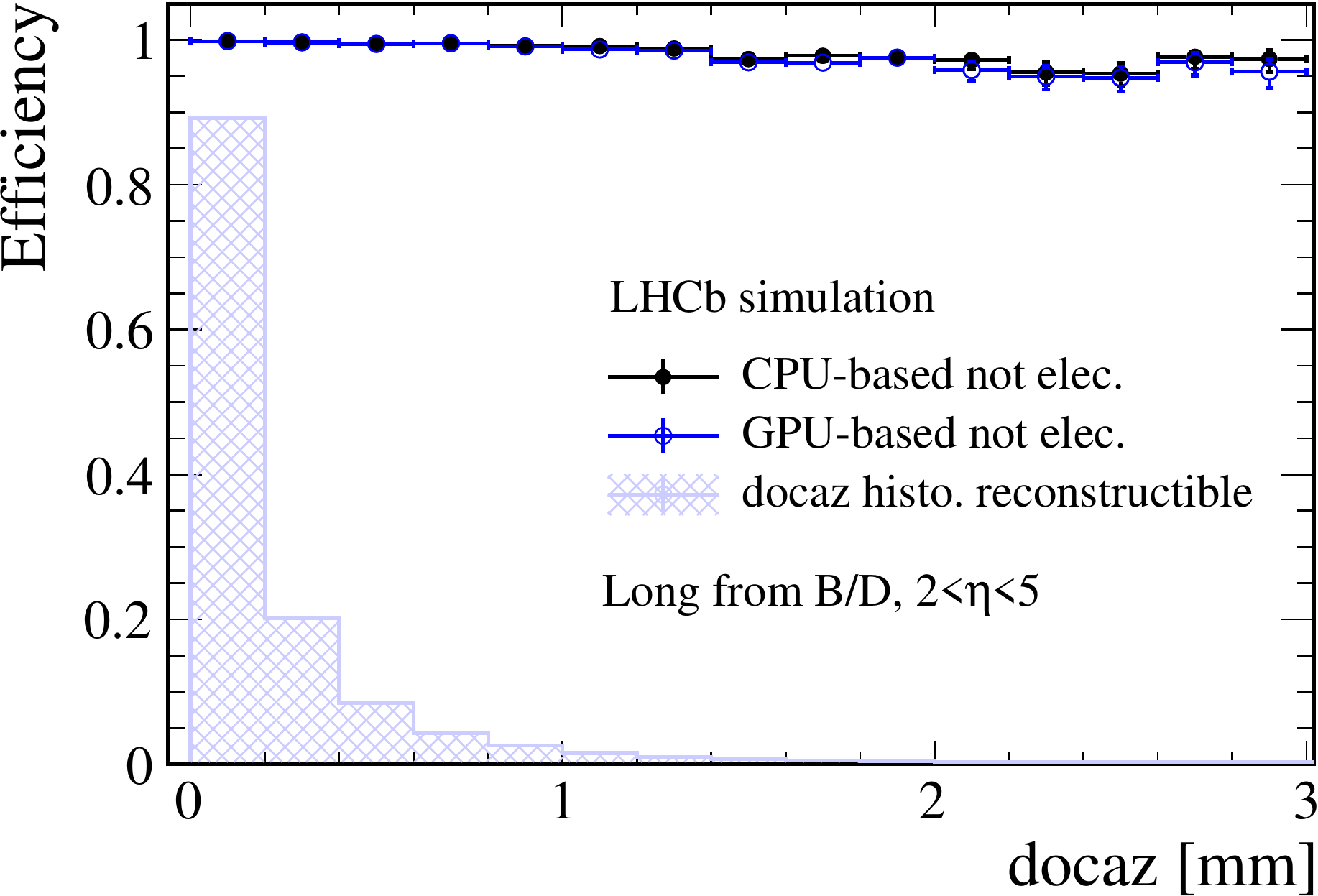}
\includegraphics[scale=0.27]{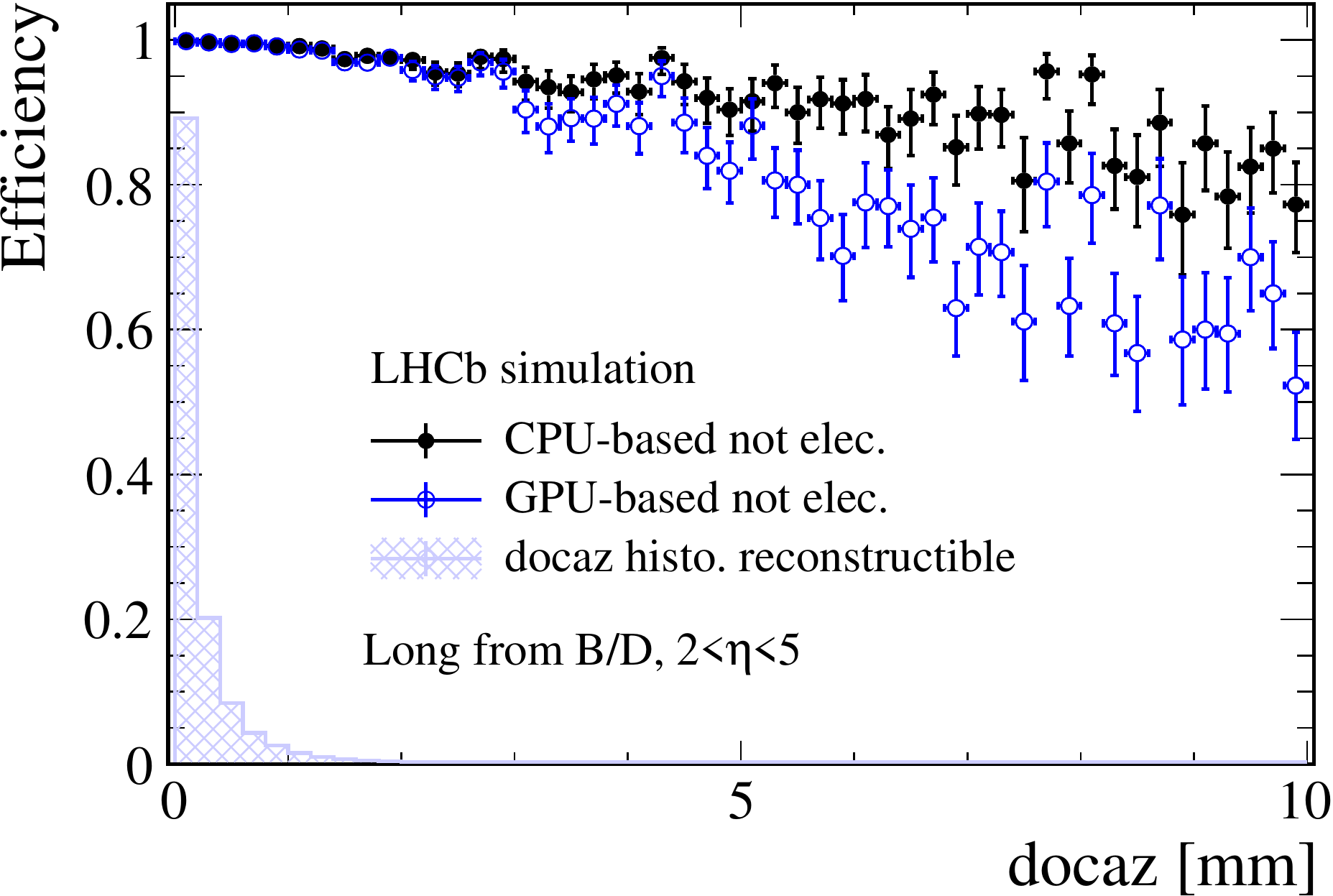}
\includegraphics[scale=0.27]{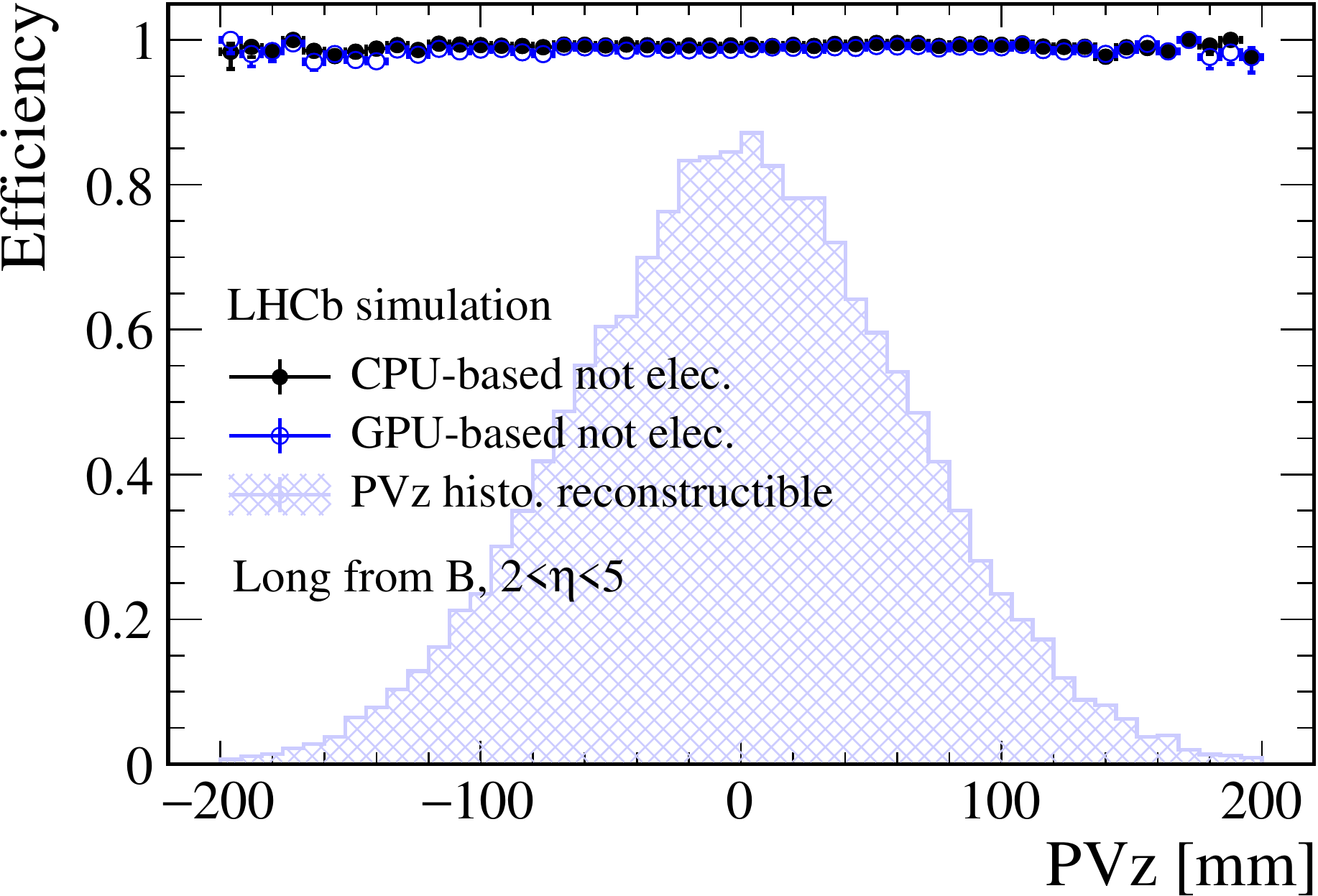}
}
\caption[]{Reconstruction efficiency for long reconstructible particles from $B$ and $D$ decays within $2<\eta<5$ in the Velo as a function of distance of closest approach to the beam axis ($docaz$) and the $z$ position of the primary vertex (PVz). The $docaz$ and $z$ distribution of
the long reconstructible particles are shown as well. }
\label{fig:velo_eff}
\end{figure}

\begin{figure}[htb]
\centering
\mbox{
\includegraphics[scale=0.4]{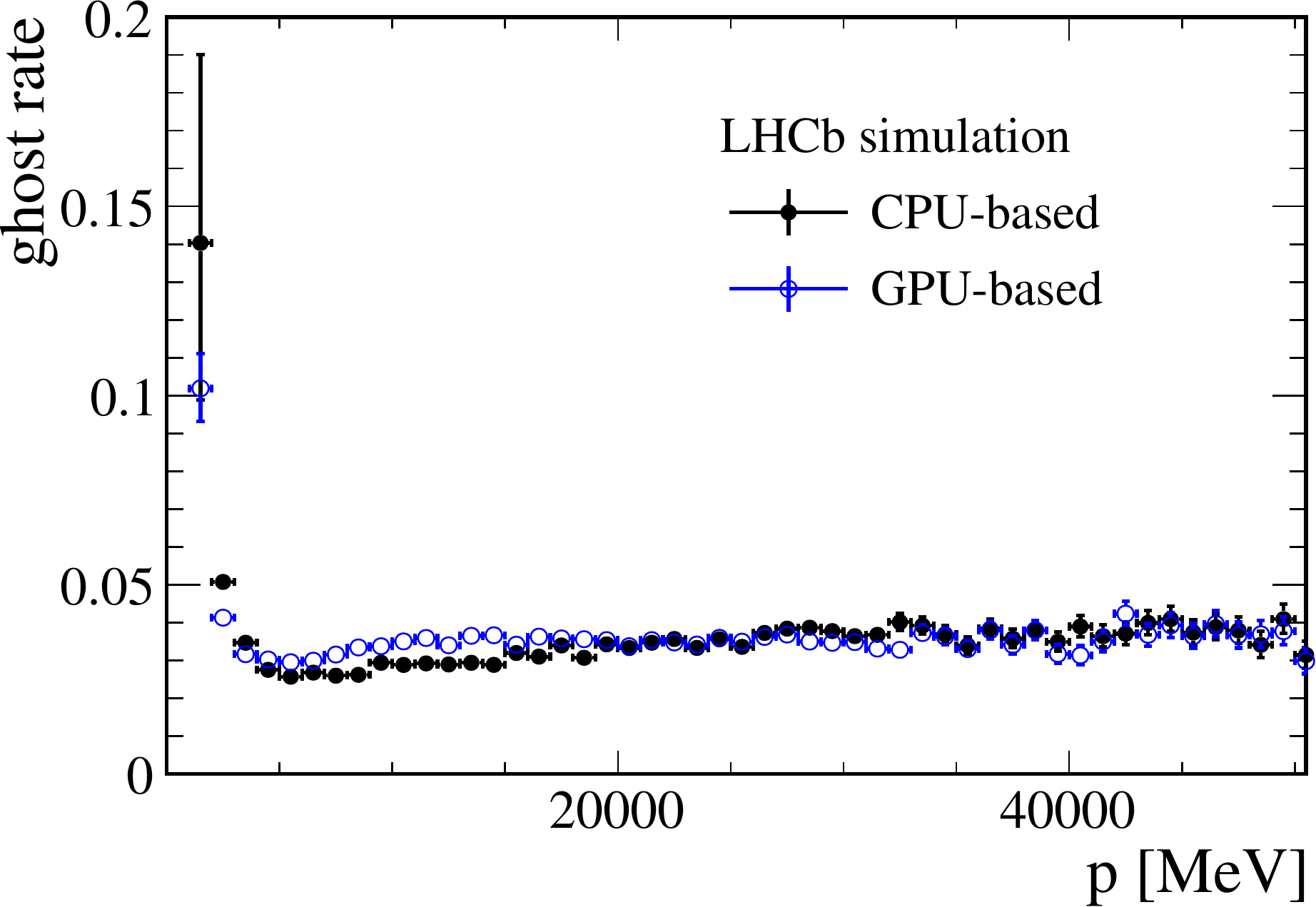}
\includegraphics[scale=0.4]{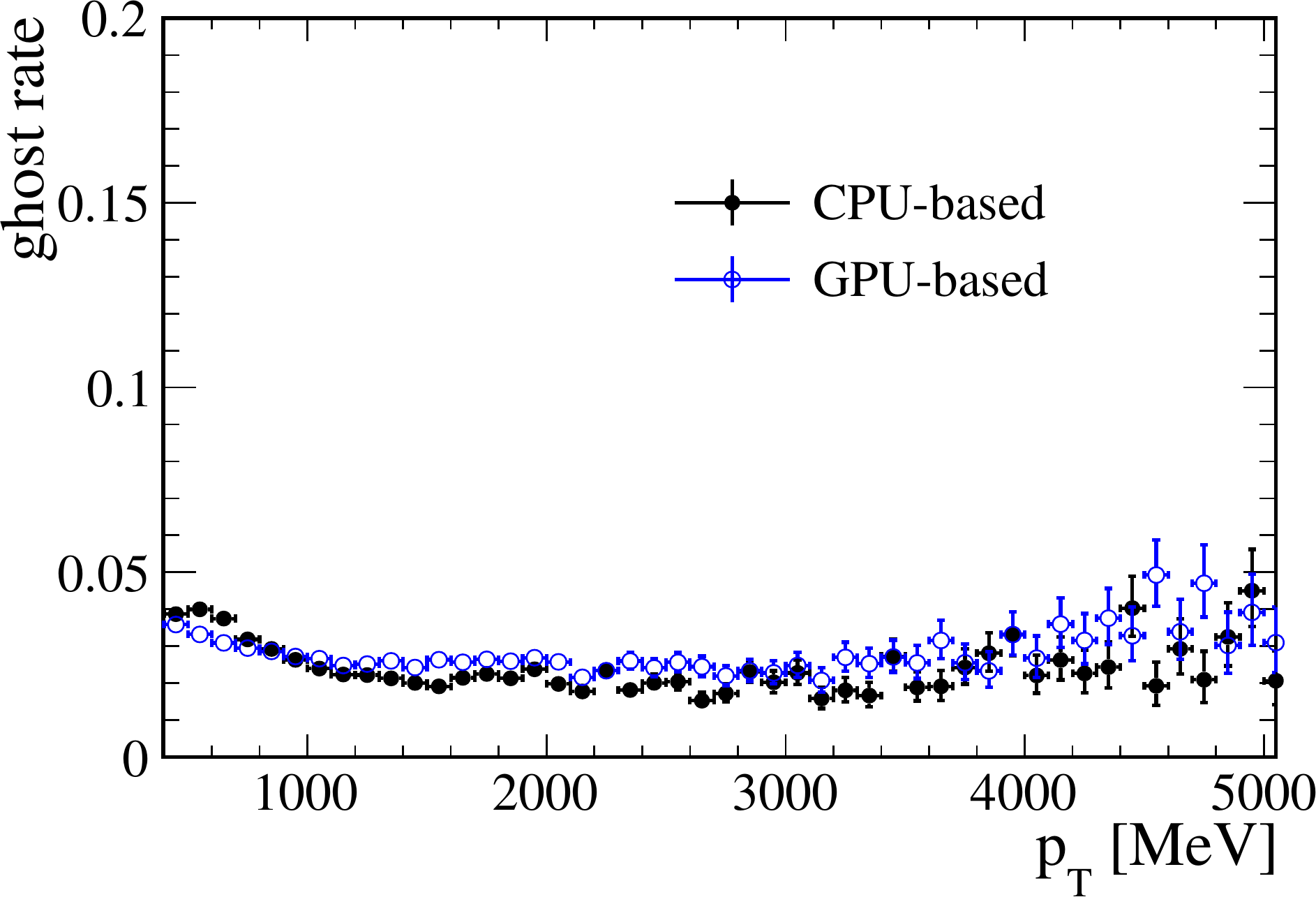}
}\\
\mbox{
\includegraphics[scale=0.4]{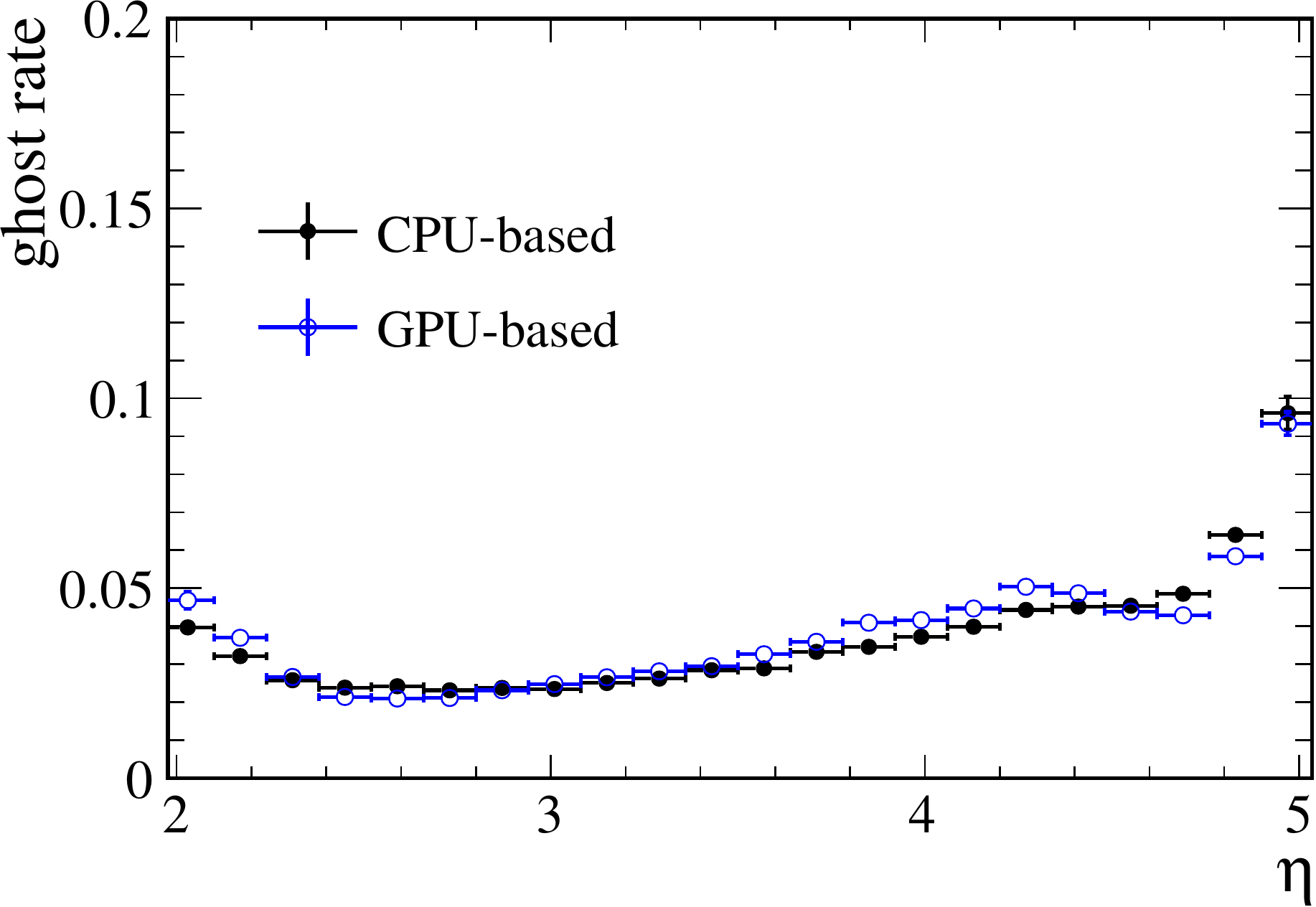}
\includegraphics[scale=0.4]{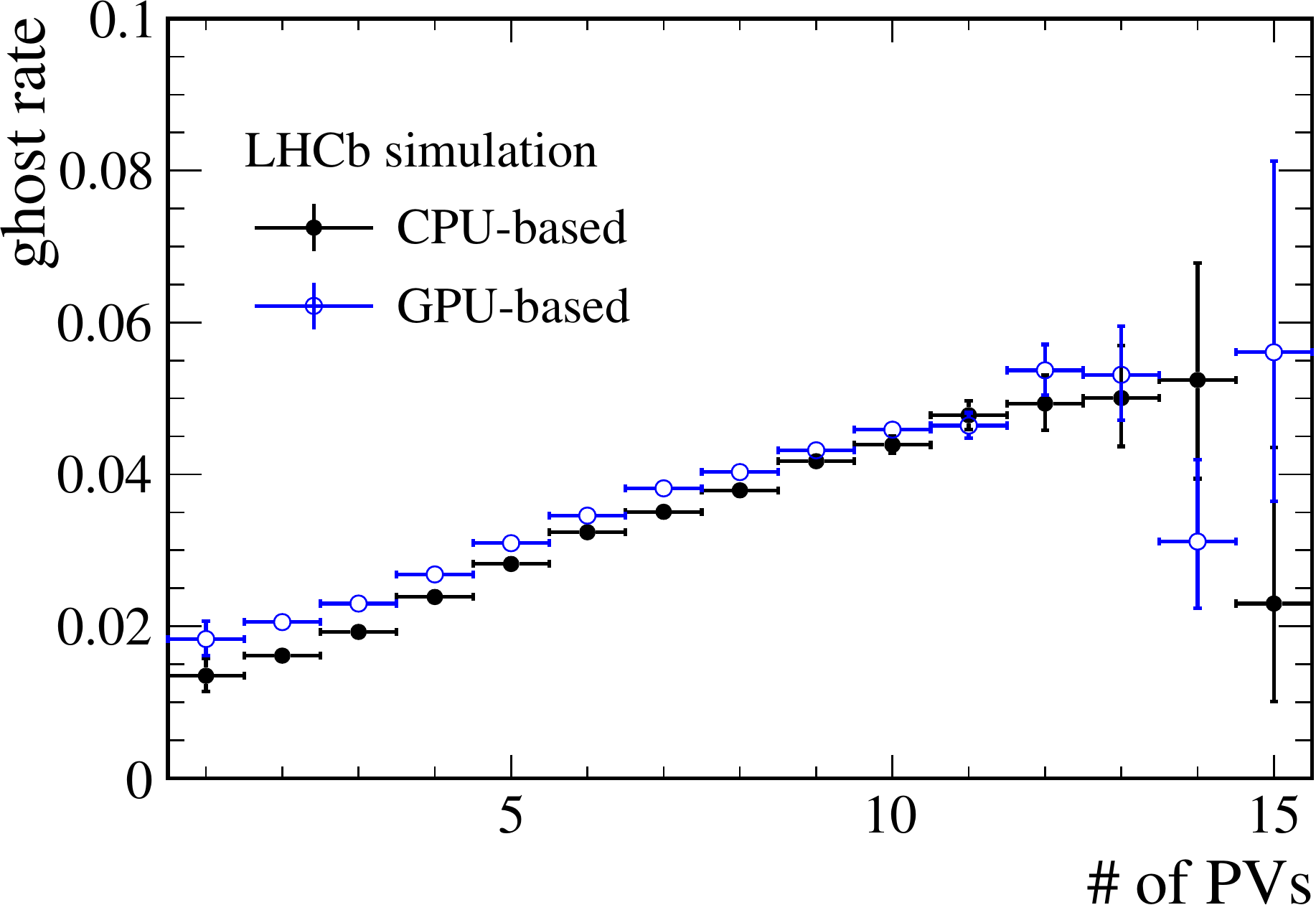}
}
\caption[]{Ghost rate of long tracks as a function of  momenta, $p$, transverse momenta, $p_{T}$, and pseudo-rapidity, $\eta$, of the reconstructed tracks and the number of primary vertices, $\#~of ~PVs$, in the event.}
\label{fig:ghost_rate}
\end{figure}

All plots shown in this section are based on the $B$ and $D$ signal samples in Tab.~\ref{tab:MCsamples}.
Tracking efficiencies are defined as the number of reconstructed tracks out of the number of reconstructible tracks in a given sub-sample.
In this section only particles which are reconstructible as long tracks are studied, which essentially means that they have left signals on at least 3 pixel sensors in the Velo and one x and one stereo cluster in each of the 3 fibre tracker (FT) stations. Furthermore, the
studies are restricted to $B$ and $D$ daughter particles which are in the range $2<\eta<5$, which have true transverse momenta $p_T \geq 0.5$ GeV/c and true momenta $p \geq 3$ GeV/c. Electrons are explicitly excluded because their performance has not yet been optimised to the same extent as that of hadrons and muons, however we have checked the electron performance for both the CPU and GPU implementations of HLT1 and it does not change our conclusions.\\
~\\
Efficiencies to reconstruct these particles in the Velo, in both the Velo and UT and in the Velo, UT, and FT are shown as functions of the true $p$ and $p_T$ of the particles in Fig.~\ref{fig:tracking_efficiency}. The performances of both technologies are very similar, the only difference is seen in the momentum dependence of the long track reconstruction efficiency. This difference is minor in the context of the overall trigger efficiency since the majority of HLT1 triggers use only a subset of tracks coming from a given signal decay to select an event, and are therefore inherently robust to differences in single-track reconstruction efficiencies at the level of a few percent. While the CPU implementation is more efficient at low momenta, the GPU implementation is better at high momenta and the overall single-track efficiency integrated over the kinematic range of interest agrees to better than $1\%$ between the two algorithms. This difference is expected to be entirely related to the tuning of the algorithms and not to the underlying technology.\\
~\\
To check for a possible decay time bias in the Velo reconstruction, the Velo efficiency for long reconstructible tracks is studied as a function of the distance of closest approach to the beamline, $docaz$ and as a function of the $z$ position of the primary vertex in the event (Fig.~\ref{fig:velo_eff}). Again the performance of both implementations is very similar. The loss in efficiency at large $docaz$ is caused by the use of search windows which favour tracks coming from the beamline, and can be recovered at a moderate throughput cost.\\
~\\
Finally, Fig.~\ref{fig:ghost_rate} shows the fraction of ghost tracks among all long tracks. Ghost tracks are tracks which can not be assigned to a true particle, thus are fake combinations of signals in the detector. Again, the performance of both implementations is very similar.

\subsection{Track parameter resolution}
\label{sec:TrackRes}
\begin{figure}[htb]
\centering
\mbox{
\includegraphics[scale=0.37]{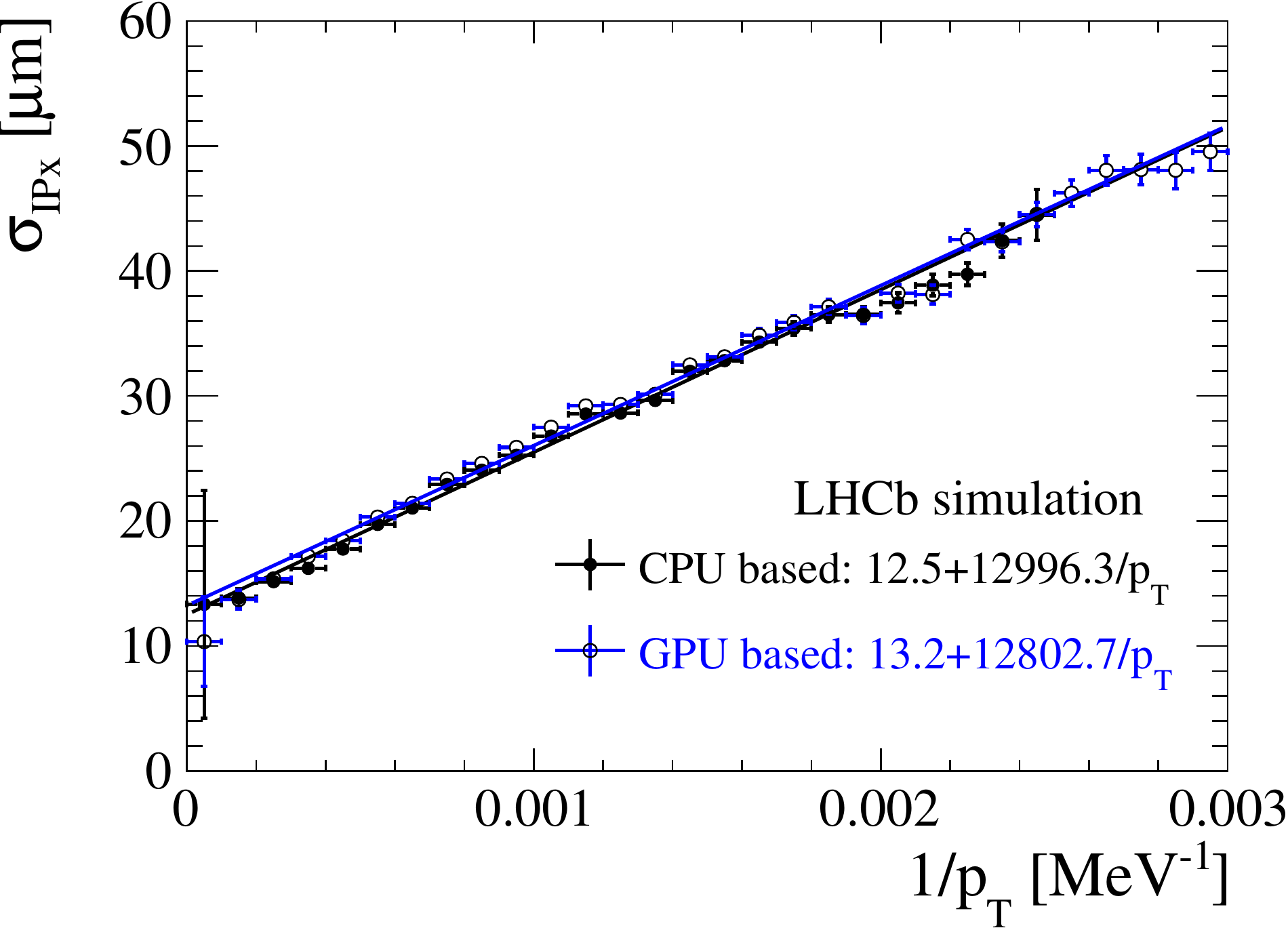}
\includegraphics[scale=0.37]{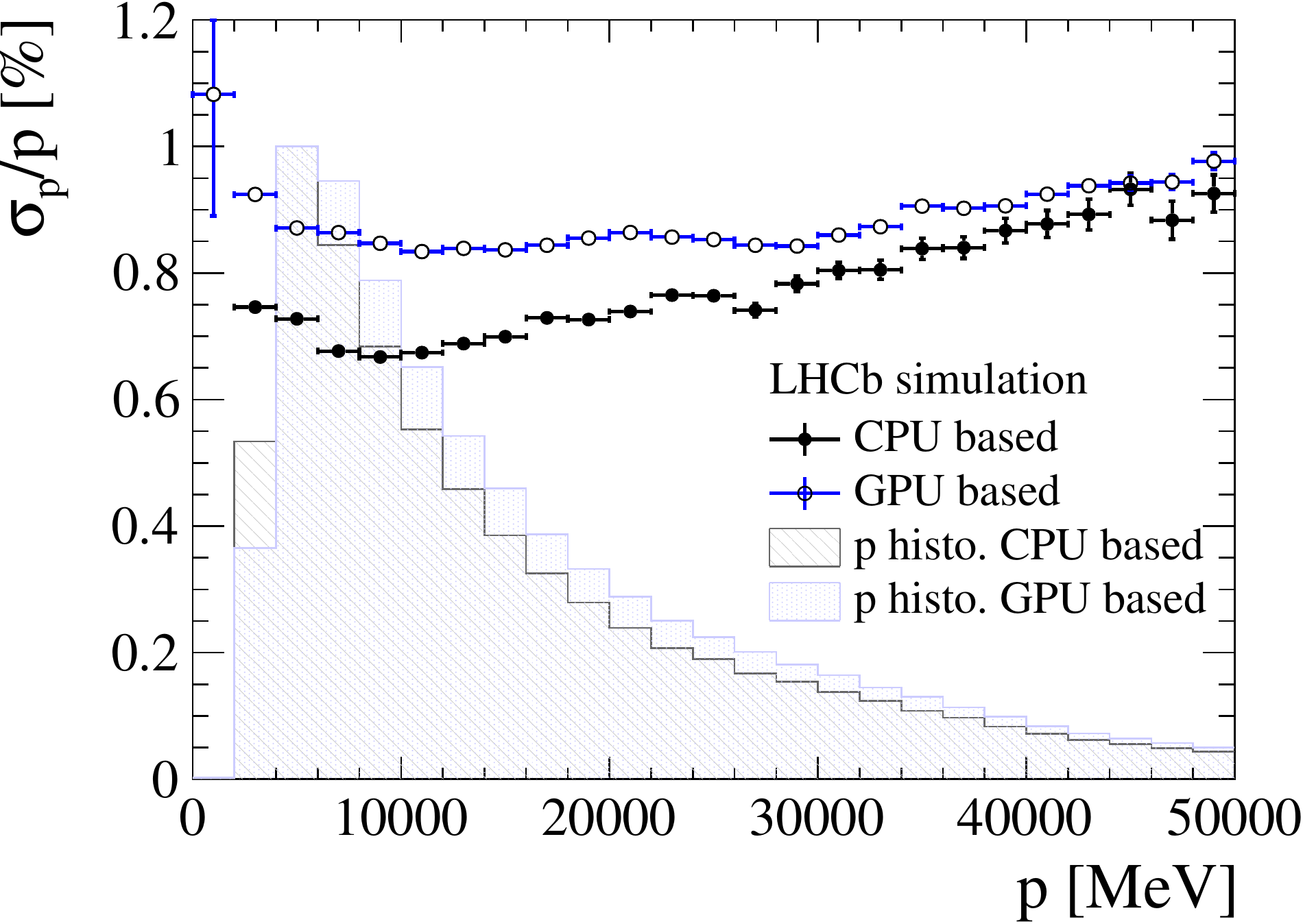}

}
\caption[]{Resolution of the $x$ projection of the impact
parameter, $\sigma_{IPx}$ as function of the inverse transverse momentum, $1/p_T$,  and the relative momentum resolution $\sigma_p/p$ as function of the momentum, $p$.}
\label{fig:resolution}
\end{figure}

For the comparison of the impact parameter  resolution, $\sigma_{IPx}$, and the momentum resolution, $\sigma_p/p$, the minimum bias sample (Tab. \ref{tab:MCsamples}) is used. The results are shown in Fig.~\ref{fig:resolution}. Note that the $x$ and $y$ component of the impact parameter have very similar resolution, therefore only  $\sigma_{IPx}$ is shown here. 
The impact parameter resolution is very similar for both 
technologies. The momentum resolution is worse in the GPU framework, with a maximum absolute resolution difference of $0.15-0.2\%$ at low momenta. This difference is caused by a suboptimal tuning of the parametrization used to derive the momenta of the particles in the GPU algorithm. However, since the computational cost of this parametrization is negligible compared to the track finding itself, this difference in performance can be recovered without any significant change in its throughput.

\subsection{Primary vertex reconstruction efficiency and resolution}
\label{sec:PVPerf}
\begin{figure}[htb]
\centering
\mbox{
\includegraphics[scale=0.37]{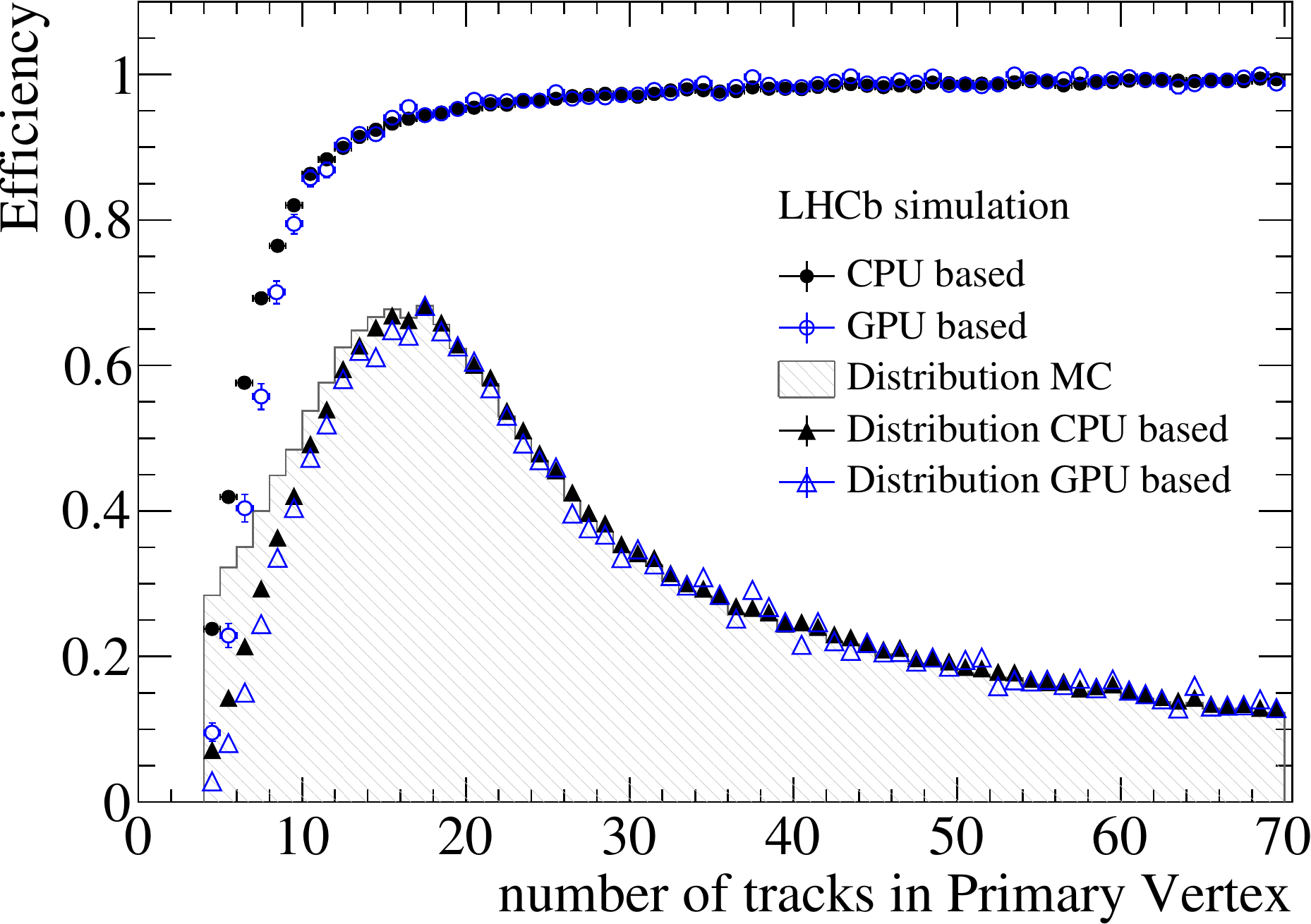}
\includegraphics[scale=0.37]{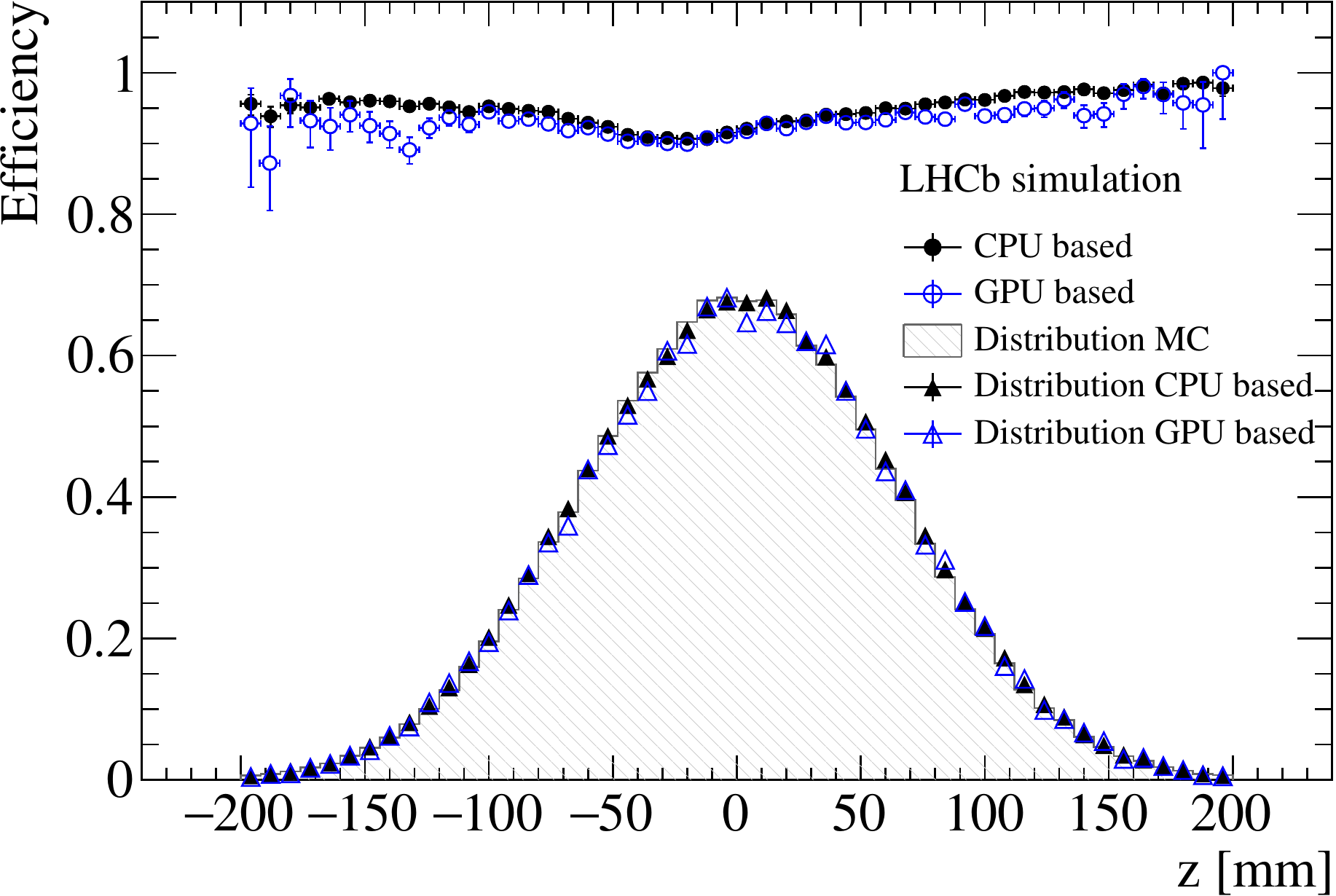}
}
\caption[]{Efficiency to reconstruct primary vertices as function of the number of reconstructed Velo tracks associated to the simulated primary vertex and as function of the true vertex $z$ position.}
\label{fig:PV_Eff}
\end{figure}

\begin{figure}[htb]
\centering
\mbox{
\includegraphics[scale=0.37]{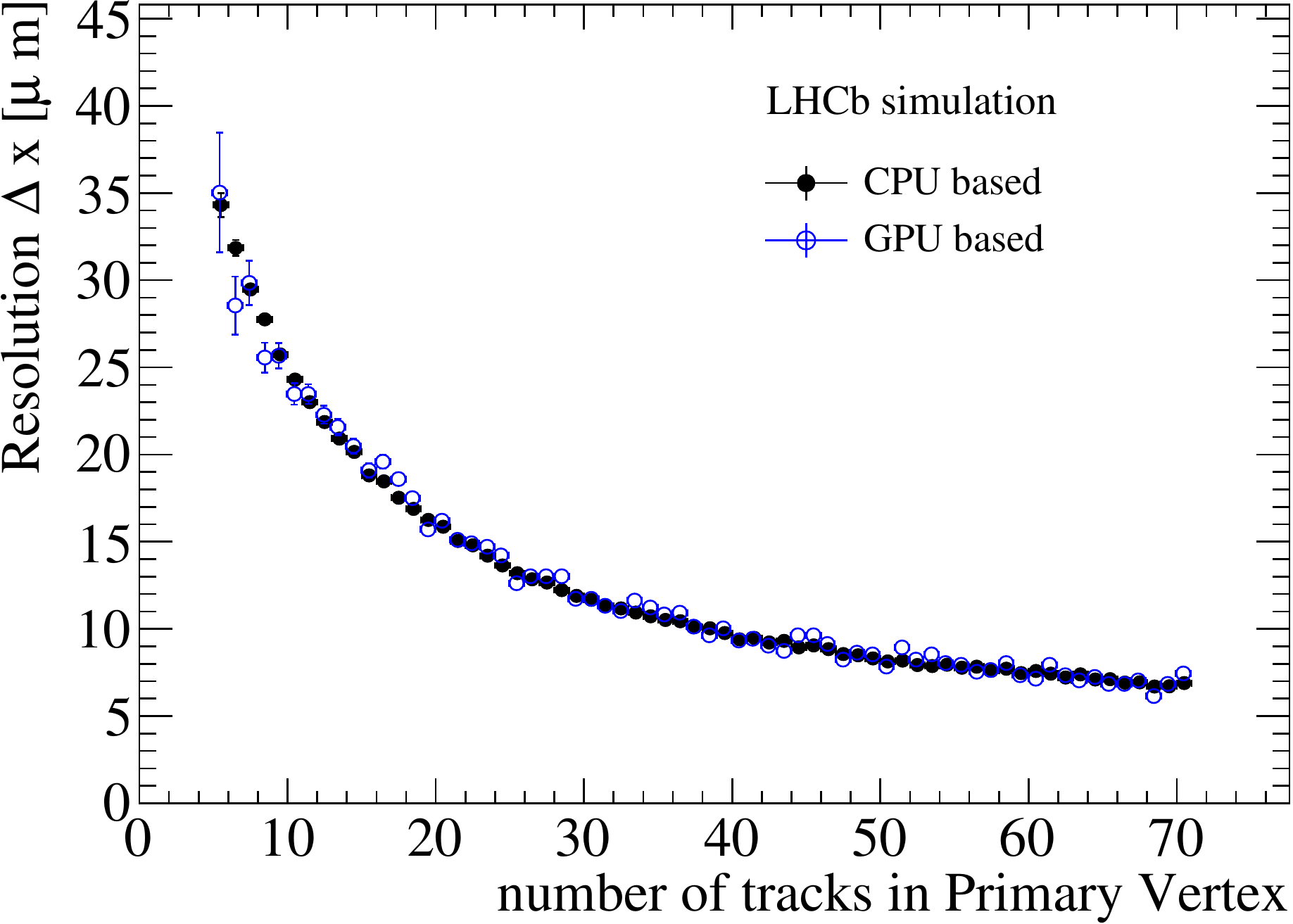}
\includegraphics[scale=0.37]{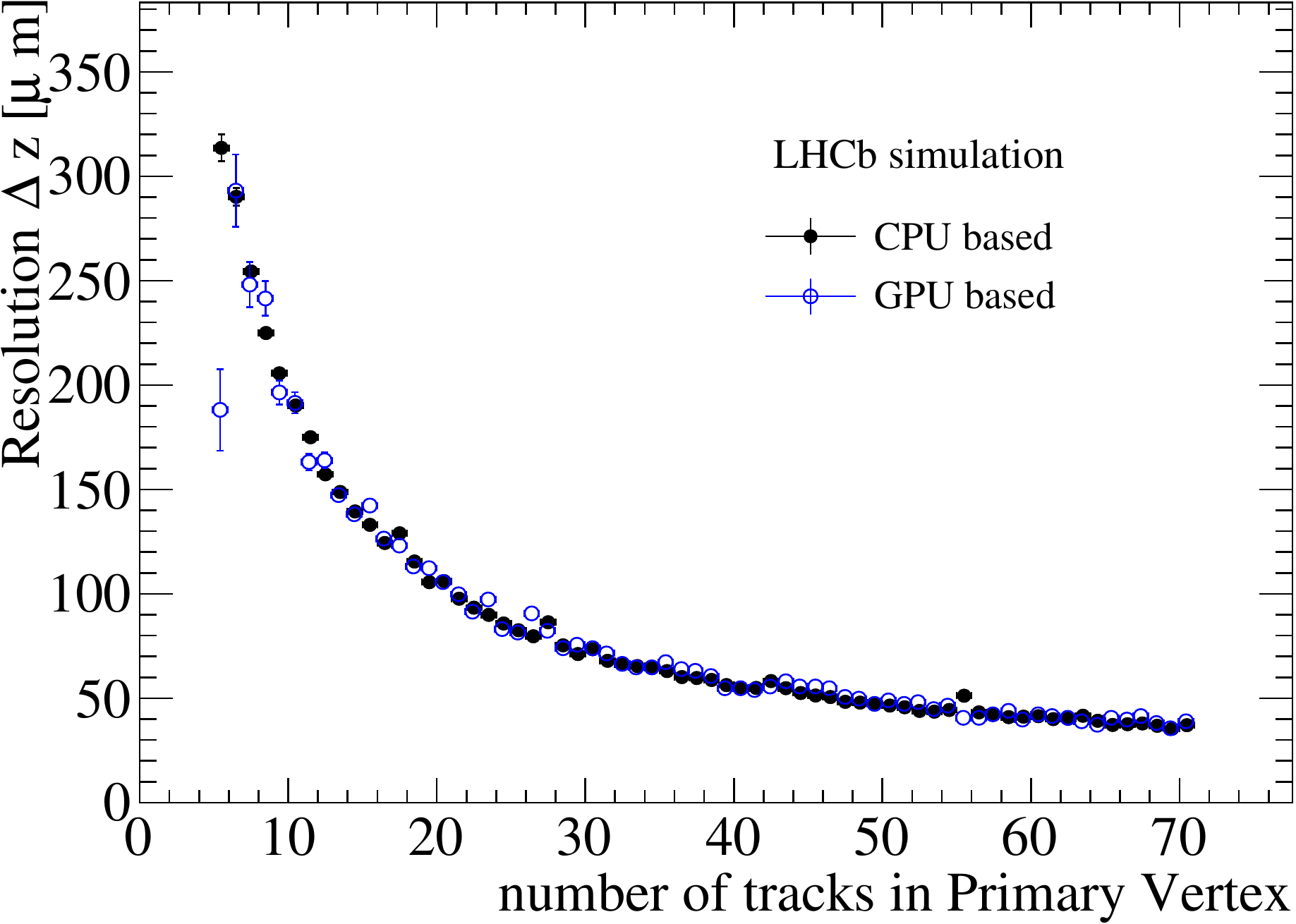}}

\mbox{
\includegraphics[scale=0.37]{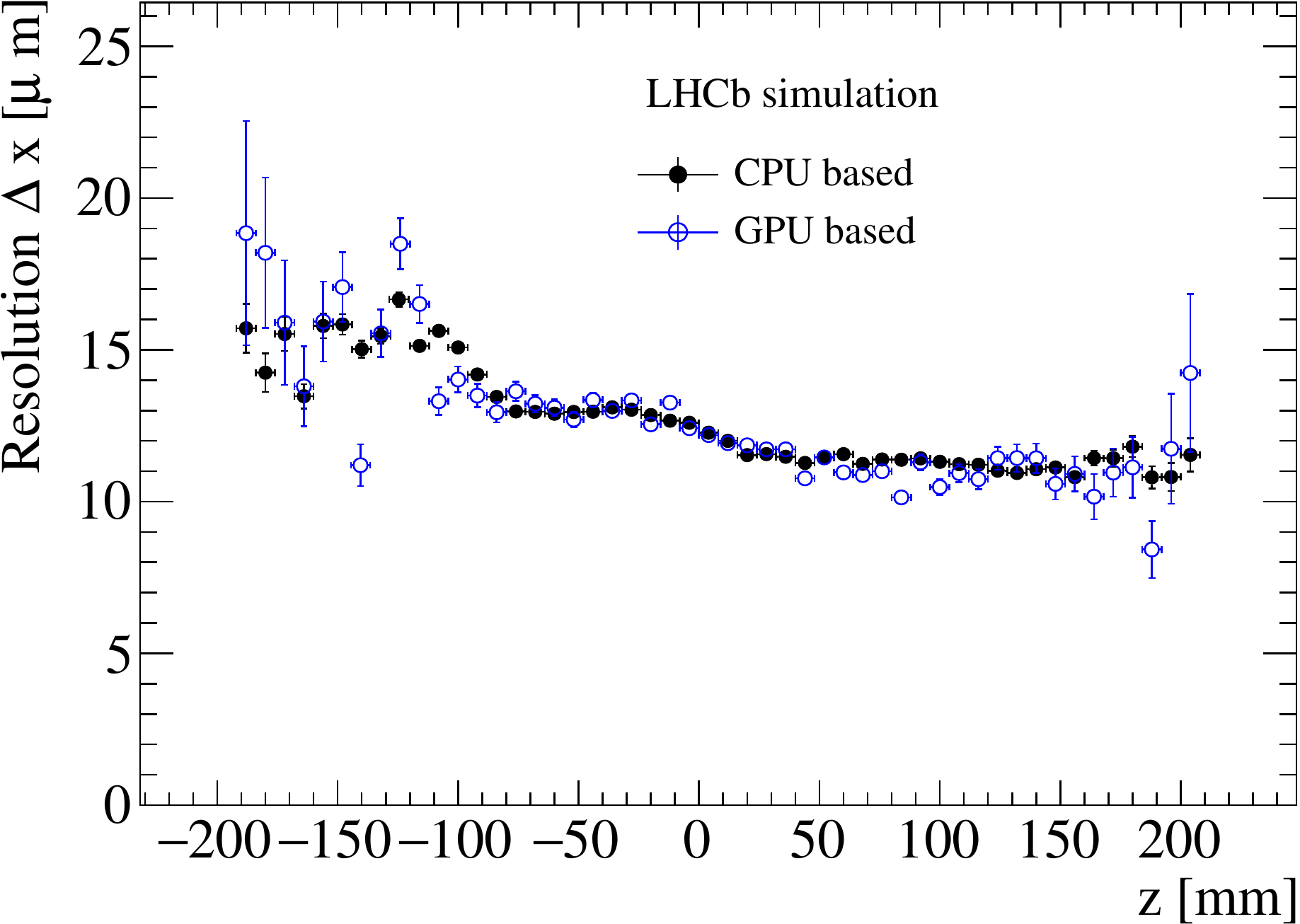}
\includegraphics[scale=0.37]{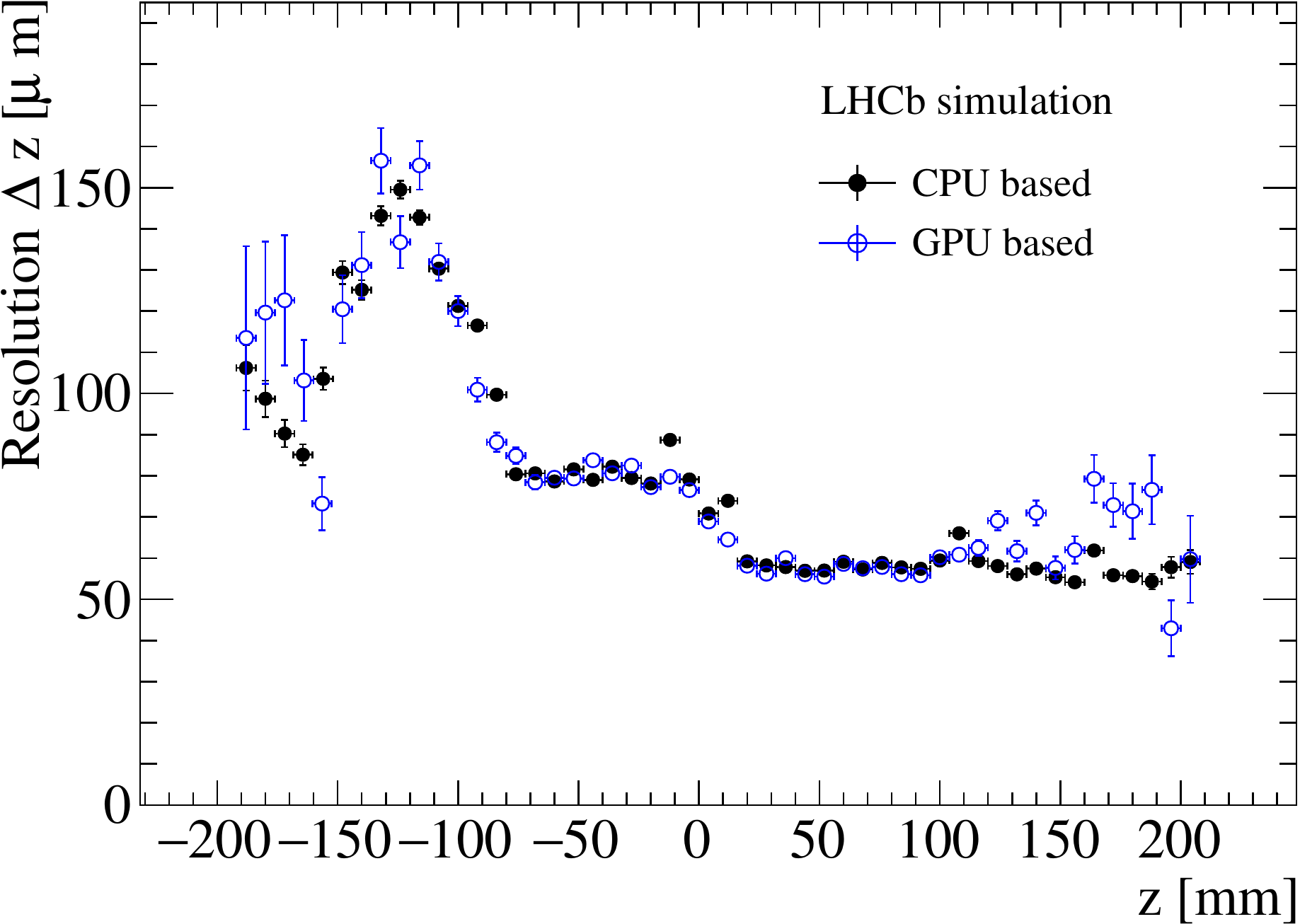}}

\caption[]{Resolution in $x$ and $z$ of all reconstructed primary vertices as function of the number of reconstructed Velo tracks  associated to the simulated primary vertex (top row) and as function of the true vertex $z$ position (bottom row).}
\label{fig:PV_resolution}
\end{figure}

A simulated primary vertex is defined as reconstructed if a primary vertex is found within 2~mm of its true position. The primary vertex efficiency as a function of the number of its reconstructed Velo tracks and of the primary vertex $z$ position is shown in Fig.~\ref{fig:PV_Eff}. All plots in this section are obtained on a sample of minimum bias events (Tab.~\ref{tab:MCsamples}).\\
~\\
The primary vertex resolution in $x$, $y$ and $z$ are studied as a function of the number of Velo reconstructible particles associated to the primary vertex and as a function of the $z$ position of the primary vertex. The results on minimum bias events are shown in Fig.~\ref{fig:PV_resolution}. 
The $x$ resolution of the primary vertex is very similar to the $y$ resolution, thus only one of them is shown. The average resolution of the minimum bias data set is $\sigma_x = \sigma_y \sim 14~\mu m$ and  $\sigma_z \sim 87~\mu m $ for both technologies.
The performances in terms of efficiency as well as in terms of resolution are close to identical for the both implementations.

\subsection{Muon ID efficiency}
\label{sec:MuonPerf}
\begin{figure}[t]
\centering
\mbox{
\includegraphics[scale=0.37]{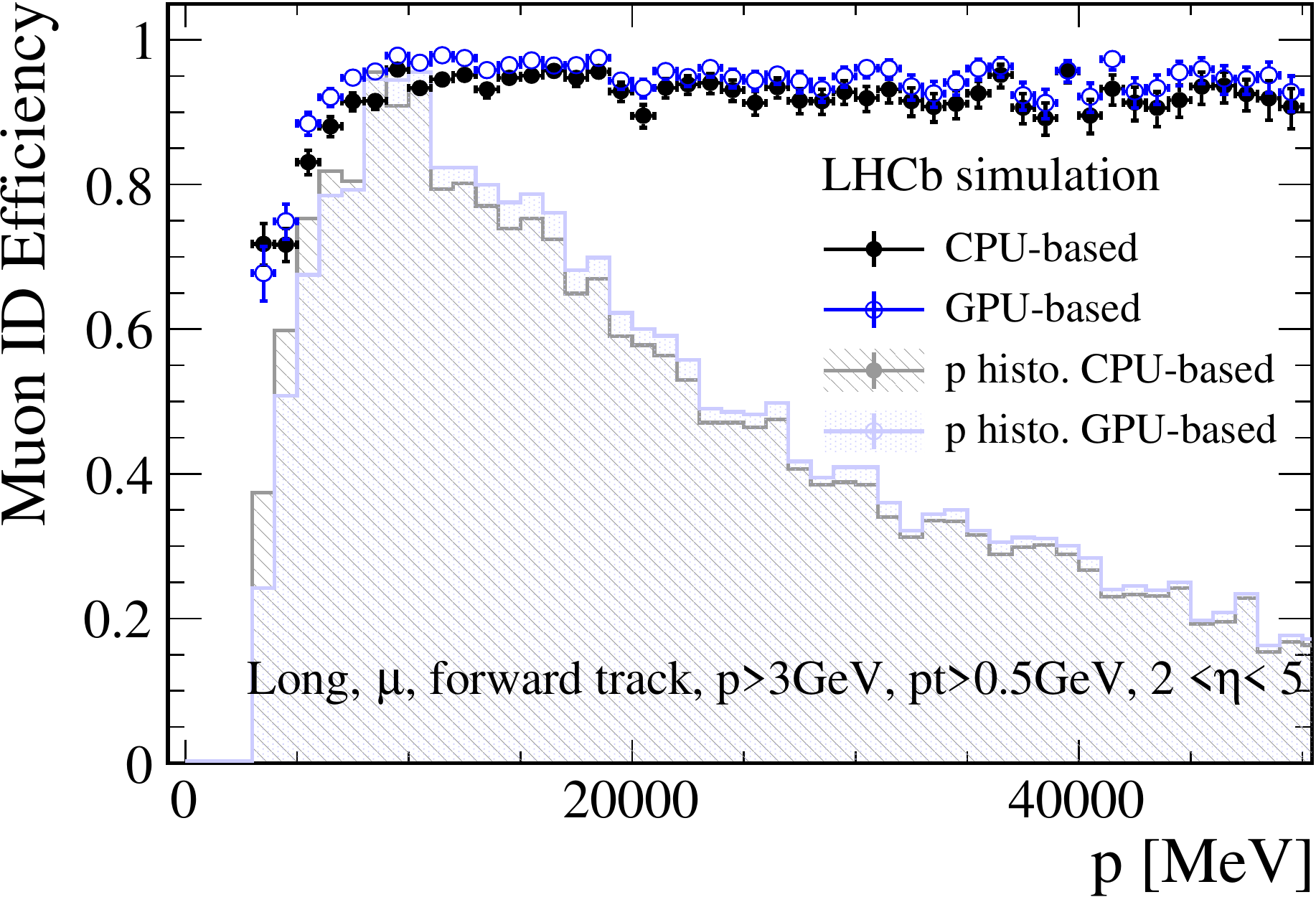}
\includegraphics[scale=0.37]{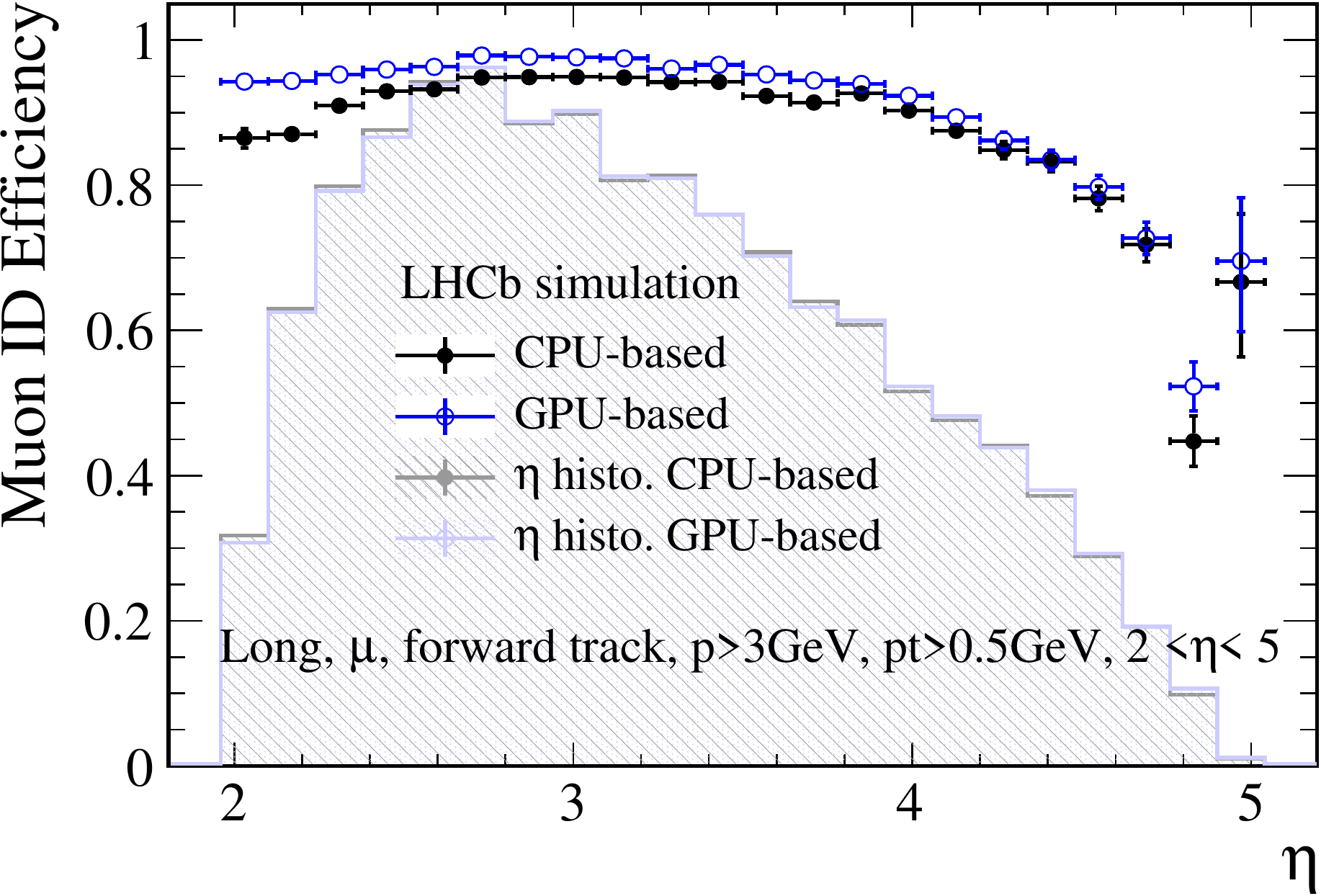}
}\
\caption[]{Efficiency to identify true muons reconstructed as long muon tracks as muons as function of the muon momentum and pseudorapidity.}
\label{fig:muon_eff}
\end{figure}

\begin{figure}[t]
\centering
\mbox{
\includegraphics[scale=0.37]{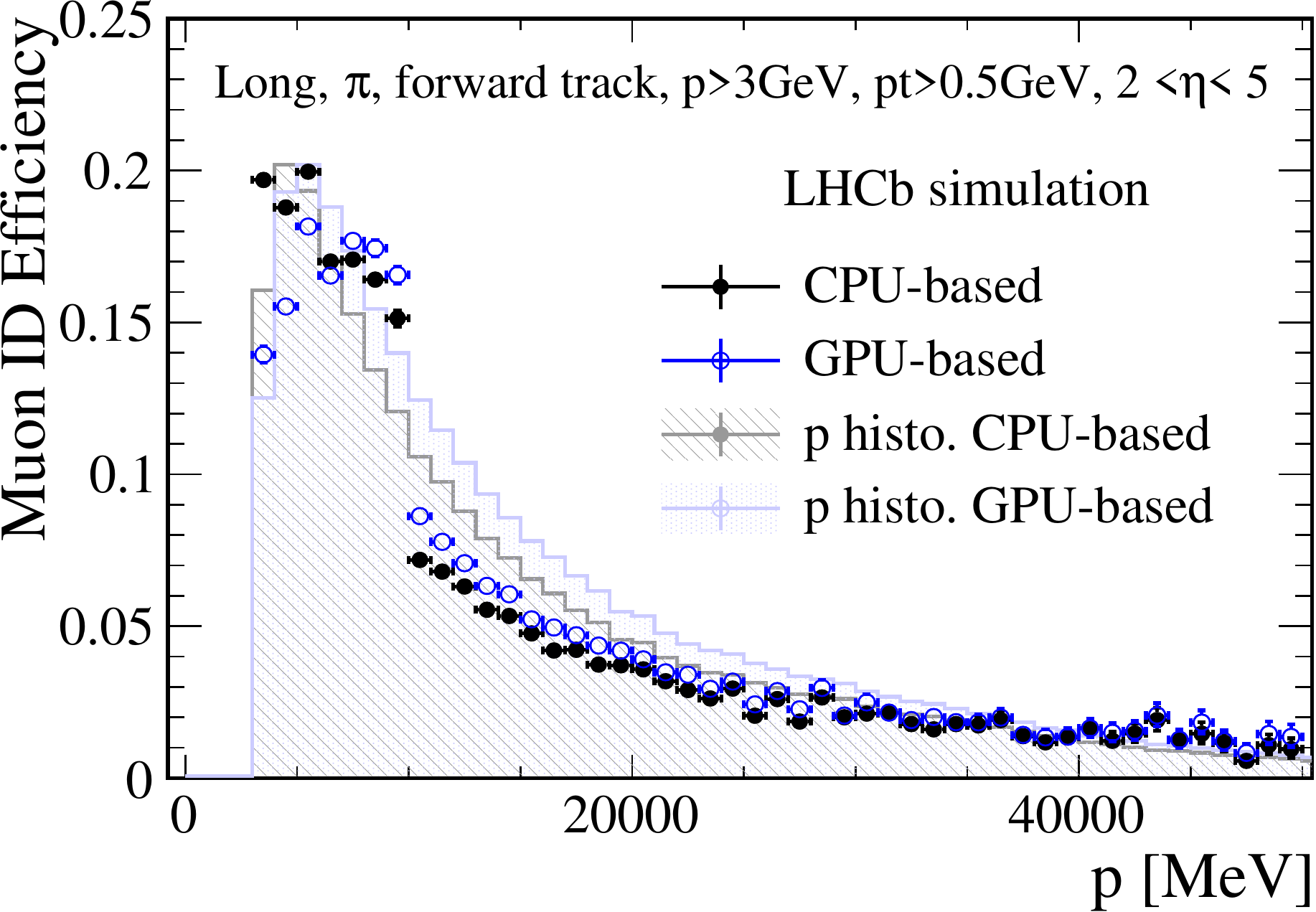}
\includegraphics[scale=0.37]{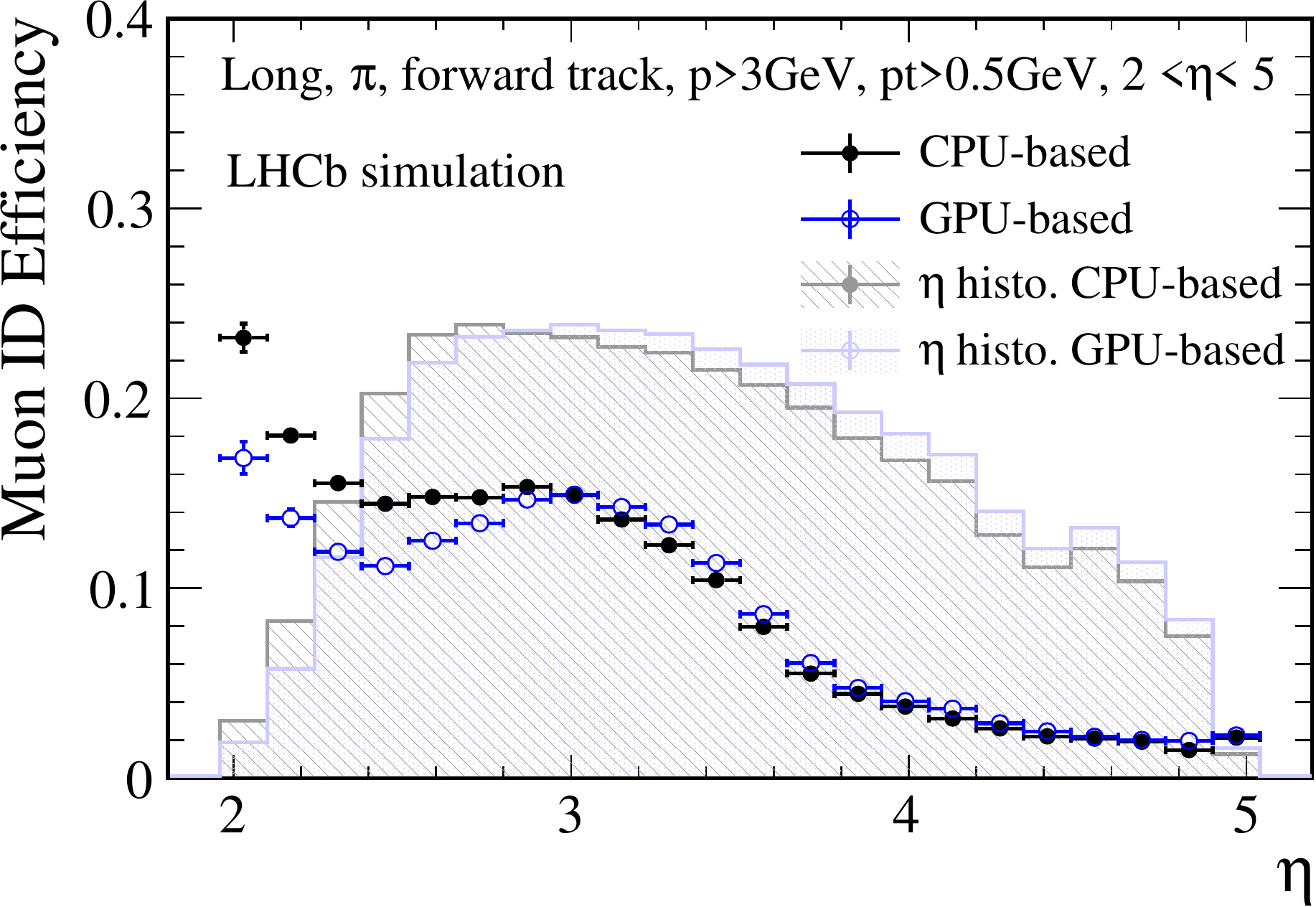}
}\
\caption[]{Efficiency to falsely identify true pions reconstructed as long muon tracks as muons as function of the pion momentum and pseudorapidity.}
\label{fig:muon_misID}
\end{figure}

The efficiency for the muon identification has been measured using the $J/\Psi \rightarrow \mu^+ \mu^-$,  $Z \rightarrow \mu^+\mu^-$ and $B^0 \rightarrow K^{*0}\mu^+\mu^-$ samples (Tab. \ref{tab:MCsamples}). The denominator of the efficiency term counts muons with a minimum momentum of $p \geq 3$ GeV/c and a minimum transverse momentum of $p_T \geq 0.5$~GeV/c in the pseudorapidity range $2 \leq \eta \leq 5$. Furthermore, they must be reconstructed as long tracks.
The numerator requires additionally these tracks to be identified as a muon in the detector. The efficiency is displayed in Fig.~\ref{fig:muon_eff} as a function of the momentum of the muon and as function of its pseudorapidity. A second performance criterion is the purity of the muon reconstruction. For this we count which fraction of pions with  a minimum momentum $p \geq 3$~GeV/c and a minimum transverse momentum of $p_T \geq 0.5$~GeV/c in the pseudorapidity range $2 \leq \eta \leq 5$ are misidentified as muons by the reconstruction algorithm (Fig. \ref{fig:muon_misID}). 
The GPU implementation of the muon identification gives better performance than the CPU version, with an absolute efficiency improvement of up to $10\%$ at small pseudorapidities. The misidentification rate is similarly better by up to $5\%$ absolute at small pseudorapidities and momenta for the GPU implementation, while it is a couple of percent better at higher momenta for the CPU implemntation . It is expected that it is a matter of throughput-neutral tuning to obtain similar results for the two implementations.


\subsection{HLT1 Efficiency}
\label{sec:HLT1PErf}

\begin{figure}[tb]
\centering
\mbox{
\includegraphics[scale=0.37]{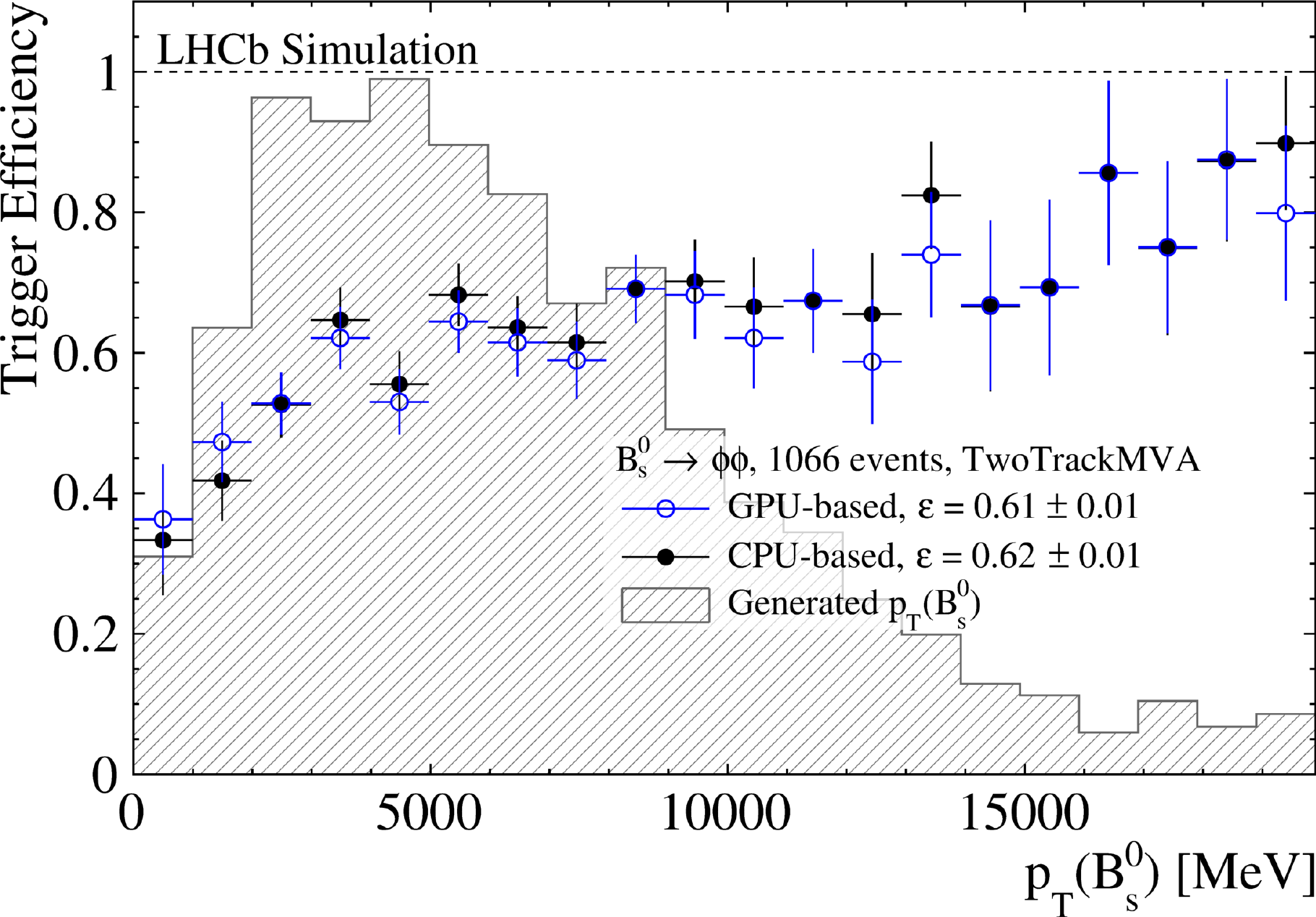}
\includegraphics[scale=0.37]{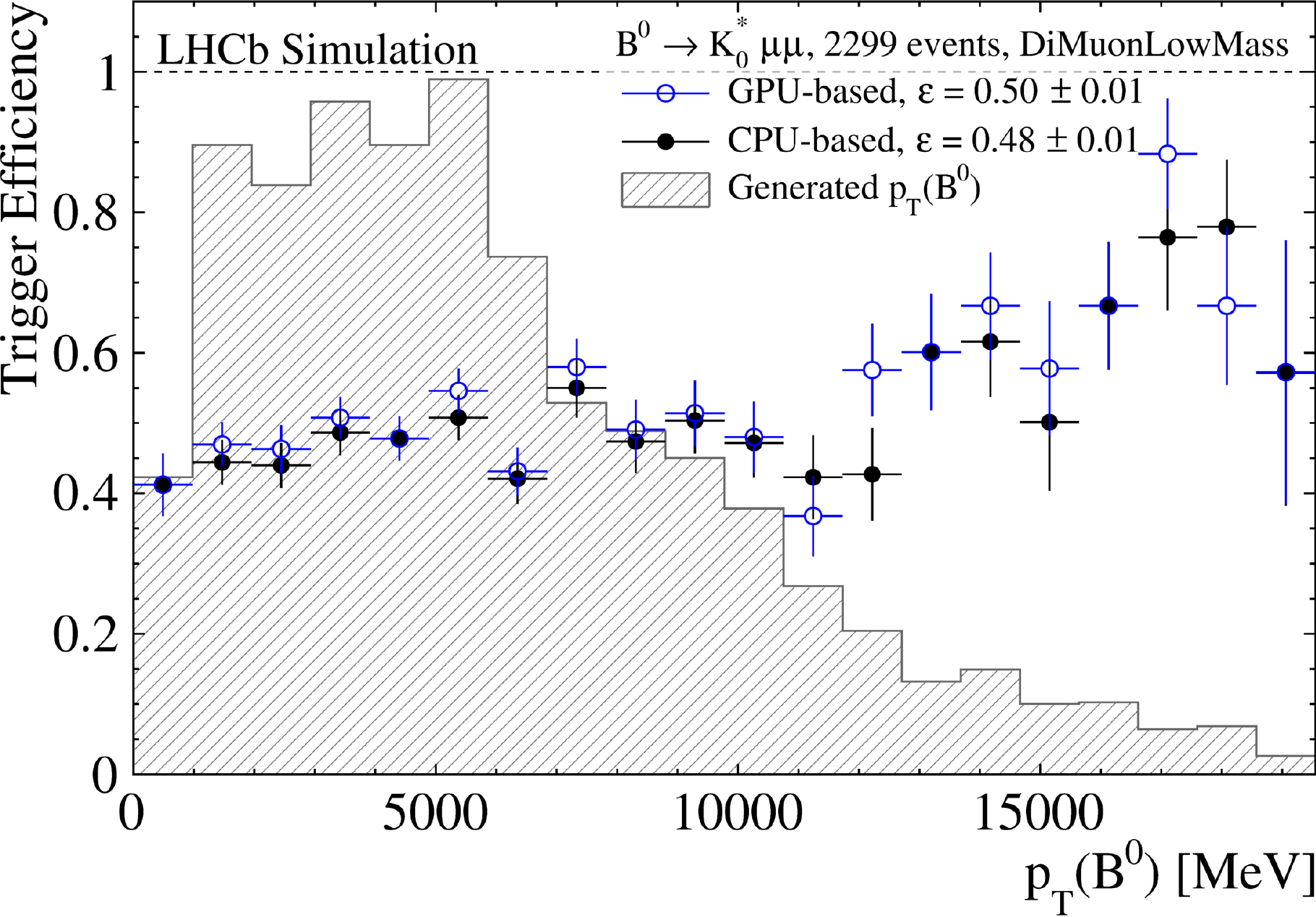}}
\mbox{
\includegraphics[scale=0.37]{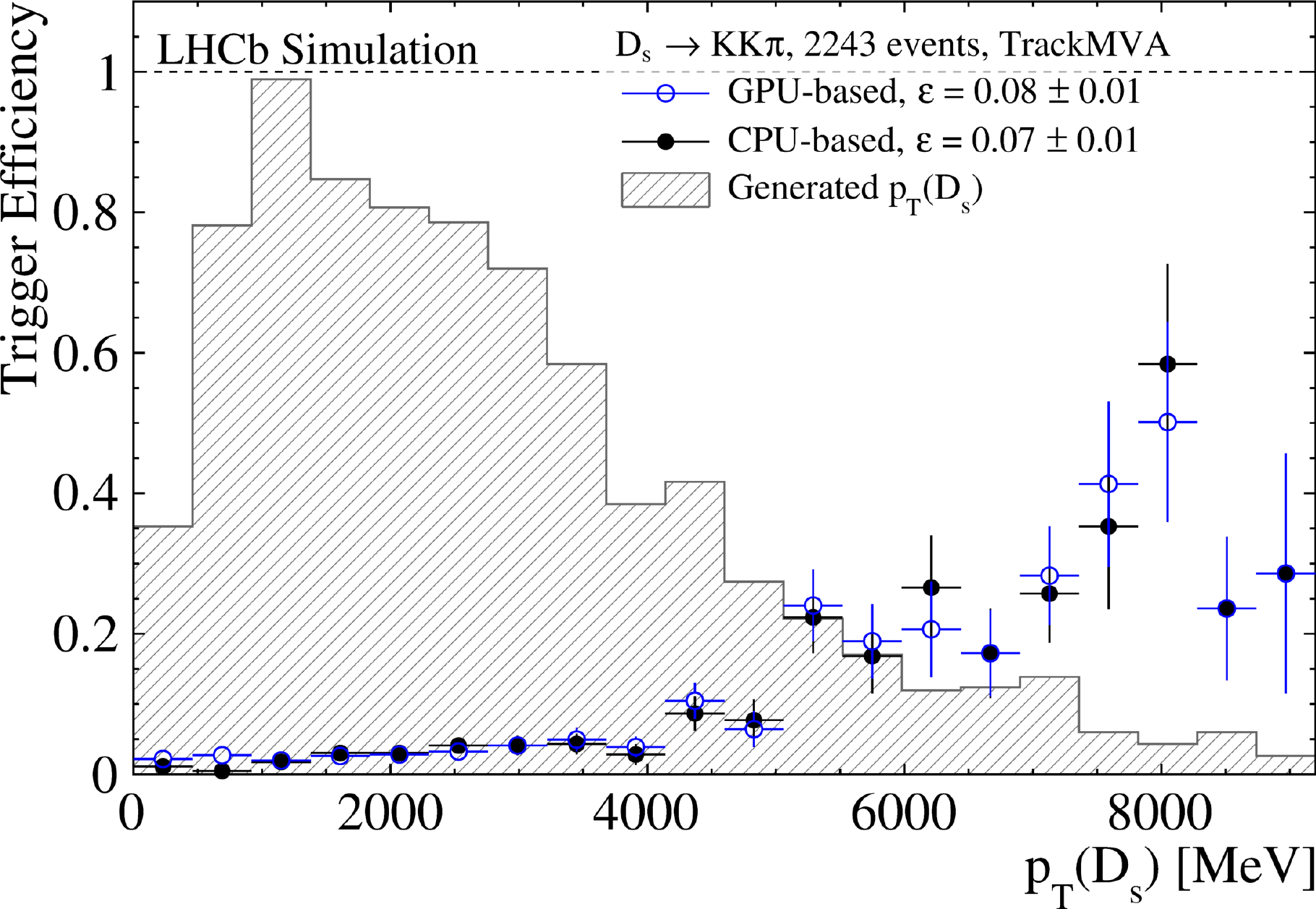}
\includegraphics[scale=0.37]{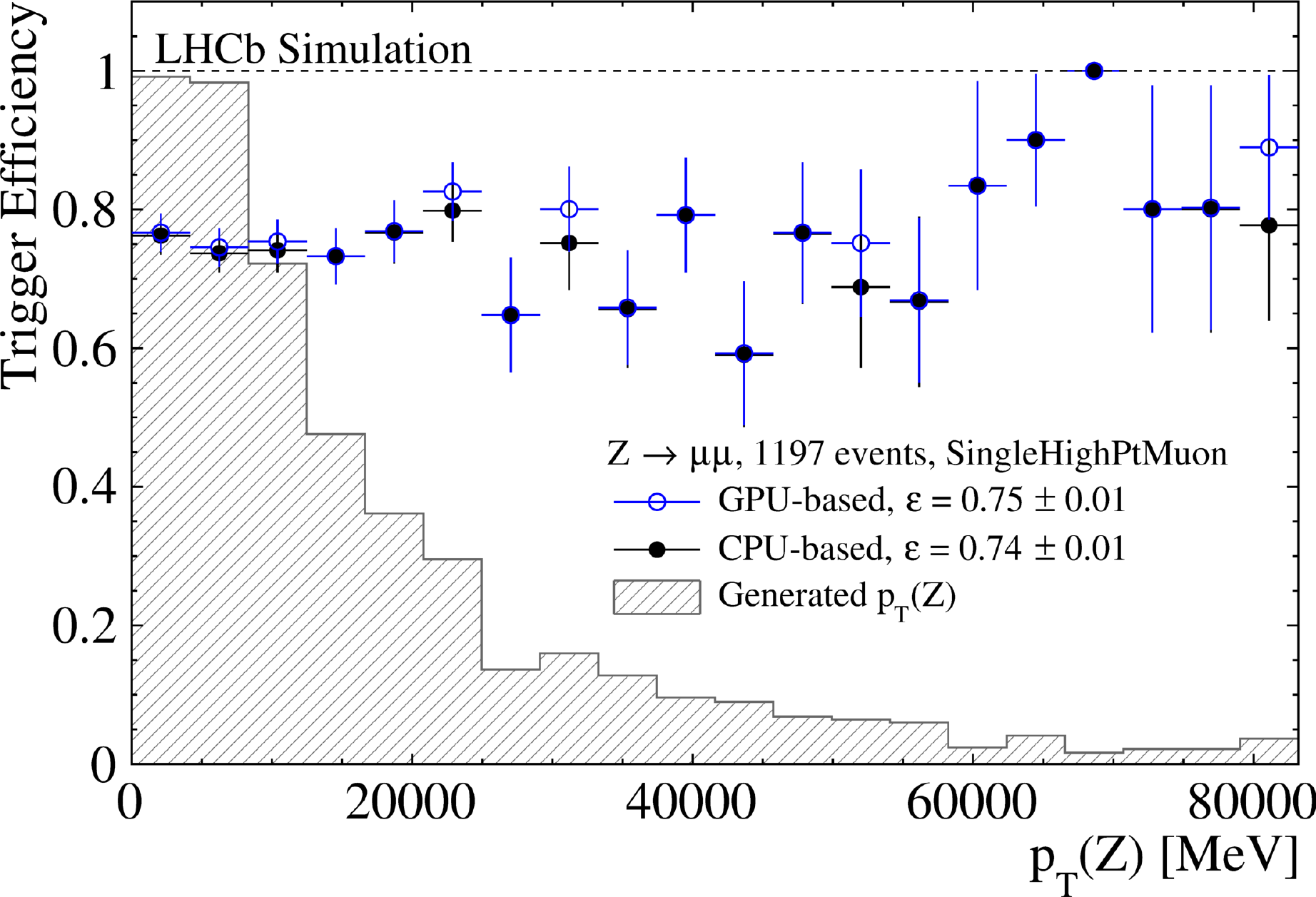}}
\caption[]{Trigger efficiencies for CPU-based and GPU-based HLT1 as a function of parent transverse momentum. Results are shown for the \texttt{TwoTrackMVA} (top left), \texttt{DiMuonLowMass} (top right), \texttt{TrackMVA} (bottom left) and \texttt{SingleHighPtMuon} (bottom right) selections firing on the $B^0_s \rightarrow \phi \phi$, $B^0 \rightarrow K^{*0} \mu^+\mu^-$, $D_s \rightarrow KK\pi$ and $Z\rightarrow \mu^+ \mu^-$ signal samples, respectively. The generated parent transverse momentum distribution is also shown for all events passing the denominator requirement.}
\label{fig:HLTEffs_CanRecoChildren_PT}
\end{figure}

\begin{figure}[htb]
\centering
\mbox{
\includegraphics[scale=0.37]{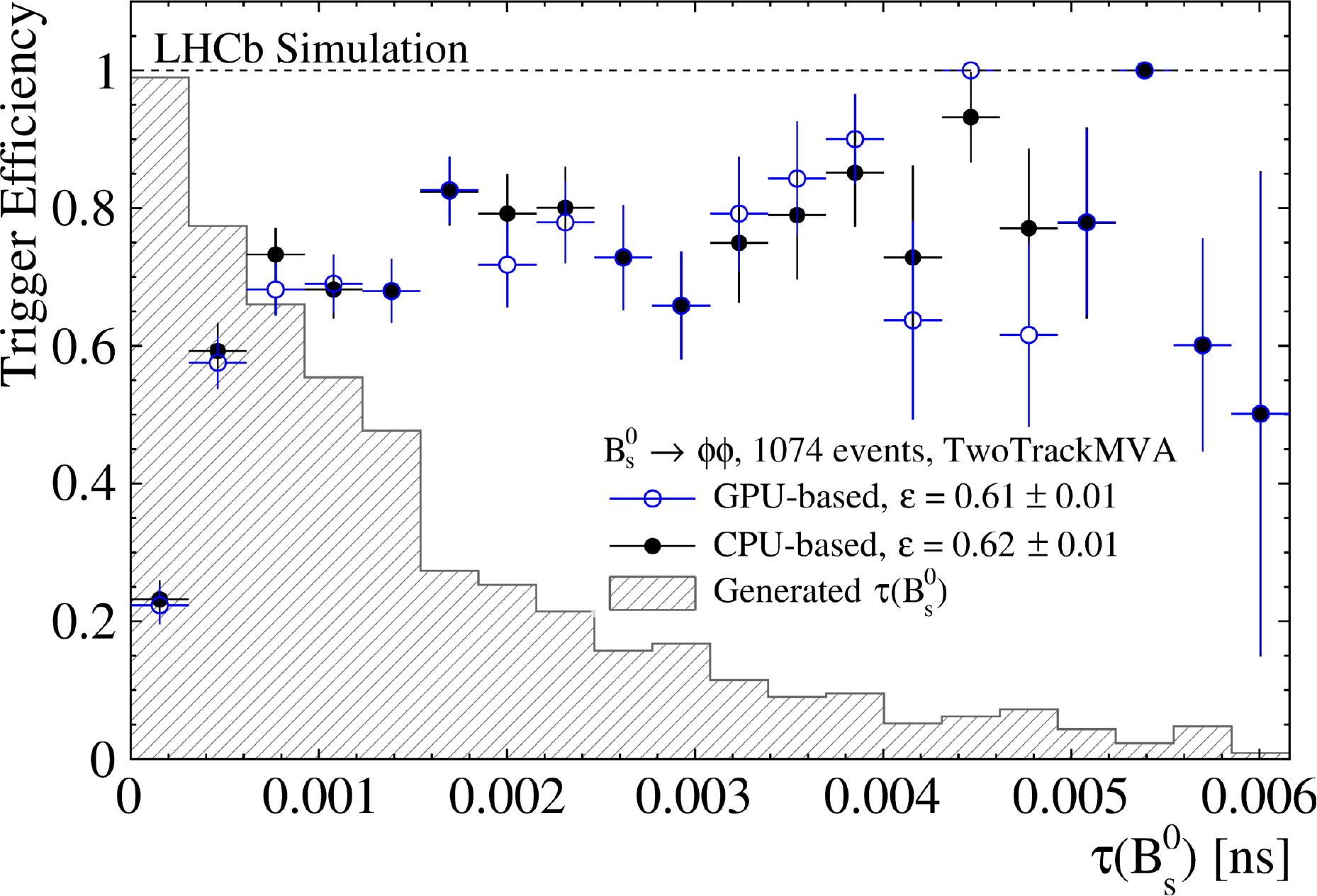}
\includegraphics[scale=0.37]{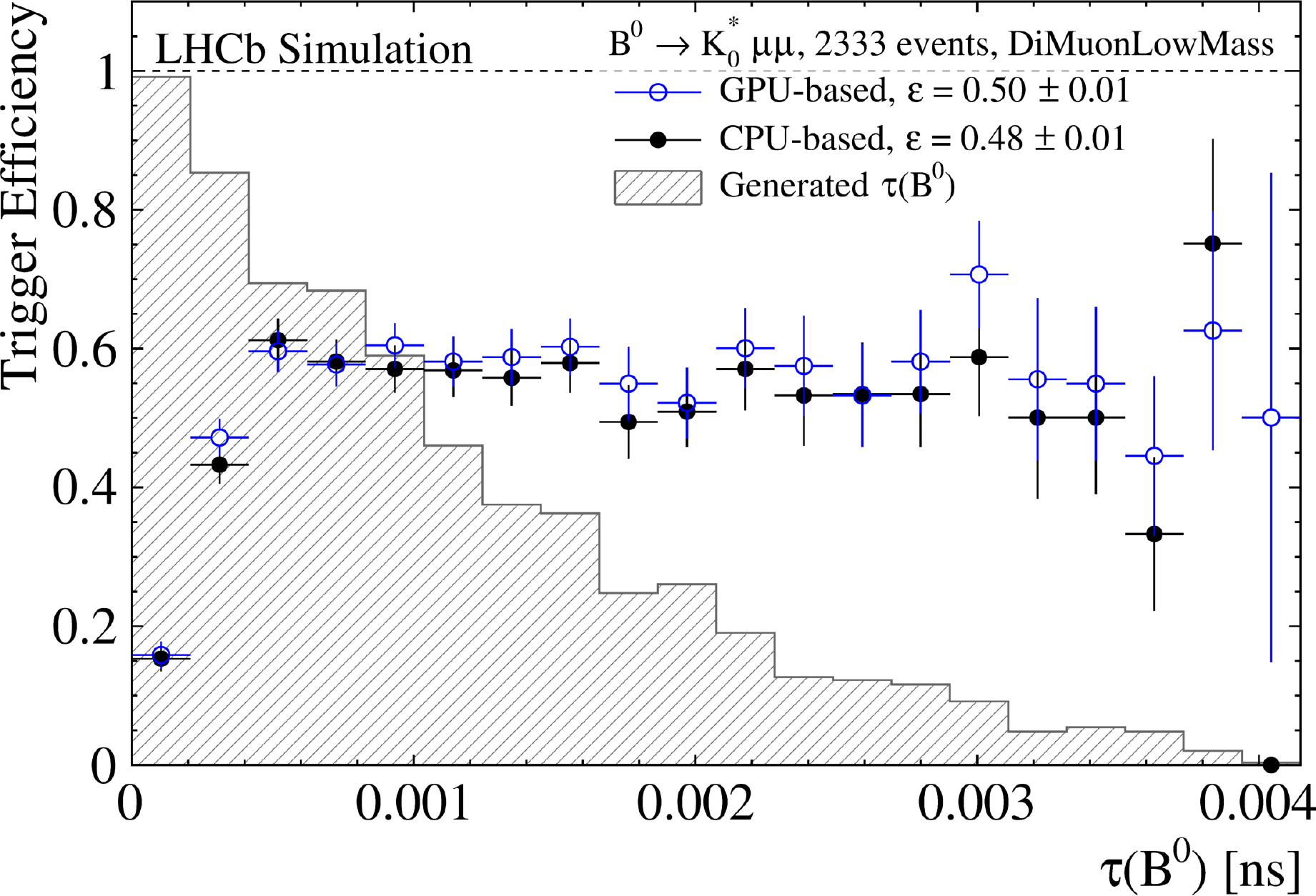}}
\mbox{
\includegraphics[scale=0.37]{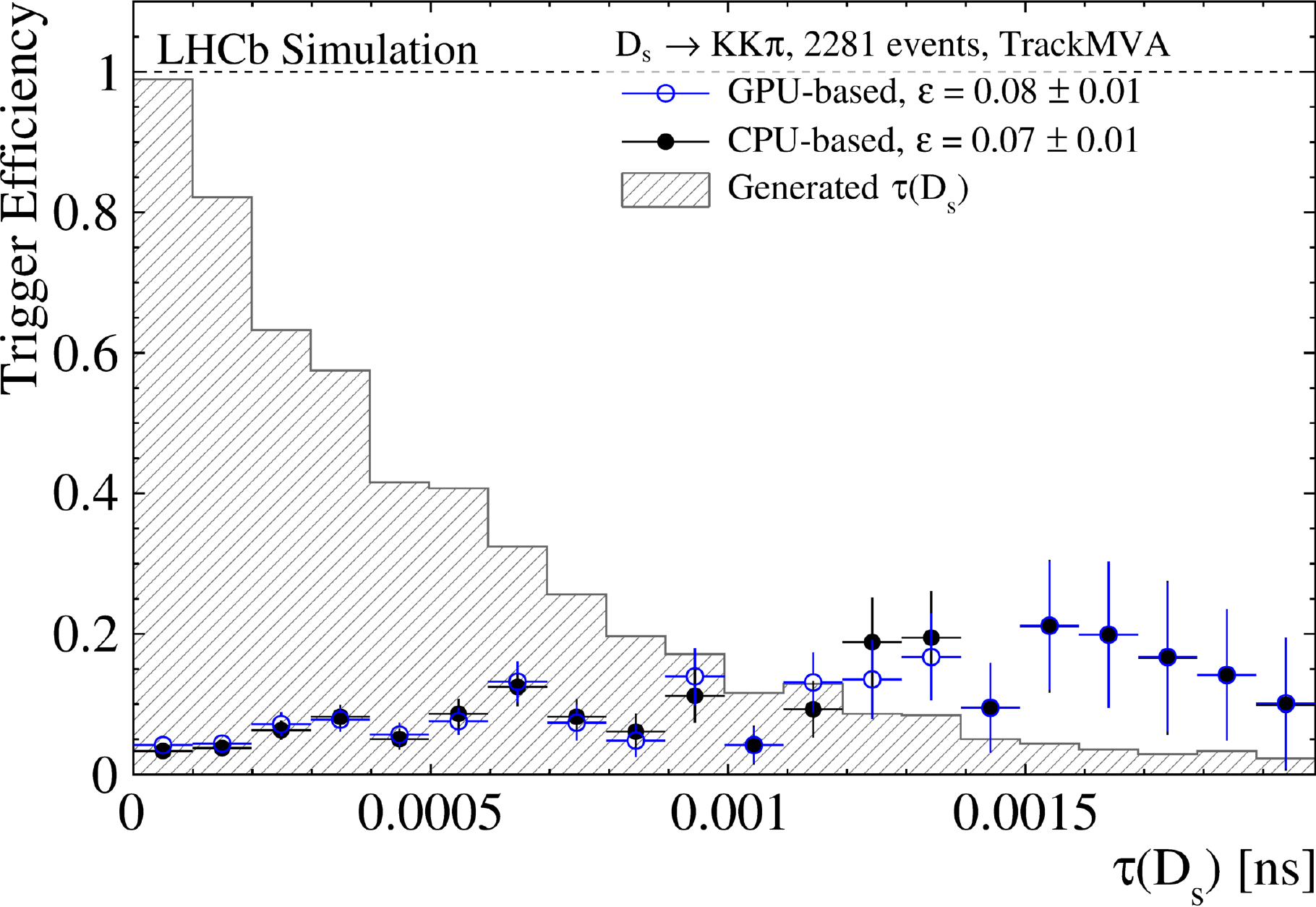}
}
\caption[]{Trigger efficiencies for CPU-based and GPU-based HLT1 as a function of parent decay time. Results are shown for the \texttt{TwoTrackMVA} (top left), \texttt{DiMuonLowMass} (top right) and \texttt{TrackMVA} (bottom) selections firing on the $B^0_s \rightarrow \phi \phi$, $B^0 \rightarrow K^{*0} \mu^+\mu^-$ and $D_s \rightarrow KK\pi$ signal samples, respectively. The generated parent decay time distribution is also shown for all events passing the denominator requirement.}
\label{fig:HLTEffs_CanRecoChildren_TAU}
\end{figure}

In this part of the study, the full HLT1 sequence of both the CPU and the GPU implementations are run on the simulated signal samples listed in Tab.~\ref{tab:MCsamples}, and the trigger efficiencies are then compared. The efficiency denominator contains all signal candidates for which
all stable charged daughter particles are reconstructible as long tracks within the detector
acceptance of $2 <  \eta  < 5$.
Thus, the  trigger efficiency is the number of selected events divided by the total number of events. We choose four generic trigger selections to perform the comparison: a one-track MVA line (\texttt{TrackMVA}), a two-track MVA line (\texttt{TwoTrackMVA}), a low-mass dimuon line (\texttt{DiMuonLowMass}) and a single, high-$p_{\textrm{T}}$ muon line (\texttt{SingleHighPtMuon}). The former two are inclusive selections designed to select generic $B$ and $D$ decays, the third is specialized for semileptonic $B$ decays, whereas the latter is more suitable for high-$p_\textrm{T}$ electroweak physics such as $Z \rightarrow \mu^+\mu^-$. The selections that each trigger line performs are as follows:
\begin{itemize}
    \item \textbf{TrackMVA}: requires a track with a minimum transverse momentum and a minimum significance of its impact parameter with respect to any primary vertex. These requirements are the same for the CPU- and GPU-based implementation, thus very similar performance is expected for this line.
    \item \textbf{TwoTrackMVA}: combines two displaced tracks of good quality with a minimum momentum to a vertex. In the CPU-based implementation a multivariate analysis (MVA) classifier is trained to identify those vertices which originate from a decay of a long-lived particle such as a beauty or charmed hadron. An MVA is not yet in place for the GPU-based implementation, so percent level differences in the signal efficiencies are expected for a given background retention.
    \item \textbf{DiMuonLowMass}: requires a two track combination of displaced tracks with very low momentum and transverse momentum. This trigger line is implemented identically for the CPU- and GPU-based HLT1. 
    \item \textbf{SingleHighPtMuon}: selects tracks which are identified as a muon and fulfill a minimum momentum and transverse momentum requirement. This trigger line is also implemented identically for both architectures.
\end{itemize}
In Fig.~\ref{fig:HLTEffs_CanRecoChildren_PT} and \ref{fig:HLTEffs_CanRecoChildren_TAU}, the trigger efficiencies of the CPU- and GPU-based HLT1 implementations are plotted against the parent transverse momentum, $p_{\textrm{T}}$, and parent decay time, $\tau$ (where applicable), respectively, for four signal samples. For ease of interpretation, only the efficiency of one suitably chosen trigger line per sample is shown. The trigger efficiencies  of the two implementations are found to be comparable. 
In Tab.~\ref{tab:IntEffsMVA} and \ref{tab:IntEffsMuons} the trigger efficiencies, integrated across the kinematic phase space of the samples, are compared for all four selections and various simulated signal samples. They are found to be comparable for the GPU and CPU-based implementation.

\begin{table}
  \centering
  \begin{tabular}{|c|c|c|c|c|}
    \hline
     &  \multicolumn{2}{c|}{\texttt{TrackMVA}} &
    \multicolumn{2}{c|}{\texttt{TwoTrackMVA}} \\
    Signal & GPU & CPU & GPU & CPU \\
    \hline
    \multirow{1}{*}{$B^0_s \rightarrow \phi \phi$} & 0.340(14) & 0.332(14) & 0.606(15) & 0.621(15) \\ \hline
    \multirow{1}{*}{$J/\psi \rightarrow \mu^+ \mu^-$} & 0.034(4) & 0.031(3) & 0.049(4) & 0.042(4) \\ \hline
    \multirow{1}{*}{$B^0 \rightarrow K^{*0} e^+e^-$} & 0.276(10) & 0.278(10) & 0.439(12) & 0.473(12) \\ \hline
    \multirow{1}{*}{$B^0 \rightarrow K^{*0} \mu^+\mu^-$} & 0.391(10) & 0.385(10) & 0.554(10) & 0.582(10) \\ \hline
    \multirow{1}{*}{$D_s \rightarrow KK\pi$} & 0.076(5) & 0.073(5) & 0.178(8) & 0.193(8) \\ \hline
    \multirow{1}{*}{$Z\rightarrow \mu^+ \mu^-$} & 0.051(6) & 0.040(6) & 0.024(4) & 0.028(5) \\ \hline
  \end{tabular}
  \caption{\label{tab:IntEffsMVA}Comparison of trigger efficiencies integrated over the kinematic phase space of the candidates, for each of the six simulated signal samples for the \texttt{TrackMVA} and \texttt{TwoTrackMVA} selections. Statistical uncertainties are indicated in parentheses.}
\end{table}

\begin{table}
  \centering
  \begin{tabular}{|c|c|c|c|c|c|}
    \hline
     & \multicolumn{2}{c|}{\texttt{DiMuonLowMass}} &
    \multicolumn{2}{c|}{\texttt{SingleHighPtMuon}} \\
    Signal & GPU & CPU & GPU & CPU \\
    \hline
    \multirow{1}{*}{$B^0_s \rightarrow \phi \phi$} &  0.025(5) & 0.024(5) & 0.005(2) & 0.004(2) \\ \hline
    \multirow{1}{*}{$J/\psi \rightarrow \mu^+ \mu^-$} & 0.078(5) & 0.067(5) & 0.048(4) & 0.045(4) \\
    \hline
    \multirow{1}{*}{$B^0 \rightarrow K^{*0} e^+e^-$} & 0.024(4) & 0.027(4) & 0.0011(8) & 0.0011(8) \\ \hline
    \multirow{1}{*}{$B^0 \rightarrow K^{*0} \mu^+\mu^-$} & 0.502(10) & 0.482(10) & 0.091(6) & 0.088(6) \\ \hline
    \multirow{1}{*}{$D_s \rightarrow KK\pi$} &  0.018(3) & 0.019(3) & 0.0013(7) & 0.0013(7) \\ \hline
    \multirow{1}{*}{$Z\rightarrow \mu^+ \mu^-$} &  0.033(5) & 0.036(5) & 0.749(12) & 0.740(13) 
    \\ \hline
  \end{tabular}
  \caption{\label{tab:IntEffsMuons}Comparison of trigger efficiencies integrated over the kinematic phase space of the candidates, for each of the six MC signal samples and the \texttt{DiMuonLowMass} and \texttt{SingleHighPtMuon} selections. Statistical uncertainties are indicated in parentheses.}
\end{table}

\subsection{HLT1 rates}

\begin{figure}[htb]
\centering
\includegraphics[scale=0.7]{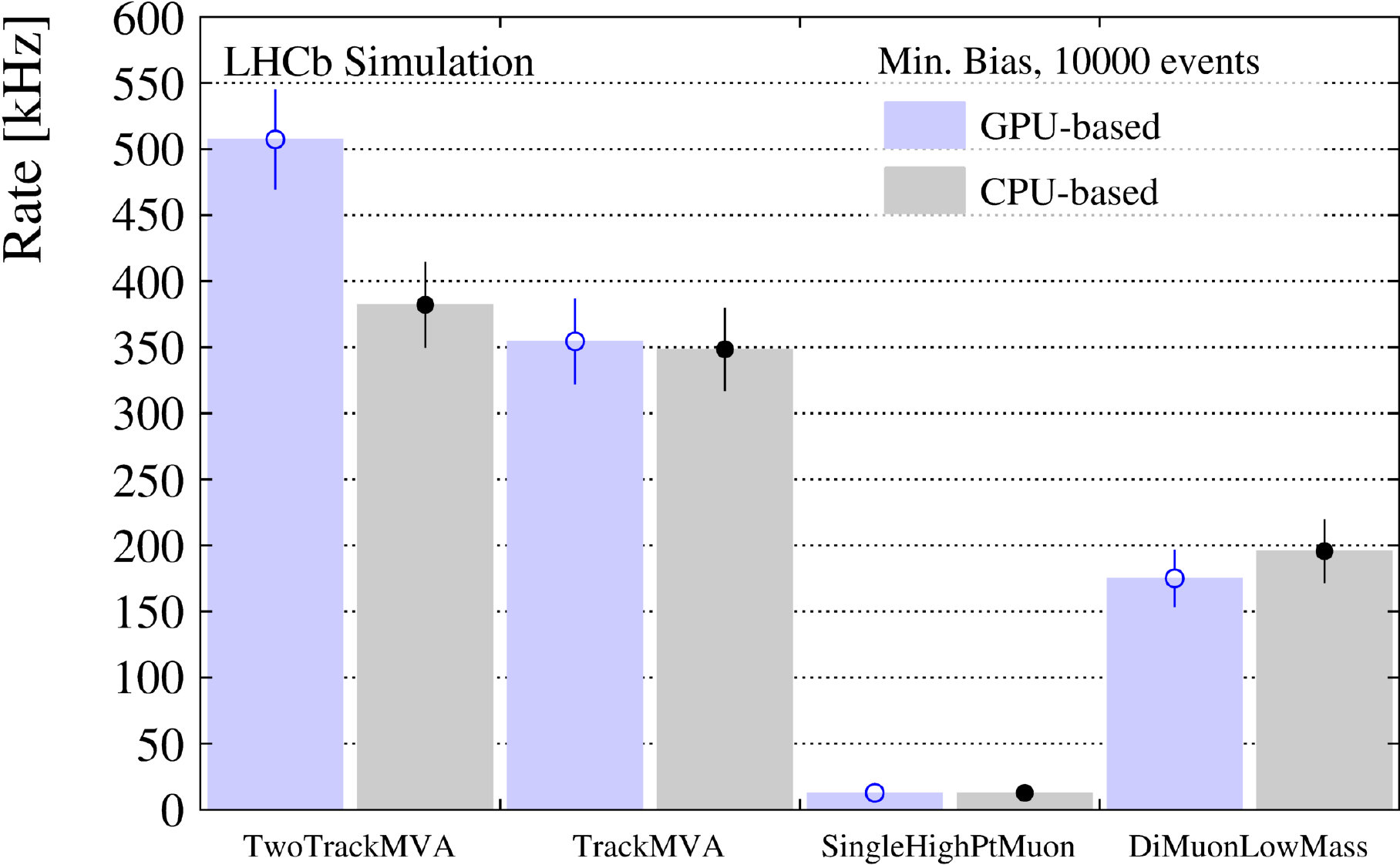}

\caption[]{Trigger rates of the CPU- and GPU-based implementation for the four trigger selections of interest. The difference in the rate of the TwoTrackMVA selections can be explained by the differing implementation, which is detailed in Section \ref{sec:HLT1PErf}.}
\label{fig:HLTRates}
\end{figure}

HLT1 rates are calculated in a similar way to the HLT1 efficiencies in the previous subsection. Both the CPU- and GPU-based implementation are run over 10k minimum bias events, and the positive decisions on each of the four selections defined in the previous subsection are counted. The rate for each line is defined as the number of events that fire that line, divided by the number of minimum bias events that are sampled, multiplied by the LHCb non-empty bunch crossing rate (30 MHz). Note that, contrary to the efficiency studies on MC signal samples of the previous subsection, no preselection (or ``denominator'' requirement) is applied. The comparison of the HLT1 rates is shown in Fig.~\ref{fig:HLTRates}. The \texttt{TrackMVA}, \texttt{DiMuonLowMass} and \texttt{SingleHighPtMuon} selections have comparable rates, although there is a discrepancy in the rates of the respective \texttt{TwoTrackMVA} selections. This discrepancy can be explained by the different implementation of this line across the two projects, as detailed in the previous subsection. The inclusive rate for these four selections is found to be 912 $\pm$ 52 kHz for the GPU-based implementation, and 798~$\pm$~48~kHz for the CPU-based one, largely due to the different implementation of the  \texttt{TwoTrackMVA} trigger line. These numbers are well within the requirements of HLT1 to output between $1-2$~MHz of events for further processing.

\subsection{Global Event Cut efficiency}
\label{sec:geceffs}
Both HLT1 implementations apply an identical global event cut (GEC) requiring fewer than 9750 total SciFi and UT clusters. Consequently, the GEC efficiencies for the GPU- and CPU-based implementation are found to be identical across all samples. The integrated efficiency of $Z \rightarrow \mu^+\mu^-$ is found to be 0.75 $\pm$ 0.01. The other $B$ and $D$ decay samples under study have GEC efficiencies of about 85\%, with statistical uncertainties of $\sim1\%$. The efficiency on minimum bias events is 0.931 $\pm$ 0.003.

\subsection{Summary of the physics performance}
The GPU- and CPU-based implementations presented here result in close to identical performance in all aspects. The observed differences are considered to be a matter of tuning the algorithms to balance between efficiency and fake rate, or misidentification rate, and are expected to be neutral in terms of throughput. We therefore conclude that only the economic costs and the costs in terms of developer time need to be further considered in the cost-benefit calculation for the two architectures.

\section{Cost benefit analysis}
\label{sec:CostBenefit}
In this section the information of the previous sections is brought together to estimate the cost required for each
architecture to process 30 MHz of data at the nominal instantaneous luminosity of $2\cdot 10^{33}$~cm$^{-2}$s$^{-1}$. As computing prices are in rapid flux, and any real purchase would be subject to tendering, we give here relative ``costs'' in units of the reference ``Quanta'' CPU server node used for the HLT during Run~2 datataking.
\subsection{Nominal cost-benefit}
\label{sec:sub:CostBenefitNominal}

\begin{table}[]
    \centering
    \begin{tabular}{|l|r|r|r|}
    \hline
    \textbf{Item} & \textbf{CPU-only} & \textbf{hybrid} & \textbf{Difference} \\
    \hline
        Event Builder nodes & 1000 & 1000 & 0 \\
        HLT1 network & 275 & 25 & 250 \\
        HLT1 compute & 450 & 125 & 325 \\
        Storage for 1 MHz output & 575 & 575 & 0 \\ 
        \textbf{Sub-total} & 2300 & 1725 & 575 \\
        Storage add. cost 2 MHz output &  575 & 575 & 0 \\ 
        \textbf{Total} & 2875 & 2300 & 575 \\
    \hline
    \end{tabular}
    \caption{Indicative overall cost of the  HLT1 implementations including contingency in units of the reference ``Quanta'' CPU server node used for the HLT during Run~2 datataking. Numbers have been rounded to reflect inevitable order(10\%) fluctuations in real-world costs depending on the context of any given purchase.
    \label{tab:cost}}
\end{table}

The nominal cost-benefit analysis is based on the assumption that the LHC will run at 30~MHz with full luminosity from the start of data taking. It is summarised in Tab.~\ref{tab:cost}. 
The difference in capital expenditure costs for the CPU-only and the hybrid scenario is the sum of the difference in HLT1 compute costs and the HLT1 network costs. 
For the HLT1 compute costs of the CPU-only scenario, it should be noted that additional usage of the EFF is foreseen, namely for processing simulation when the LHC is in a technical stop or end-of-year shutdown (``out-of-data-taking'') and for processing HLT2 when the LHC is between fills (``out-of-fill''), respectively. While the out-of-fill use of the CPU nodes for HLT2 processing is taken into account in the following studies, the usage for simulation out-of-data taking is not.

\subsection{Disk buffer simulation}
\label{sec:sub:disk_buffer_sim}
Based on the boundary conditions outlined in 
Section~\ref{sec:BoundaryConditions}, studies of the expected maximum HLT1 output rate that can be processed within the budget envelope have been performed. The studies are performed as follows: 
\begin{enumerate}
\item Fill lengths and inter-fill gaps are randomly sampled from 2018 data, with three machine development and technical stop periods and an average machine efficiency of 50\%;
\item The fills and gaps are grouped into pairs before sampling to capture potential correlations between them;
\item A thousand such toys are randomly generated, and are shown to have a residual distribution compatible with that of the 2018 machine efficiency and a width of $\pm1.7\%$; \item The disk buffer required to ensure sufficient space for 95\% of these 1000 toys is then determined using as input an event size of 100~kB, a chosen HLT1 output rate, and a chosen HLT2 throughput in and out of fill;
\item These values are scanned, generating 1000 toys per datapoint, over  a range of HLT1 output rates and HLT2 throughput rates resulting in a 3 dimensional distribution of required disk buffer as a function of HLT1 and HLT2 rates. 
\end{enumerate}
For each datapoint in the distribution, the cost of the HLT2 throughput and cost of the disk buffer are determined. For combined costs greater than that of the overall budget in the CPU-only and hybrid scenarios, the datapoint is rejected. This leaves a distribution of valid points for which LHCb could purchase the necessary resources. The optimal working point is the one which maximises the HLT1 output rate. The inputs to this procedure are summarized in Tab.~\ref{tab:bufferinputs} and described in more detail in the following sections.\\
~\\
The simulation studies assume that in both the CPU-only and hybrid scenarios, any remaining budget after attaining a 30~MHz throughput at HLT1 and a sufficient buffer is used to buy CPU to provide additional HLT2 throughput. This  corresponds to an increased HLT1 output rate allowing more, or more efficient selections at HLT1.

\subsubsection{Quantification of HLT1 and HLT2 througputs}

\begin{table}[htb]
    \centering
    \begin{tabular}{|l|c |c|c|}\hline
         Node type & HLT1 Throughput/node & HLT2 Throughput/node  & minimum nodes\\\hline
       AMD 7502  &171~kHz & 471~Hz & 175\\
       Quanta & n/a & 134~Hz & 2050\\
       \hline
    \end{tabular}
    \caption{Summary of HLT1 and HLT2 throughputs for the EFF nodes described in Sec.~\ref{sec:Throughput}. A minimum of 175 AMD 7502 nodes would be required to sustain 30~MHz of HLT1 throughput, corresponding to 82.5~kHz of out-of-fill HLT2 throughput. This is supplemented by 2050 Quanta equivalent nodes which are only used for HLT2 in-fill and out-of-fill, providing an additional 275kHz of HLT2 throughput.}
    \label{tab:costbenefitminimum}
\end{table}

The planned EFF nodes are equivalent to dual-socket AMD 7502 servers.
The HLT1 throughput for this node was given in Section~\ref{sec:Throughput} requiring 175 nodes to sustain 30~MHz.
As described in Section~\ref{sec:Throughput} these nodes are also capable of an HLT2 throughput of 471~Hz. The 
baseline scenario then results in an out-of-fill HLT2 throughput of 82.5~kHz and the remaining funds can be used to purchase disk or additional CPU. 
In the GPU scenario the purchase of these nodes for HLT1 is not necessary, but as a result no additional out-of-fill rate is available for HLT2 beyond the free resources described in the next subsection. The throughputs and quantities of these nodes before cost-benefit optimisation are presented in Table~\ref{tab:costbenefitminimum}.

\subsubsection{Additional free computing resources for HLT2}
\label{sec:sub:budget}
The Run 2 EFF Quanta nodes currently sustain a throughput of  134~Hz for HLT2, as documented in Section~\ref{sec:Throughput}. In addition to the most recently purchased Run 2 EFF nodes, older nodes are available which correspond to 1450 quanta-equivalents, with up to 1200 additional nodes. These nodes will only be used for HLT2. In this study we take the 1450 nodes and assume a conservative 50\% of the additional nodes, totalling 2050 quanta equivalents, or 275~kHz of in- and out-of-fill HLT2 throughput.

\begin{table}[]
    \centering
    \begin{tabular}{|l|c|c|}\hline
        Scenario & CPU-only & hybrid 
        \\\hline
        Maximum usable disk (PB) & 28 & 28 \\
        In-fill min. HLT2 rate (kHz) & 275 & 275 \\
        Out-of-fill min. HLT2 rate (kHz) & 358 & 275 \\\hline
    \end{tabular}
    \caption{Input parameters to the toy studies used to determine the maximum HLT1 output rate in each scenario.}
    \label{tab:bufferinputs}
\end{table}

\subsubsection{Results}
\begin{figure}
    \centering
    \includegraphics[width=0.45\textwidth]{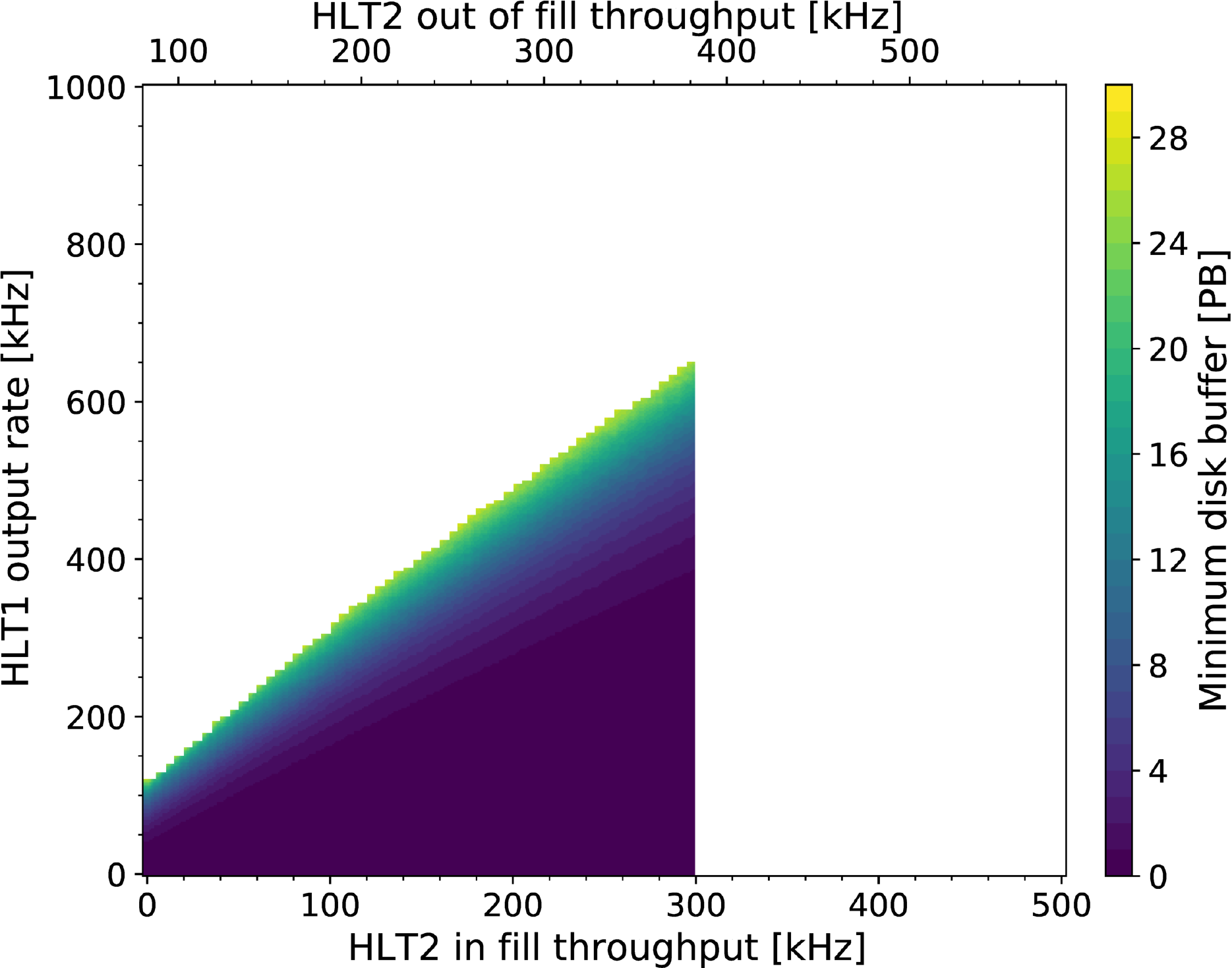}
    \includegraphics[width=0.45\textwidth]{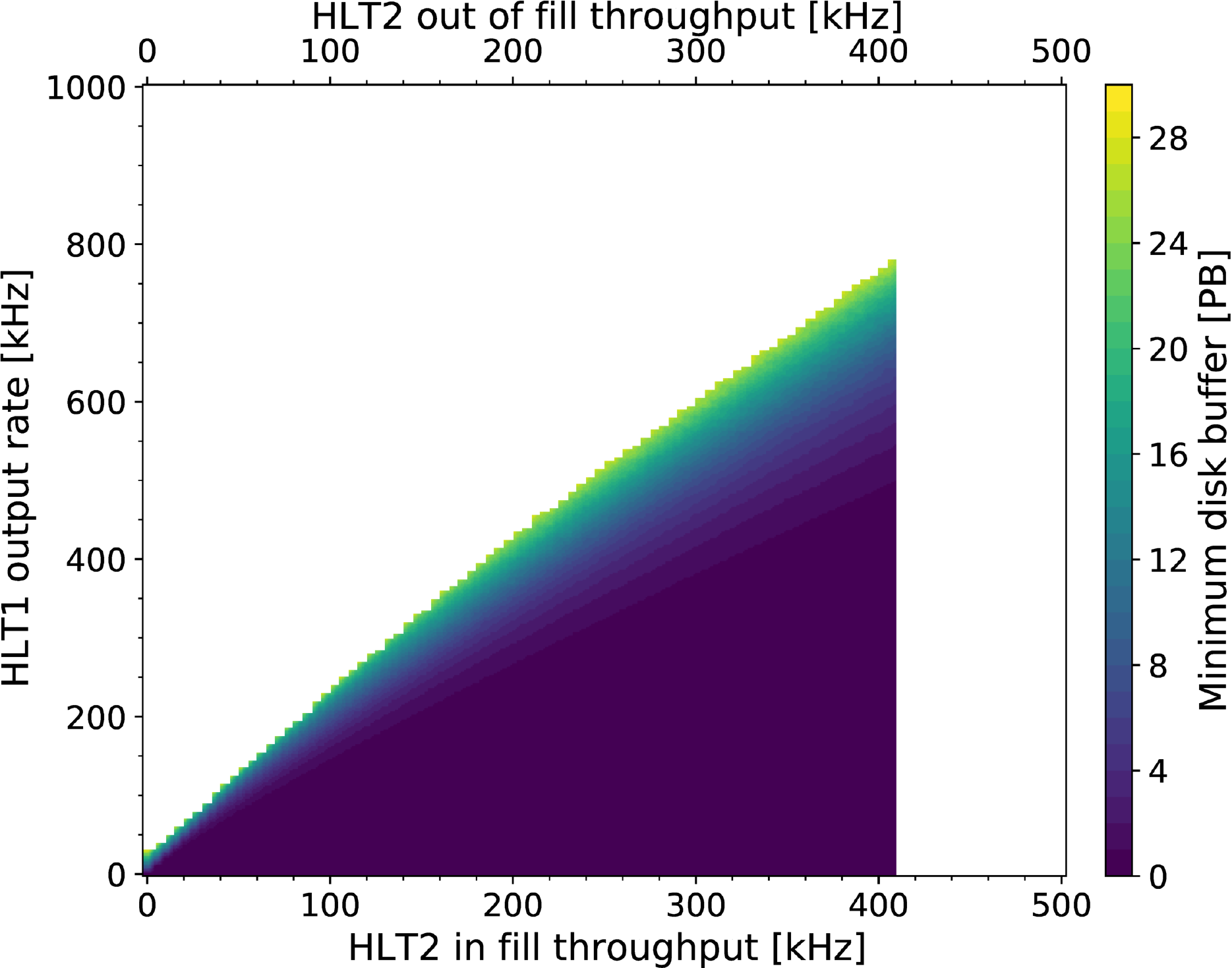}
    \caption{Region of affordability for the CPU-only scenario (left)  and the hybrid scenario (right). The maximum affordable HLT1 throughput is 660~kHz and 790~kHz respectively.}
    \label{fig:buffer_currentHLT2_baseline}
\end{figure}

\begin{table}[]
    \centering
    \begin{tabular}{|c|c|} \hline
        Scenario & Maximum HLT1 output rate (kHz) \\\hline
        CPU-only  &  660 \\
        hybrid &  790 \\ \hline
    \end{tabular}
    \caption{Results of the toy studies indicating the maximum affordable throughput in the both scenarios. The disk buffer is used to its maximum of 28~PB. 
    }
    \label{tab:buffer_maxHLT1}
\end{table}
The two scenarios are scanned according to the method described in Sec.~\ref{sec:sub:disk_buffer_sim}. The results are shown in Fig.~\ref{fig:buffer_currentHLT2_baseline}. In these figures, any combination of additional CPU and disk buffer that is cheaper than the total budget for the two scenarios is shown, defining a region of affordability, with the necessary disk buffer capacity represented by the z-axis colour. 
Unsurprisingly the maximum HLT1 throughput that can be sustained arises when the buffer is fully used and the remaining resources are spent exclusively on HLT2. The maximum HLT1 sustainable throughput in these scenarios is provided in Tab.~\ref{tab:buffer_maxHLT1}, corresponding to the maximum y-axis extent of the region of affordability.

\subsubsection{Results assuming a factor two performance increase in HLT2}
\begin{figure}
    \centering
    \includegraphics[width=0.45\textwidth]{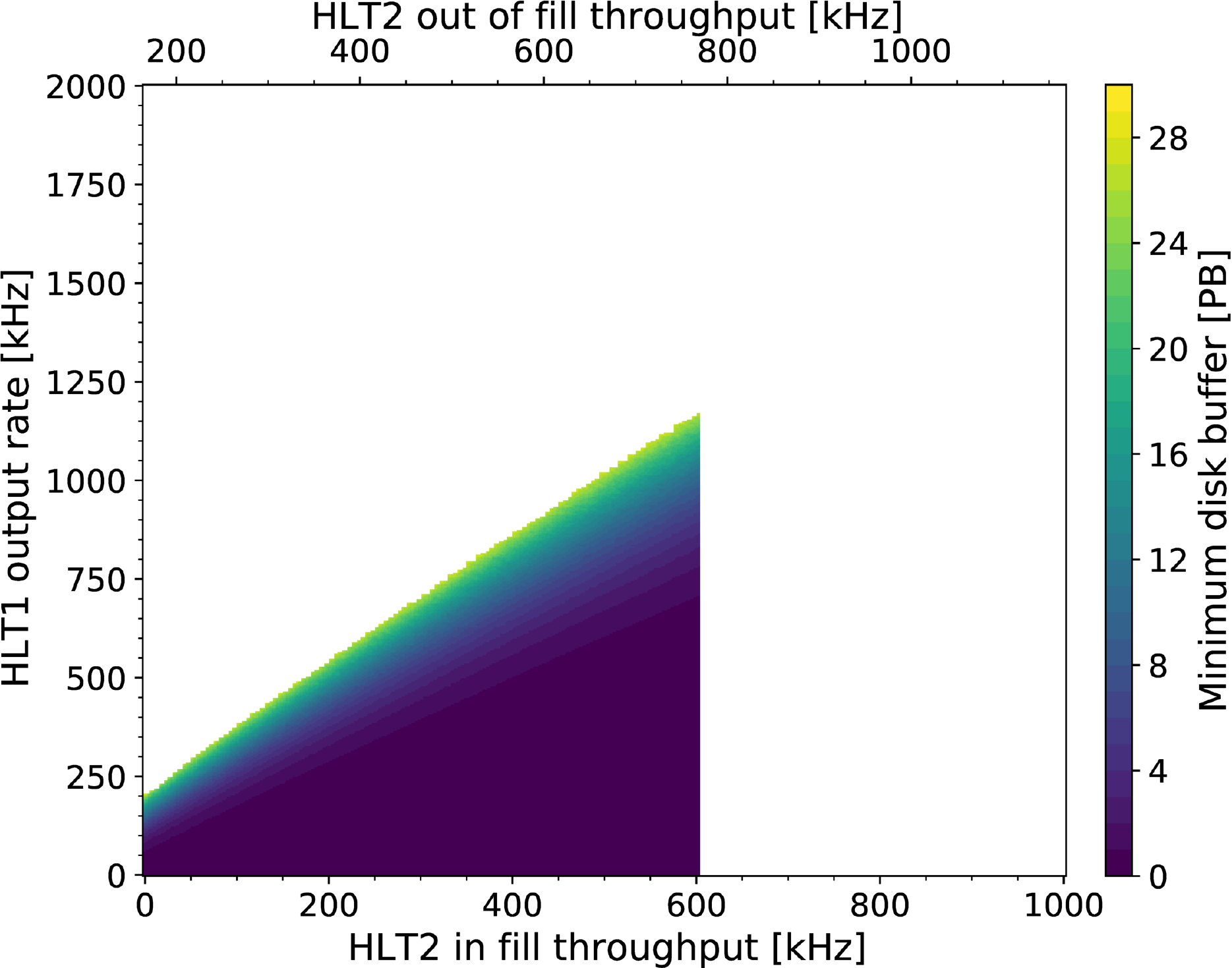}
    \includegraphics[width=0.45\textwidth]{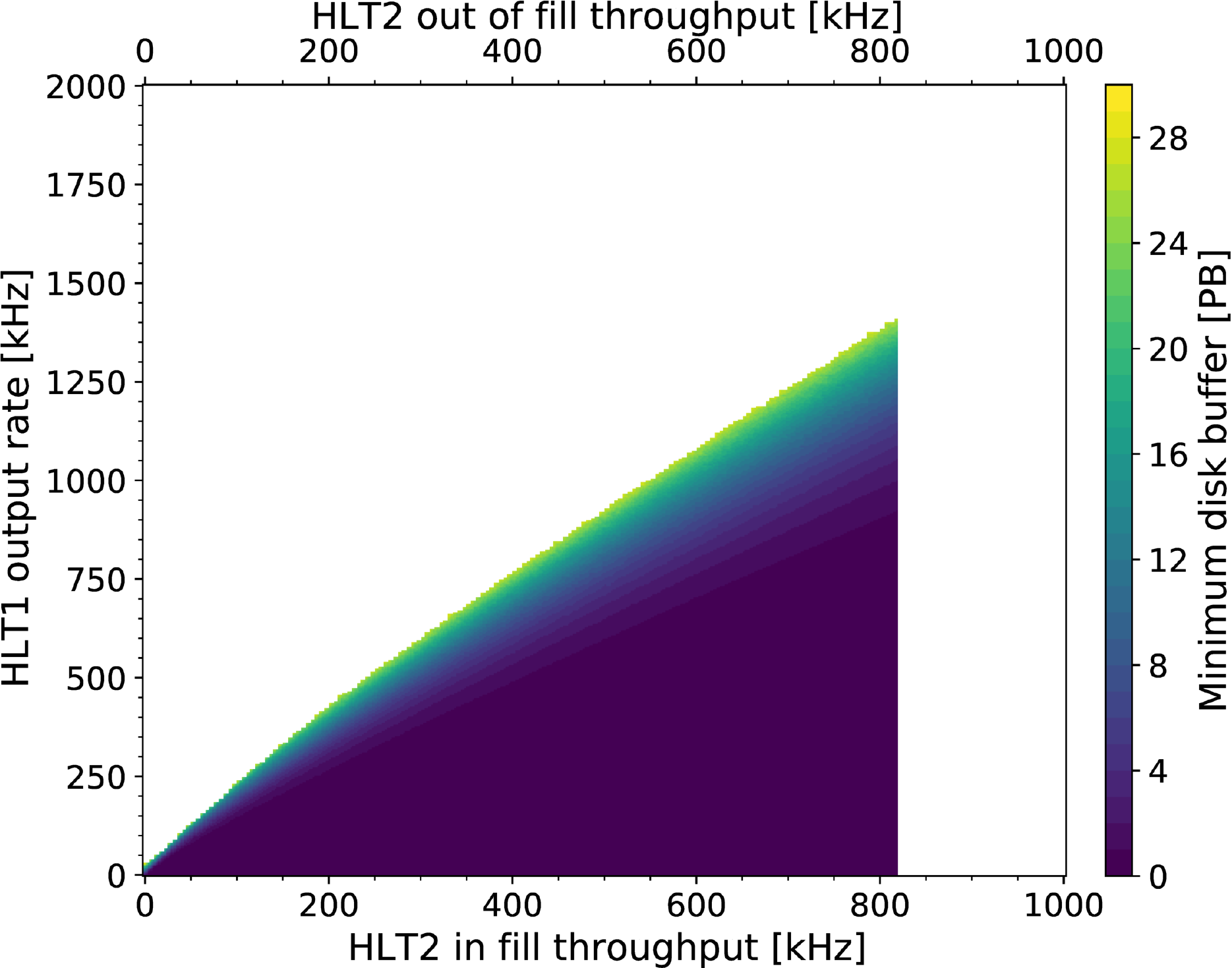}
    \caption{Region of affordability for the CPU-only scenario (left) and the hybrid scenario (right) when HLT2 throughput is two times higher than in the current best performance. The maximum affordable HLT1 throughput is 1185~kHz and 1425~kHz, respectively.}
    \label{fig:buffer_2xHLT2_baseline}
\end{figure}

HLT2 has not been optimised to the same extent as HLT1, therefore it is expected to improve its performance. To highlight the importance of doing so, the previous studies are repeated assuming a factor two improvement in the HLT2 throughput using the same architectures. This is motivated by preliminary studies using optimisations to the reconstruction sequence that are pending validation.\\
\begin{table}[]
    \centering
    \begin{tabular}{|c|c|} \hline
        Scenario & Maximum HLT1 output rate (kHz) \\\hline
        CPU-only &  1185 \\
        hybrid &  1425 \\ \hline
    \end{tabular}
    \caption{Results of the toy studies indicating the maximum affordable throughput in the four scenarios in which HLT2 throughput has increased by a factor of two. In each case the maximum throughput arises when the full remaining spend is used on HLT2 and the disk buffer is used to its maximum of 28~PB respectively.}
    \label{tab:buffer_2xHLT2_maxHLT1}
\end{table}
~\\
The two scenarios with double the HLT2 performance result in the regions of affordability shown in Fig.~\ref{fig:buffer_2xHLT2_baseline}. 
As before, the maximum HLT1 throughput that can be sustained arises when the buffer is fully used and the remaining resources are spent exclusively on HLT2.  The maximum HLT1 sustainable throughput in these scenarios is provided in Tab.~\ref{tab:buffer_2xHLT2_maxHLT1}.

\subsection{Deferred purchasing}
\label{sec:sub:CostBenefitDeferred}
It is expected that the first year of Run 3 will be commissioning year, with stable pp running at lower luminosities than nominal early in the year. This is expected to be followed by a period of running at nominal luminosity. There is general agreement that if possible hardware purchasing should be deferred to make best use of the collaboration's financial resources. For this reason, the plan is to buy a system for 2022 which can handle half the expected nominal processing load. As the throughput of both the considered HLT1 architectures scales linearly (or worse) with detector occupancy, this implies that buying half the number of HLT1 processing units is sufficient.  Many of the relevant costs from Tab.~\ref{tab:cost} can therefore be divided by two.\\
~\\
 In the case of the CPU-only architecture, the disk storage required for 1~MHz output involves a great deal of additional mechanical and cabling work if purchased in two parts, so it is preferable to buy the full amount already for the commissioning year. However it is technically possible to divide this purchase also, in case there are good financial reasons to expect that it saves the collaboration a significant amount of money. 
Similar arguments also apply to the network between 
the EB and the CPU HLT1.\\
~\\
In the hybrid case, deferred purchasing allows HLT1 to be implemented with one GPU card per EB node in the first year. This results in maximum flexibility to benefit from the next generation of GPU cards which are expected to be released over the next two years, and mitigates the risk arising from the limited number of EB slots available to implement the GPU HLT1.

\section{Conclusion}
\label{sec:Conclusion}
LHCb has investigated two viable proposals for its first level trigger to be deployed for Run 3 onwards. They are  characterised by similar performances: The choice  between x86-based and GPU-based technologies is far from obvious when viewed solely in terms of raw throughput and physics performance, and as a result the decision over which option to use has come down to a detailed cost/benefit assessment, presented in this paper.  \\
~\\
In addition to certain immediate cost savings the GPU-based solution offers more opportunities for future performance gains, as a more forward-looking longer-term investment; it is anticipated that GPU performance will advance at a more rapid pace, as evidenced by the new generation of GPU cards released while this paper was being prepared. This increased growth rate presents an opportunity to do more physics in Run 3 and beyond; and provides LHCb with the expertise and increased flexibility to exploit fully heterogeneous computing environments in future upgrades.  The overhead of maintaining different technologies on HLT1 and HLT2 is not negligible and has
been taken into account in the decision making process as well. On balance however, as a result of all these considerations, the collaboration has ultimately decided to implement a Hybrid HLT with a GPU-based first stage, in Run 3. The studies presented in this paper show that GPUs are increasingly a viable alternative to CPUs as general purpose, quasi-standalone processors.

\section*{Data availability statement}
The datasets generated during and/or analysed during the current study are not publicly available as LHCb collaboration policy is to make data fully available 5 years after it has been analysed. They are available from the corresponding author on reasonable request.

\section*{Conflict of interest statement}
No author of this paper has a financial or personal relationship with a third party whose interests could be positively or negatively influenced by the article’s content.

\section*{Acknowledgements}
This version of the article has been accepted for publication, after peer review, but is not the Version of Record and does not reflect post-acceptance improvements, or any corrections. 
The Version of Record is available online at: \href{https://doi.org/10.1007/s41781-021-00070-2}{https://doi.org/10.1007/s41781-021-00070-2}.
\\

\noindent We
thank the technical and administrative staff at the LHCb
institutes.
We acknowledge support from CERN and from the national agencies:
CAPES, CNPq, FAPERJ and FINEP (Brazil); 
MOST and NSFC (China); 
CNRS/IN2P3 (France); 
BMBF, DFG and MPG (Germany); 
INFN (Italy); 
NWO (Netherlands); 
MEiN and NCN UMO-2018/31/B/ST2/03998 (Poland);
MEN/IFA (Romania); 
MSHE (Russia); 
MICINN (Spain); 
SNSF and SER (Switzerland); 
NASU (Ukraine); 
STFC (United Kingdom); 
DOE NP and NSF (USA).
We acknowledge the computing resources that are provided by CERN, IN2P3
(France), KIT and DESY (Germany), INFN (Italy), SURF (Netherlands),
PIC (Spain), GridPP (United Kingdom), RRCKI and Yandex
LLC (Russia), CSCS (Switzerland), IFIN-HH (Romania), CBPF (Brazil),
PL-GRID (Poland) and NERSC (USA).
We are indebted to the communities behind the multiple open-source
software packages on which we depend.
Individual groups or members have received support from
ARC and ARDC (Australia);
AvH Foundation (Germany);
EPLANET, Marie Sk\l{}odowska-Curie Actions and ERC (European Union);
A*MIDEX, ANR, IPhU and Labex P2IO, and R\'{e}gion Auvergne-Rh\^{o}ne-Alpes (France);
Key Research Program of Frontier Sciences of CAS, CAS PIFI, CAS CCEPP, 
Fundamental Research Funds for the Central Universities, 
and Sci. \& Tech. Program of Guangzhou (China);
RFBR, RSF and Yandex LLC (Russia);
GVA, XuntaGal and GENCAT (Spain);
the Leverhulme Trust, the Royal Society
 and UKRI (United Kingdom).
The authors would like to thank the LHCb computing and simulation teams for their support and for producing the simulated LHCb samples used to benchmark the performance of RTA software.

\addcontentsline{toc}{section}{References}
\bibliographystyle{LHCb}
\bibliography{main,standard,LHCb-PAPER,LHCb-CONF,LHCb-DP,LHCb-TDR}
\end{document}